%% file: ms_arXiv_revised2.tex
\documentclass[times,twocolumn]{aastex63}

\usepackage[figure,figure*]{hypcap}

\usepackage{natbib}

\usepackage{amsmath}

\usepackage{graphics,graphicx}
\usepackage{epsfig}

\newcommand\olivine{\mbox{[Mg$_y$,Fe$_{1-y}$]$_2$SiO$_4$}}
\newcommand\pyroxene{\mbox{[Mg$_x$,Fe$_{1-x}$]SiO$_3$}}

\newcommand\fcryst{\mbox{$f_{cryst}$}}

\newcommand{\chisq}{$\chi^2$}
\newcommand{\chisqnu}{$\chi_\nu^2$}

\newcommand{\ltsimeq}{\raisebox{-0.6ex}{$\,\stackrel
	{\raisebox{-.2ex}{$\textstyle <$}}{\sim}\,$}}
\newcommand{\gtsimeq}{\raisebox{-0.6ex}{$\,\stackrel
	{\raisebox{-.2ex}{$\textstyle >$}}{\sim}\,$}}

\usepackage{color,soul}

\usepackage{datetime2}

\shorttitle{Comets Observed by Spitzer}
\shortauthors{Harker, D. E. et al.}

\begin{document}
\title{Dust Properties of Comets Observed by Spitzer}

\author[0000-0001-6397-9082]{David E.\ Harker}
\affiliation{Center for Astrophysics and Space Sciences, University of
California, San Diego, 9500 Gilman Drive, La Jolla, CA 92093-0424, USA}

\author[0000-0002-1888-7258]{Diane H. Wooden}
\affiliation{NASA Ames Research Center, Space
Science Division, MS~245-1, Moffett Field, CA 94035-1000, USA}

\author[0000-0002-6702-7676]{Michael S.~P.\ Kelley}
\affiliation{University of Maryland, Department of Astronomy,
4254 Stadium Dr, College Park, MD 20742-2421, USA}

\author[0000-0001-6567-627X]{Charles E.\ Woodward}
\affiliation{Minnesota Institute for Astrophysics, School of
Physics and Astronomy, 116 Church Street, S.E., University of
Minnesota, Minneapolis, MN 55455, USA}

\correspondingauthor{Charles E. Woodward}
\email{mailto:chickw024@gmail.com}
\received{2023-06-12}
\revised{2023-09-24}
\accepted{2023-09-26}
\submitjournal{To Appear in The Planetary Science Journal} 

\begin{abstract}
As comets journey into the inner solar system, they deliver particulates and volatile 
gases into their comae that reveal the most primitive materials in the solar system. Cometary 
dust particles provide crucial information for assessing the physico-chemical conditions in the 
outer disk from which they formed. Compared to the volatiles and soluble organics, the 
refractory dust particles are more robust and may be traceable to other small bodies. Using data from the
Spitzer Heritage Archive, we present thermal dust models of 57 observations of 33 comets observed 
spectroscopically with the NASA Spitzer Space Telescope. This comet 
spectral survey offers the opportunity to study comets with data from the same instrument, reduced 
by the same methods, and fitted by the same thermal model using the same optical constants. 
The submicron dust tends to be dominated by amorphous carbon, and
the submicron silicate mass tends to be dominated by amorphous
silicate materials. We discuss the implications of these findings as they relate to 
Mg-rich crystalline silicates, which are high-temperature condensates, as well as to 
potential ion irradiation of amorphous Mg:Fe silicates prior to their incorporation into 
comets. These results impact our understandings of the protoplanetary disk 
conditions of planetesimal formation. Lastly, we cannot definitively conclude that a distinct 
difference exists in the dust composition between Oort cloud and Jupiter-family comet dynamical
population as a whole. 
\end{abstract}


\section{Introduction}
\label{sec-Intro}

Solar system formation is an engine that simultaneously preserves and transforms interstellar
(ISM) dust grains into planetesimals and, ultimately, planets.  Being a mixture of ISM dust 
and solar nebula processed material, comets allow us to investigate both the inputs 
and outputs of dust transformation in the young solar system.  Dust grains in the 
ISM are dominated by ``amorphous'' silicates, i.e., they have chemical composition (stoichiometry) 
similar to pyroxene (\pyroxene) and olivine (\olivine).  Studies of silicates in the galaxy measured 
the silicate crystalline fraction to be  $f_{\mathrm{cryst}}\leq5\%$ by mass in the ISM
\citep{ {2004ApJ...609..826K}}. In contrast, dust ejected by comets, such as C/1995~O1 
(Hale-Bopp) and 9P/Tempel~1, have significant crystalline mass fractions, both near 30\% 
\citep{ 2007Icar..190..432H, 2005Sci...310..278H, 2021PSJ.....2...25W}.
Moreover, laboratory studies of GEMS (glass with embedded metal and sulfides),  an abundant 
amorphous silicate material in interplanetary dust particles thought to be derived from 
comets, have isotopic  composition indicative of formation the protosolar
nebula \citep{2011GeCoA..75.5336K}. They also have elemental compositions and 
lower densities that make them candidates of processed ISM silicates \citep{2022GeCoA.335..323B}. 
In aggregate, observations of ISM and of comets strongly suggest that 
ISM amorphous silicates were extant in the icy outer-disk as well as were destroyed 
and re-condensed as Mg-rich crystals in the inner protoplanetary disk, or annealed, prior 
to accretion of comets in the icy outer-disk. Thus, determination of comet dust 
characteristics enables deeper understanding of  a)~how dust was processed in the inner-solar 
system, and b)~how that dust was transported to the outer-solar system; processes that are 
apparently ubiquitous in observations of external protoplanetary 
disks \citep[e.g.,][]{ 2010A&A...520A..39O, 2023ApJ...945L...7K}.

Common taxonomy schemes are based on the abundances of primary volatiles 
\citep[water, CO$_{2}$, methanol, ethane, etc.;][]{2011LPI....42.2428M}, molecular daughter 
species \citep[C$_{2}$, CN, etc.;][]{1995Icar..118..223A, 2016Icar..278..301D, 2017RSPTA.37560252B, 2021AJ....162...74L},  
or the physical properties of molecules \citep[isotopic ratios, atomic spin states;][]{2009E&PSL.288...44B, 2007ApJ...661L..97B}. 
However, a dust based taxonomy utilizes species that are resistant to chemical alteration 
up to at least the melting point of water, and complements studies of the more volatiles species.
Development of a comet dust taxonomy classification could provide a powerful 
constraint for solar system formation scenarios more robust than any conclusion based 
on derived inferences from a single taxonomy in isolation. For example, identification of a 
relationship between organic D-to-H ratios and silicate dust crystallinity ($f_\mathrm{cryst}$) may 
delineate the location and epoch in the disk evolution during which different comet nuclei 
formed \citep{2009E&PSL.288...44B}.

In this paper, we investigate the dust characteristics of comets uniformly observed 
with the same platform, the NASA Spitzer Space Telescope 
\citep[hereafter Spitzer,][]{2004ApJS..154....1W}, with the goal to assess whether 
classification of comets into dynamical families is possible using dust taxonomy. A companion
manuscript explores deeper nuances of the statistical, machine-learning methodology 
and interpretations. This spectroscopic survey of comets presented herein serves
as a guide to potential JWST studies, although the total number of JWST-observed 
comets to date does not yet eclipse those observed by Spitzer.

\section{Archival selection} 
\label{sec-obs_archive}

Grain characteristics of dust in comet comae can be determined from moderate resolution 
IR spectroscopy covering wavelength ranges from 8.0 to 28~\micron{} where 
broad and narrow band features from mineral solid state resonances 
occur \citep{2023arXiv230503417E, 2017RSPTA.37560260W}. These materials are lofted 
into the coma from nuclei surface regions through sublimation, volatile out 
gassing, and other processes. Within the Spitzer Heritage 
Archive\footnote{\url{https://irsa.ipac.caltech.edu/Missions/spitzer.html}} lies a 
substantial set of comet observations from which one can construct a sample of objects to study.


Comets selected for analysis in this manuscript were observed with the Spitzer Infrared 
Spectrograph \citep[IRS;][]{2004ApJS..154...18H} using the short-wavelength low-resolution 
(SL) and long-wavelength low-resolution (LL) modules ($\Delta\lambda/\lambda\sim$64--128).  
The short-low modules cover 5.1--14.3~\micron, split between two separately observed 
spectral regions named SL2 (covering 5.1--7.6~\micron) and SL1 (covering 7.4--14.3~\micron).  
The long-low modules cover 13.9--39.9~\micron, similarly named LL2 
(covering 13.9--21.3~\micron), and LL1 (covering 19.9--39.9~\micron).  Each spectral 
region is observed separately through its own entrance slit: 3\farcs6$\times$17\arcsec{} 
and 3\farcs7$\times$17\arcsec{} for SL2 and SL1, respectively, and 
10.5\arcsec$\times$52\arcsec{} and 10.7\arcsec$\times$52\arcsec{} for LL2 and LL1, respectively.

We use the work of \citet{kelley-pds2021} as the basis for our analysis.
They presented a uniform reduction of 57 observations of 33 comets. 
Their methodology is detailed in the data set documentation,
as described in the NASA PDS Small Bodies Node.\footnote{ \url{https://pdssbn.astro.umd.edu/holdings/pds4-spitzer:spitzer-spec-comet-v1.0/SUPPORT/dataset.shtml.} }   
In summary, observations were: (a)~background subtracted; (b)~spectra extracted from 
individual exposures using constant width aperture sizes; (c)~exposures were combined 
with outlier rejection; (d) spectra were cleaned for bad data (e.g., radiation hits, and 
the scattered light artifact at 14-\micron); (e)~a model nucleus subtracted when possible; 
(f)~spectral order-to-order discontinuities removed by scaling orders as needed; and 
(g)~a 1.8\% relative calibration uncertainty was added in quadrature.  Three aspects of 
the reduction present potential issues for spectral analysis: (1)~corrections for spectral 
fringing were not attempted for these data, and fringing will be most apparent at 21 to 25~\micron; 
(2)~the edges of some spectral orders were trimmed to avoid potentially unphysical features 
in regions most likely to be affected by data calibration or background subtraction issues, and 
the trimmed data tends to vary by observation; (3)~as a consequence of the spectral trimming, 
the order-to-order scaling near 19 to 22~\micron{} could potentially remove or enhance real 
spectral structure present in cometary data.  Details on these points and all aspects of the 
data selection and reduction are given in the archive documentation \citep{kelley-pds2021}.  

We have significantly improved reductions for four comets.  Here, comets 17P/Holmes and
C/2003 K4 on 2005 September 13 have improved background removals, and comet C/2003 T4 on 
2006 January 29 has an improved SL to LL scaling. The archive will be updated accordingly 
(Kelley et al.~in prep.). Table~\ref{tab:obssum} lists the observational summary of each 
spectrum analyzed in this work including the comet name, UT date, the heliocentric ($r_{h}$) 
and geocentric ($\Delta$) distances, and phase angle of the observations.  Not every comet 
observation in our sample covers both the short-low (SL) and long-low (LL) modules.  
In Section~\ref{sec:model_fits}, we discuss the model fits in relation to the different wavelength 
coverage between spectra of a given comet and the ensemble of the survey comets. \pagebreak

\input{table_1}


\section{Thermal Dust Model}
\label{sec:models}

Derivations of comae dust properties are achieved by computing and fitting thermal models 
to observed IR spectral energy distributions
\citep[SEDs,][see sections 2.5.1-2.5.5]{2007Icar..190..432H, 2011AJ....141...26H, 2023arXiv230503417E}. 
Dust properties include the relative mass fractions of five primary dust compositions: amorphous 
Mg:Fe olivine (ao50), amorphous Mg:Fe pyroxene (ap50), Mg-rich crystalline olivine (forsterite, co), 
Mg-rich crystalline pyroxene (enstatite, cp), and amorphous carbon (ac), as well as the 
particle differential size distribution parameters and the particle porosity
\citep{2002ApJ...580..579H, 2004ApJ...615.1081H, 2004ApJ...612L..77W, 2007Icar..190..432H, 2011AJ....141...26H, 2021PSJ.....2...25W}.

\subsection{Dust Composition}
\label{sec:dust_comp}
The dust composition used in thermal models is derived from laboratory studies of interplanetary dust 
particles \citep[IDPs,][]{2000Icar..143..126W}, micrometeorites \citep{1999Sci...285.1716B}, 
the NASA Stardust mission \citep{2006Sci...314.1711B} and other remote sensing 
modeling efforts \citep[e.g.,][]{1994ApJ...425..274H, 2002ApJ...580..579H, 2004ApJ...612L..77W}. 
Our selection of basic dust grain components is consistent with the major mineral groups used by other
modelers to generate synthetic SEDs arising from dust thermal emission in
the 10~\micron\ region \citep{, 2005Icar..179..158M, 2005Sci...310..274S, 2006Sci...313..635L, 2006ARA&A..44..269W}.
Amorphous silicates with chemical composition (stiochiometry) similar to
olivine (Mg$_{y}$,Fe$_{(1-y)}$)$_{2}$SiO$_{4}$ and pyroxene
(Mg$_{x}$,Fe$_{(1-x)}$)SiO$_{3}$ with $x = y = 0.5$ (i.e., Mg/(Mg+Fe) = 0.5)
reproduce the broad width of the 10~\micron{} feature.  Mg-rich orthopyroxene is 
detected through its 9.3 and 10.5~\micron{} features \citep{, 1999ApJ...517.1034W, 2002ApJ...580..579H}. 
Mg-rich crystalline olivine is uniquely identified through its distinct, relatively
narrow 11.2~\micron{} silicate feature \citep{1994ApJ...425..274H}. Mg-rich crystalline
species are defined as grains with a stoichometry of $0.9~\le~x~\ltsimeq~1.0$
(Mg/(Mg+Fe) = 0.9) \citep{2002A&A...391..267C, 2003A&A...399.1101K, 2008SSRv..138...75W, 2017RSPTA.37560260W}.

The model does not include grain species that have either not been well constrained through 
remote sensing, or are found only through laboratory analyses.  Several species thought to 
possibly exist within comets do not have detected spectral signatures in the mid-IR region in order
to be adequately constrained through thermal emission modeling.  Such species are found in 
significant amounts through the laboratory analysis of Stardust return samples \citep{2014AREPS..42..179B},
interplanetary dust particles (IDPs) \citep{ 1994aidp.work...54Z, 2017M&PS...52..471B}, 
and ultra-carbonaceous micro-meteorites (UCAMMs) \citep{2012GeCoA..76...68D}. They
include metal sulfides (e.g., submicron sized FeS grains which produce a broad 23~\micron{} 
feature), refractory organics found in cometary IDPs \citep{2005A&A...433..979M} with features 
between 3.3--3.6~\micron \, and Fe-rich crystalline olivine (Fe/(Mg+Fe)$>$90 to Fe/(Mg+Fe)$\approx$50) 
which produce a distinct feature at 11.35--11.4~\micron{} as well as features in the mid-infrared, all of 
which are shifted to wavelengths longer than those produced by Mg-rich crystalline olivine.  

A similar argument is made for species detected in trace amounts in laboratory spectra, 
including Mg-carbonates, PAHs, and Fe metal:~spectral features from Mg-carbonates and 
polycyclic aromatic hydrocarbons (PAHs) have not been detected incontrovertibly  
and Fe metal is featureless.  Lastly, phyllosilicates are not detected in laboratory 
examinations of Stardust return samples and their stronger broad far-infrared 
features are not centered at the wavelengths of observed resonances.

Comparing various techniques to thermal emission modeling of cometary IR spectral 
energy distributions (SEDs) shows that different approaches applied to the 
same comet yield the same composition, i.e., the same relative abundances of 
the primary grain components, as long as the same range of particle sizes are 
employed. Small grains tend to dominate the resonant features and the short wavelength 
spectral flux whereas the mass is dominated by the larger grains.  Generally, a size
distribution from 0.1--100~\micron{} is considered because grains larger than 100~\micron{} 
are presumed to not contribute significantly to Spitzer IRS spectra. However, this interpretation 
is a consequence of the use of Mie theory in the model.

Recent observations of comet 67P/Churyumov-Gersimenko with 
the Rosetta spacecraft show a phase effect in the thermal emission from dust indicating 
that large dust grains can support a thermal gradient \citep{2019A&A...630A..22B}, and may 
be hotter and thereby brighter, than the isothermal temperatures assumed in the model.
The relative abundances of grain components derived from thermal emission models of 
cometary comae, which compute composition-dependent radiative equilibrium temperatures, 
yield similar compositions for the same comet for the major species that include olivine and 
pyroxene, and a dark absorbing carbonaceous grain material. However, challenges in 
outcomes and interpretations arise when inter-comparing models that use considerably 
different maximum grain sizes in their size distributions, $n(a)da$, or when grain temperatures 
are modified by highly absorptive coatings. The relative abundances of the grain components depend on 
their relative temperatures, and in turn, on the optical constants employed in the calculations. 
Our methods and optical constants are carefully documented so that our results can be clearly 
compared to those of other  investigators.

\subsection{Absorption Efficiency}
A grain's composition, mineral structure (crystalline or amorphous), porosity, and radius 
determine its absorption and emission efficiency, $Q_{abs}$  (a grain's absorption efficiency 
and emission efficiency are equivalent at any given wavelength by Kirchoff's Law). 
The model assumes individual porous grains having a single mineral composition 
(i.e., no mixtures of carbon and silicate). Mie theory combined with effective medium 
approximation is sufficient to calculate $Q_{abs}$ for amorphous porous 
grains \citep[Bruggeman mixing formula;][]{1983asls.book.....B}.  
Amorphous porous grains are modeled with an increasing vacuous content as expected 
for hierarchical aggregation, using the fractal porosity prescription or fractional filled volume given by
$f = 1 - (a/0.1~\mu m)^{D-3}$ with the fractal dimension parameter $D$ ranging
from $D=3$ (solid) to $D=2.5$ \citep[fractal porous but still spherical enough
to be within the applicability of Mie scattering computations;][]{2002ApJ...580..579H}.
Grain porosity is a critical parameter that uniquely affects the observed spectra from comets
\citep{2008SSRv..138...75W, 2013ApJ...766...54L, 2019A&A...630A..24G}. 
A large porous grain and a small compact grain of the same composition can have
the same temperature. A size distribution of porous grains compared to compact
grains produces enhanced emission at longer wavelengths while producing the
silicate emission features. Hence, the shape of the observed spectrum
constrains the porosity and slope of the grain size distribution of amorphous
grains.

Crystals, however, are not well modeled by Mie Theory nor by mixing theory
because of their anisotropic optical constants and irregular shapes. Discrete solid crystals 
from 0.1--1~\micron{} in radius are considered in the coma, since crystals of larger 
sizes do not fit the observed SEDs.  Discrete solid crystals are better computed using 
the continuous distribution of ellipsoids (CDE) approach \citep{2007Icar..190..432H} or discrete dipole
approximation \citep[DDA;][]{2013ApJ...766...54L}. CDE reasonably reproduces laboratory spectra of 
crystalline powders and is a good starting point for our models \citep{2001A&A...378..228F}.

\subsection{Grain-size Distribution}
For each grain, its flux density emitted at the heliocentric distance ($r_{h}$[au]) of the comet
at the time of observation is the product of $Q_{abs}$ with a Planck function evaluated at 
the radiative equilibrium temperature $T_{d}$ of the dust grain, $B_\lambda (T_{d})$. The 
total flux from a particular mineral species is calculated by integrating over the grain size
distribution.  The differential grain size distribution we use in the model is the Hanner 
modified power law \citep[hereafter HGSD;][]{1994ApJ...425..274H} in which the relative number 
of each grain {\it size} is computed from: $n(a) = (1 - a_{\circ}/a)^{M}$ $(a_{\circ}/a)^{N}$, 
where $a$ is the grain radius, $a_{\circ} = 0.1$~\micron, the minimum grain radius, and 
$M$ and $N$ are independent parameters.  $N$ is considered the differential size 
distribution slope  for larger radii sized grains ($\gtsimeq 10$~\micron) and is better 
constrained at wavelengths longer than 13~\micron.  $M$ is used with $N$
to calculated the peak of the HGSD; $a_{peak} = a_{\circ} (M + N)/N.$

Finally, correcting for the distance between the observer and the comet, a linear combination 
of discrete mineral components, porous amorphous materials and solid crystals, is 
least-squares fit to the observed SED. To manage the extent of parameters 
in our thermal model grids, we choose five values for $D$ ranging from solid to
porous (3.0, 2.857, 2.727, 2.609, and 2.5), eight values for $N$ ranging from a broad
to narrow grain size distribution ($3.4-4.2$ in increments of 0.1), and increment $M$ such that
$a_{p}$ begins at 0.1~\micron{} and is incremented by 0.1~\micron. The relative mass fractions, 
their correlated errors, the porosity, and the size distribution are given as a prescription for 
the composition of coma grains. We quote the relative masses of the small grain portion 
of the size distribution (0.1--1~\micron) that includes discrete crystals, which is a standard 
practice in the literature.

\subsection{Model-fitting and Monte Carlo Uncertainties}
The model is fit to the observed spectra using a least-squares method.
If observed, emission from the $\nu_2$ water band around 6~\micron{}
is masked out of the model fit.  In addition, an emissivity feature is
seen at 13.5--15.7~\micron{} (covered by the LL2 module) that appears
to be emission from a not yet identified material.  To test whether
the 14~\micron{} feature should be masked or not during the modeling,
we applied the Akaike Information Criteria 
(Section~\ref{sec:waves_SLLL_vs_SLLLdashSL_vs_SL}, Appendix~\ref{sec:AIC-Appendix}).
Thermal models affected by this feature are 78P, 88P, C2003K4-20050913,
C2003T4-20051122, and C2006P1-20070802.

Model compositional uncertainties are estimated with a Monte Carlo
(bootstrap) analysis.  For each spectrum, 10,000 alternate
realizations are generated by adding noise to each data point, picked
from a normally distributed variate with a width equal to the
estimated spectral uncertainty.  Each spectrum is re-fit, holding $N$,
$a_p$, and $D$ constant, but allowing the relative mass fraction of each dust
component to vary.  The final compositional uncertainties are the
inner 95\% confidence intervals for each parameter.

\section{Results and discussion}
\label{sec-results}

\subsection{Silicate feature shapes}

We first present an empirical analysis of the data set, before later employing the dust 
thermal emission model to interpret spectral features.  First, the 7 to 13~\micron{} region 
of each spectrum was examined for evidence of common dust emission features, especially the broad 
amorphous silicate band and narrow emission features from Mg-rich olivine.  For each spectrum, 
a scaled Planck function is fit to the pseudo-continuum defined as spanning
$\lambda$ = 7.3--7.7~\micron{} and 12.3--13.0~\micron.  These pseudo-continuum points 
were chosen through experimentation on the data, but are similar to those used by other comet
modelers \citep[e.g.,][]{2004ApJ...612..576S}.  In particular, the shape of the best-quality 
Spitzer spectra suggests the blue edge of the 10-\micron{} feature is near 7.5~\micron{} for most 
comets.  Such a minimum would be difficult to observe at a ground-based facility due to 
strong telluric methane absorption around $\simeq 7.64$~\micron. The silicate strength
and local color temperature are summarized in Table~\ref{tab:colex}.

The ability to extract the shape of the 10-\micron{} band is primarily affected by the S/N 
ratio near 7.5~\micron{} and the relative strength of the band above the local continuum, i.e., 
comet spectra with band strengths $\lesssim5\%$, perhaps due to large grain sizes, 
and/or low silicate-to-amporhous carbon ratios tend to lack 
sufficient S/N for a band shape analysis.  Therefore, our examination of the silicate feature 
strengths may represent a biased subset of the comet population.  The 
analysis is limited to a subset of the survey comets only including:  9P, 17P, 
21P, 37P, 46P, 67P, 71P, 73P-B, 73P-C, 78P, C/2001~Q4, C/2003~K4, C/2006~P1, 
C/2007~N3, and C/2008~T2. 

\input{table_2}

\begin{figure}[!h]
\figurenum{1}
\begin{center}
\includegraphics[width=0.45\textwidth]{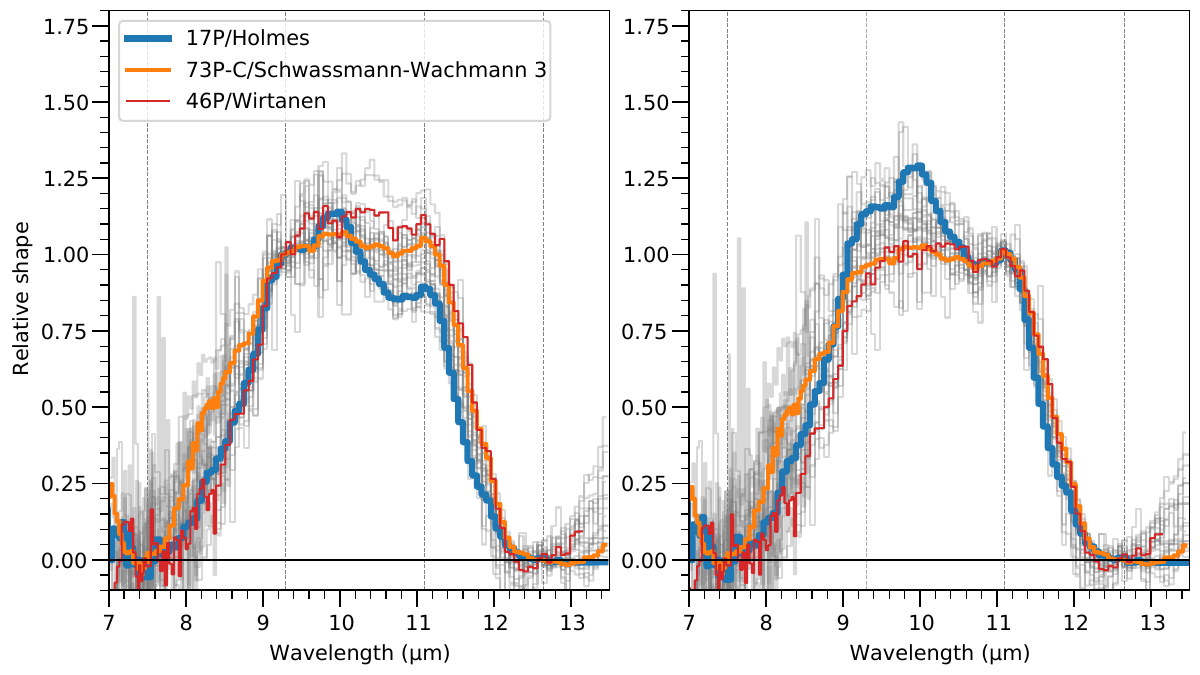}
\caption{Comparison of 10-\micron{} silicate emission feature shapes: (left)
the feature is normalized to 1.0 at 9.3~\micron{}, (right) the feature
is normalized to 1.0 at 11.1~\micron{}, regions devoid of sharp resonances. All spectra with well-defined
silicate bands are shown as thin gray lines.  Comets 9P/Tempel 1,
17P/Holmes, and 73P/Schwassmann-Wachmann 3-C are highlighted with
thick lines of varying width and color.  The complete figure set
highlighting each spectrum individually (26 spectra in all), including uncertainties, is
available in the online journal.} 
\label{fig:band-shape}
\end{center}
\end{figure}

Each spectrum was normalized by a best-fit Planck function, then offset and scaled so 
that the continuum was set to 0, and the band strength at 9.3~\micron{} scaled to 1.0. 
Figure~\ref{fig:band-shape} shows an example this band-shape analysis and the range of band-shapes 
that are possible.  Comet spectra in the 10~\micron{} region exhibit a range of silicate 
feature shapes. No single feature is representative of the whole sample ensemble, 
corroborating conclusions reached by \citet{1994ApJ...425..274H} and \citet{2004ApJ...612..576S}. 
The spectra of three comets serve to general categorize 10~\micron{} spectral shapes: 
46P, with a nearly trapezoidal feature; 17P, with strong narrow ($\sim$0.3~\micron) 
features and a stronger 9-\micron{} shoulder; and, 73P-C, with stronger emission at 8.2~\micron.

The 8.2-\micron{} excess in the spectra of 73P-C and 73P-B was noted by \citet{2017Icar..284..344K}.  
They described the spectra as being more rounded or triangular, as opposed to trapezoidal.  
To further investigate this difference, we have re-normalized the spectra at 11.1~\micron{} 
as Figure~\ref{fig:band-shape}~(right).  Here, rather than 73P-C standing out from 17P and 46P, there is
a progression in the relative 8.2-\micron{} emission strength from 46P to 17P to 73P-C.  
Ortho-pyroxene has strong emission at 9.3~\micron{} and Mg-rich crystalline olivine at 10~\micron;
however, the shape of these features varies between comets. Narrow and triangular shapes
of the 10~\micron{} feature, like those evident in comet 17P, are indicative of a preponderance of
submicron crystalline silicates (which have narrower solid state bands) being present in the coma. 
This is in contrast to comet 73P, which has a flatter, broader feature, the signature of 
amorphous silicates.  

\subsection{Thermal Model Fit Outcomes}
\label{sec:model_fits}

Thermal models (Section~\ref{sec:models}) are fitted to the observed IRS SEDs that have full 
wavelength coverage (SL+LL, hereafter called SLLL) and to those that only have short 
wavelength coverage (SL). In addition, independent thermal models are fitted to the short 
wavelength portion of the full wavelength data (hereafter designated as SLLL--SL), which is instructive 
for investigating systematic differences between thermal model parameters derived from fits 
to data limited to the mid-infrared wavelength range. The sensitivity of wavelength range 
fitting is discussed in detail in Section~\ref{sec:waves_SLLL_vs_SLLLdashSL_vs_SL}.

\begin{figure}[!h]
\figurenum{2}
\begin{center}
\includegraphics[width=0.45\textwidth]{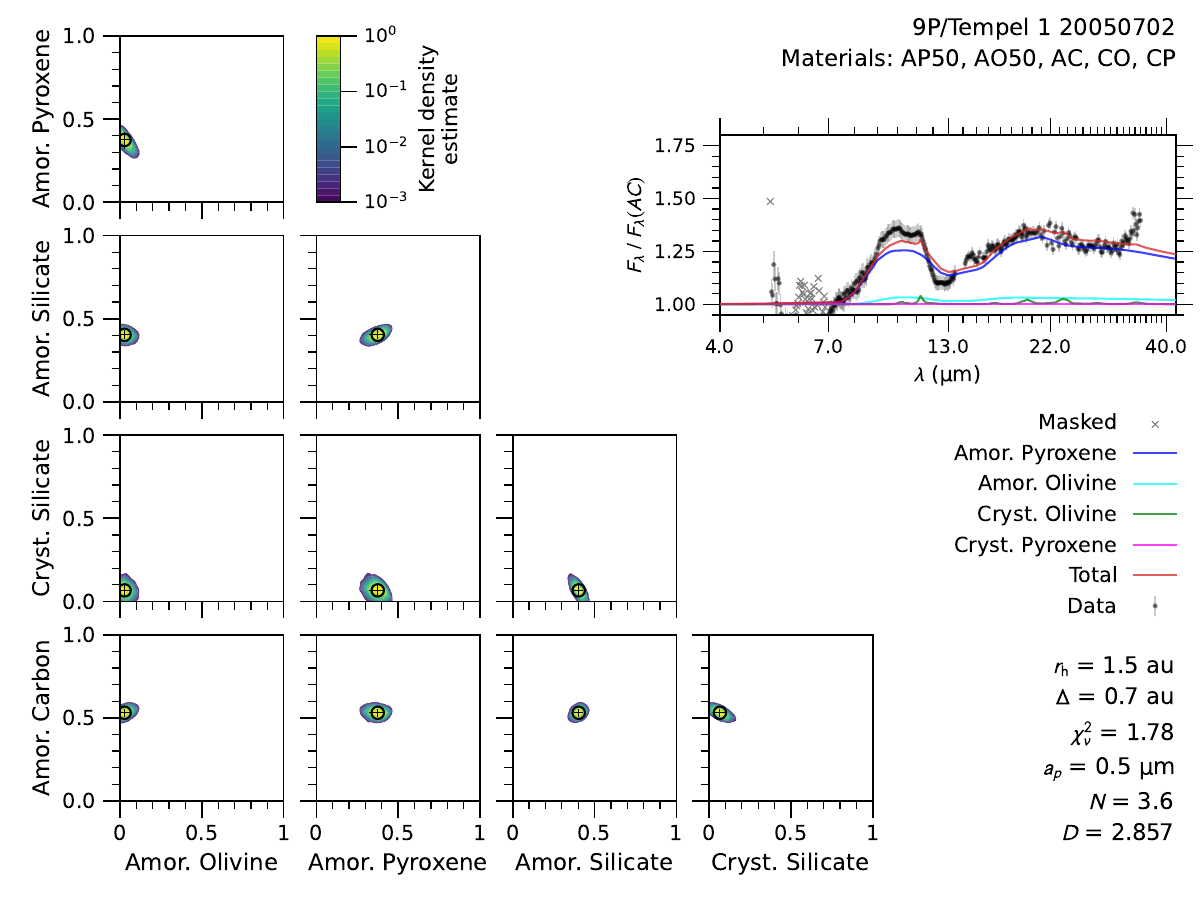}
\caption{Model best-fit and parameter correlations based on our Monte Carlo (bootstrap) spectral fit trials.  In 
the upper right spectrum, the grey filled dots are are the spectral data points used in the fit, while the grey `x' are 
the spectral data that are masked for fitting purposes. The upper-right inset is the 
spectral data normalized by the best-fit amorphous carbon model, $F_\lambda(AC)$.  Each model 
component is also shown normalized by $F_\lambda(AC)$.  The silicate models are offset by +1 for clarity.
The complete figure set (55 images for all comet spectra modeled for the survey) is available in the online journal.
(The data used to create the figure(s) in the upper right are available.) }
\label{fig:correlations}
\end{center}
\end{figure}

The parameters of the thermal model and dust composition derived from analysis of
the spectral energy distribution (SED) for each survey comet and epoch
are summarized in Tables~\ref{tab:modparams} and ~\ref{tab:minmass}. The best-fits 
obtained from the Monte Carlo (Section~\ref{sec:models}) analysis allow us to examine correlations 
between compositional abundances for each spectrum.  In Figure~\ref{fig:correlations}, the 
probability distributions for 10,000 spectral fit trials, plotted versus model composition 
is presented illustrating compositional interdependencies. For completeness, similar tables are 
provided in the appendix (Tables~\ref{tab:tab2_appendix_modparams}, \ref{tab:tab3_appendix_minmass})
for comets observed with SLLL but only modeled over SLLL--SL wavelengths.

Not every model fit is included in the subsequent analysis.  Comets only observed with the
LL modules (67P-20080629.82, 123P, C/2003T4-20060307, C/2004B1-20070609, and 
C/2006Q1-20070322) are not included because the SL1 module 
includes strong and overlapping mineral features from crystalline silicates and amorphous 
silicates that are key for mineralogical identification as well as the fluxes in 10~\micron{}
region being critical to constraining the dust temperatures. In addition, comet 132P was not 
included in the analysis due to its very low S/N which results in an ill-defined model fit. 

\begin{figure}[!ht]
\figurenum{3}
\begin{center}
\includegraphics[width=0.45\textwidth]{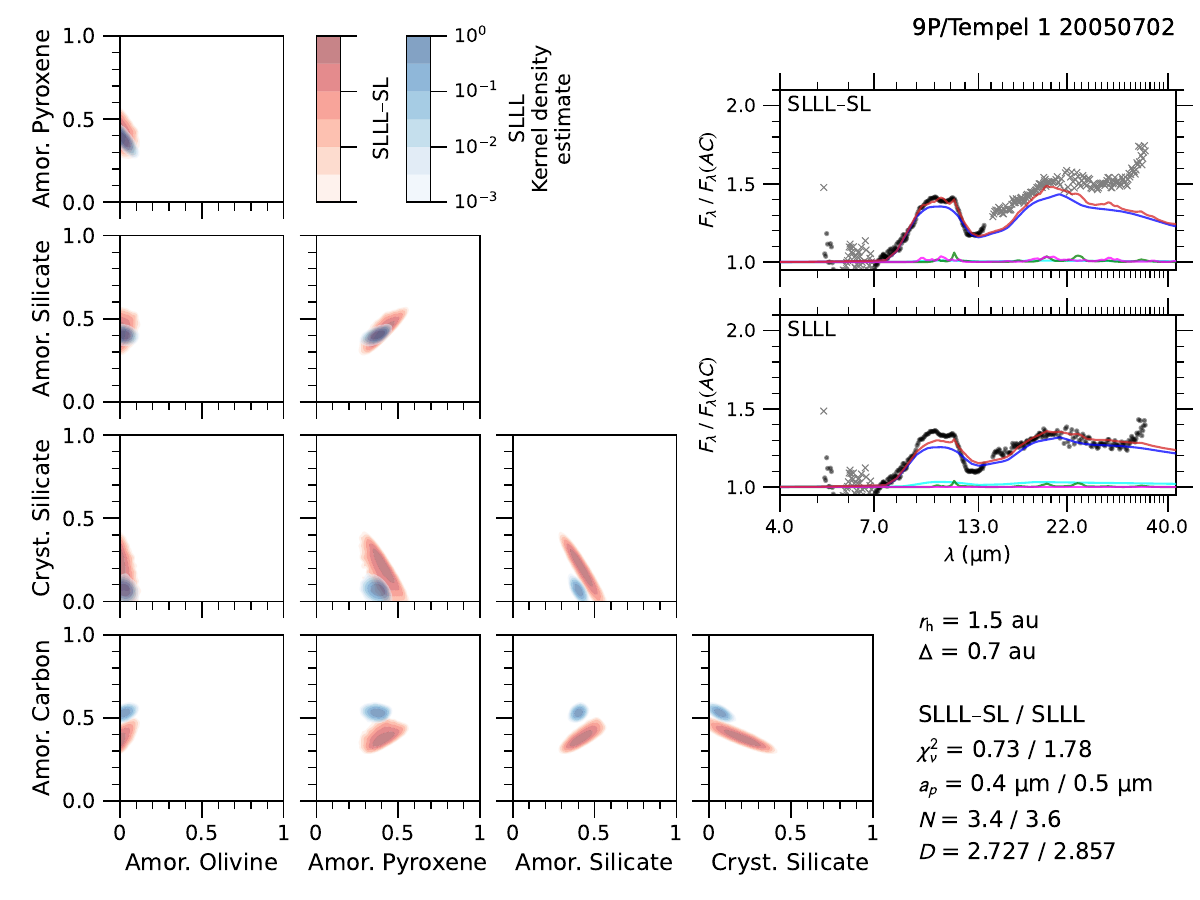}
\caption{Same as Figure~\ref{fig:correlations} but for comparisons of model compositions based on 
fits to the full-wavelength range (SLLL) to those based on fits to a limited wavelength range (SLLL--SL).  
The model results are shown as filled contours based on a kernel density estimate: SLLL--SL in red, 
SLLL in blue.  The contour shading is log-scaled as indicated by the color bars.  In the 
upper right, the spectra and their best-fits are shown after normalization with the amorphous 
carbon model spectrum in order to emphasize spectral features from silicates. In
these panels the grey filled dots are the spectral data points used in the fit, while the grey `x' are 
the spectral data that are masked for fitting purposes. For most comet 
spectra, the SLLL--SL and SLLL fits are generally comparable, with systematic decreases in 
amorphous carbon (AC), as well as anti-correlated changes in crystalline silicates (CP) and 
correlated changes in Amorphous Silicates (AS), see section~\ref{sec:restrict_rh}. 
Generally, SLLL--SL fits have larger 3$\sigma$ ranges of parameters and significantly lower 
AICc values (see text).  The complete figure set (31 images) is available in the online journal.
(The data used to create the figure(s) in the upper right are available.) }
 \label{fig:sl-v-slll}
\end{center}
\end{figure}

Comparisons of the compositions retrieved from the full wavelength range 
(SLLL, $\sim$5--35~\micron) to those retrieved from fits restricted to the 
short wavelengths (SLL--SL, $\sim$5--13~\micron) are shown in Figure~\ref{fig:sl-v-slll}.  
Rarely are the spectral features in the 15--33~\micron{} region of sufficient spectral contrast 
to allow \pagebreak 


 \clearpage
\input{table_3}


\input{table_4}

\noindent unique identifications of minerals from just the far-infrared, 
with comet C1995~O1 (Hale-Bopp) being an exception. The mineral features 
in the 10~\micron{} region are key in mineralogical identifications but also these vibrational 
stretching modes overlap such that the Mg:Fe content, crystal shape, and grain  
size all affect the range and co-dependence of fitted model relative mass 
fractions. In the far-infrared, the vibrational  bending modes of crystalline olivine and crystalline 
pyroxene, as well as amorphous Mg:Fe olivine and pyroxene, occur at more distinctive 
wavelengths than in the mid-infrared. When the full wavelength range is available 
then the mineralogical identification is more secure. Extended 
spectral coverage provided by the longer wavelengths is used to better constrain grain parameters 
such as the slope of the grain size distribution and the porosity parameter $D$. Differences between 
model results obtained by using SLLL versus SLLL--SL SED fits are evident as shown in 
Figure~\ref{fig:sl-v-slll}.  

There are 22 comet spectra with SLLL and 20 comet spectra with SL, 
and both with $r_{h} \ltsimeq 3.5$~au.  We will motivate why we analyze these models 
as 22 SLLL with $r_{h} \ltsimeq 3.5$~au. 

\begin{figure*}[!ht]
\figurenum{4}
\begin{center}
\centering
\gridline{
\hspace{-0.05cm}\rotatefig{0}{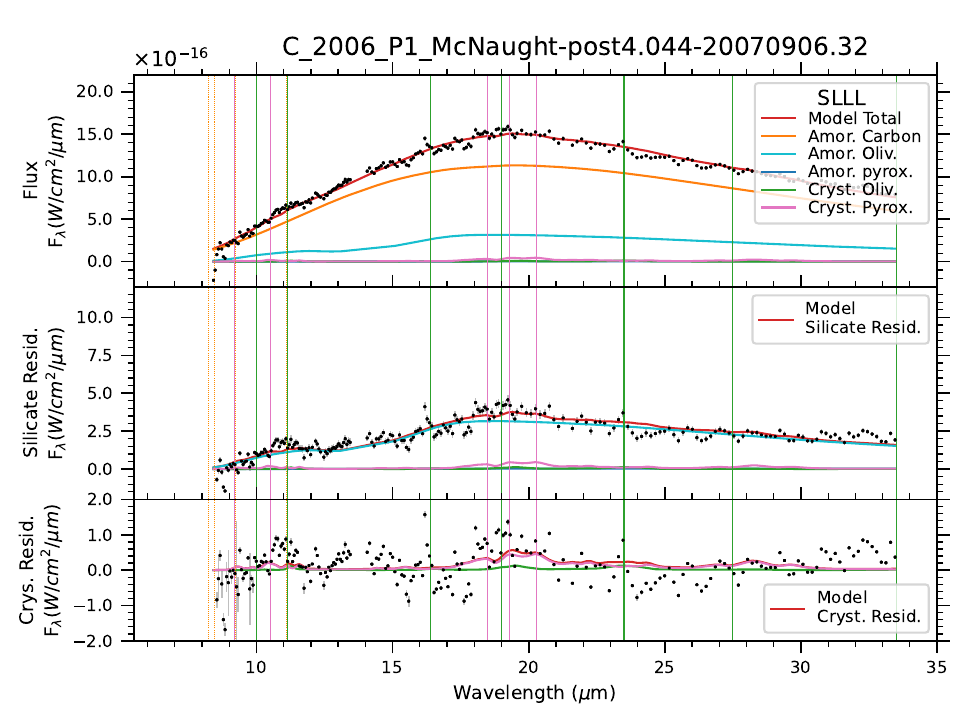}{0.48\textwidth}{(a) SLLL}
\hspace{-0.05cm}\rotatefig{0}{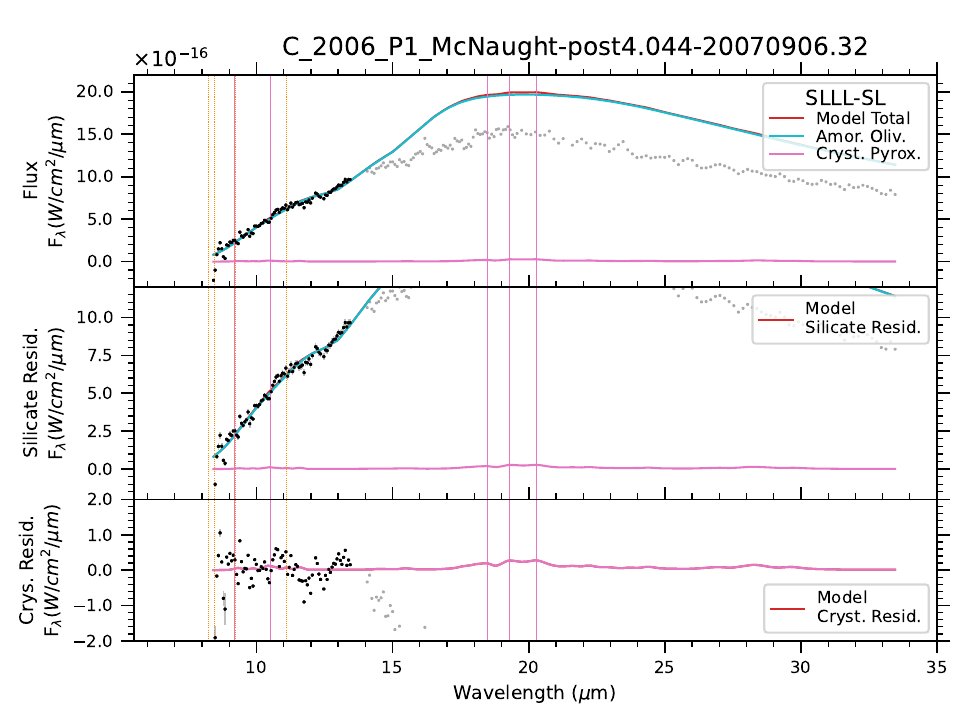}{0.48\textwidth}{(b) SLLL-SL}
}
\vspace{0.1cm}
\caption{C2006P1-20070906 (C/2006~P1~(McNaught)) at $r_{h} = 4.0$~au (two models, in flux density 
(W~m$^{2}$~\micron$^{-1}$) versus wavelength (\micron), showing how models can exhibit 
degeneracies for comets at $r_{h} \geq 3.5$~au. Degeneracies in parameters for models may 
occur between amorphous carbon and amorphous silicates at these $r_{h}$. Each figure 
consists of 3 panels, where the top is the observed spectra (black symbols) and
thermal model dust decomposition, the middle panel the residual silicate (brown solid line), 
defined as the total thermal model spectra (red solid line in each panel) minus the contribution from 
amorphous carbon (orange solid line in each panel), while the bottom panel the residual crystalline 
fraction (steel blue solid line) defined as the total thermal model flux minus the 
sum of the amorphous carbon + amorphous olivine (cyan solid line in each panel) + amorphous 
pyroxene (blue solid line in each panel) components. (a) Model fitted to SLLL, dominated 
by amorphous carbon with some contribution from amorphous silicates. (b) Model fitted to SLLL--SL, 
fitted solely by amorphous olivine. The vertical lines designate the wavelength position of features 
from various refractory species (from the left to the right) at:  9.30 (pink; crystalline pyroxene); 8.20 and 8.47 (olive; PAH features), 
10.05 (green; crystalline olivine), 10.50 (pink; crystalline pyroxene), 11.15 (green; crystalline olivine), and 
19.50~\micron{} (green; crystalline olivine). The latter feature is also near the order splicing 
of the IRS. Derived model parameters are inset top right. (The data used to create these figures are available.) }
\label{fig:model_degen_3pt5au}
\end{center}
\end{figure*}

\subsection{Analysis restricted to $r_{h} \ltsimeq 3.5$~au}
\label{sec:restrict_rh}

We restrict analyses of thermal models to $r_{h} \ltsimeq 3.5$~au. For $r_{h} \geq 3.5$~au, 
the thermal model parameters may exhibit degeneracies between amorphous carbon and 
amorphous silicates in the mid-infrared. For example, degeneracies in model 
parameters in SLLL--SL spectra are apparent when comparing SLLL spectra
to SLLL--SL spectra; at $r_{h} \geq 3.5$~au the shape of the short wavelength 
onset of the 10~\micron{} amorphous pyroxene may be degenerate with the shape of the thermal 
rise of the significantly warmer amorphous carbon emission. Representative example
of this complication is illustrated in Figure~\ref{fig:model_degen_3pt5au} for comet C2006P1-20070906
(C/2006~P1~(McNaught)) at $r_{h} = 4.0$~au,  where amorphous carbon dominates 
the model fitted to the full wavelength range (SLLL). In contrast, the models fit to solely 
the mid-infrared (SLLL--SL) is better reproduced by amorphous olivine. Thermal models for 
(SLLL) full wavelength spectra resolves this degeneracy (Sections~\ref{sec:model_fits}, \ref{sec:ratio1020_discuss}) 
by assessing the shape and contrast of the 20~\micron{} amorphous pyroxene band. 

\subsection{Analysis restricted to $2.7 \ltsimeq D \ltsimeq 3.0$ }
\label{sec:restrict_D}

Grain fractal porosity ($D$) is an important thermal model parameter. It is a parameter that
captures the fact that cometary grains are aggregate structures. Sample analysis in situ in terrestrial
laboratories and comet rendezvous encounters corroborate this physical state
\citep{2017MNRAS.469S.842B, 2017MNRAS.469S.443B, 2006Sci...314.1711B}.

Models with $D = 2.6$ and $a_{peak} \geq 1.5$~\micron{}, in our view, yield unrealistically high 
\fcryst{} or crystalline silicates (CS) values. Highly porous particles approach the behavior of 
their monomers both in their higher particle temperatures \citep{2002ApJ...580..579H} and
their higher spectral contrast features \citep{2004come.book..577K}. Highly porous particles with 
$a_{peak} \geq 1.5$~\micron{} can contribute to the SEDs and spectral features. In 
thermal models with $D = 2.6$ compared to $D = 2.7$ or $D = 2.8$, 
less of the mass fraction is required to fit the feature contrast for amorphous materials 
because highly porous particles produce higher contrast resonances in SEDs, which 
results in a significantly higher portion of the mass fraction being assigned to the solid crystals.
In our thermal models of ensembles of discrete materials, the solid crystals only 
contribute to particle radii (ellipsoidal effective radii)  of 0.1--1~\micron{} \citep[as solid crystals larger 
than this effective radii fail to match the observed spectral features,][]{2013ApJ...766...54L}. 
Solid crystals will yield a high mass fraction (f(co) or f(cp)) compared to the porous amorphous carbon (f(ac))
or amorphous silicates (f(ao50), f(ap50)). Each crystal produces relatively less thermal 
emission than porous amorphous particles of similar radii, even if the optical properties 
are similar. For instance, a representative case is provided by examining the
optical constants of crystalline Mg:Fe-olivine (fayalitic) compared to to those of amorphous 
Mg:Fe-olivine, with the caveat that spectral features of fayalitic olivine are not yet identified 
in cometary spectra despite their presence in Stardust return samples \citep{2014GeCoA.142..240F}.  
Thus, we restrict thermal models to consider amorphous materials with fractal 
dimensions in the range $2.7 \ltsimeq D \ltsimeq 3.0.$

Omitting $D = 2.6$ only omits the best-fit model for comet 17P (SLLL) and one best-fit model 
(SL) for comet 73P-C/SW3 (1.267~au). In some comets higher porosity
grains are known to exist. For comet 67P,  Rosetta's in-situ investigations suggest that
higher porosity particles comprise $\sim$25\% by number
\citep{2017MNRAS.469S.842B, 2017MNRAS.469S.443B}. The imaged extremely porous 
\citep[fractal-like,][]{2019A&A...630A..24G} hierarchical aggregates 
comprise only a handful of Rosetta particles measured
with the  Micro-Imaging Dust Analysis System (MIDAS) atomic force
microscope \citep{2016MNRAS.462S.304M}. Of the 22 SLLL models as well as those models 
constrained to fit only SLLL--SL, the majority of models are best-fit 
with $D = 2.7$ (see Table~\ref{tab:modparams}), with 10 of 22 for SLLL and 13 of 20 for SL.

\begin{figure}[!ht]
\figurenum{5}
\begin{center}
\gridline{
\hspace{-0.05cm}\rotatefig{0}{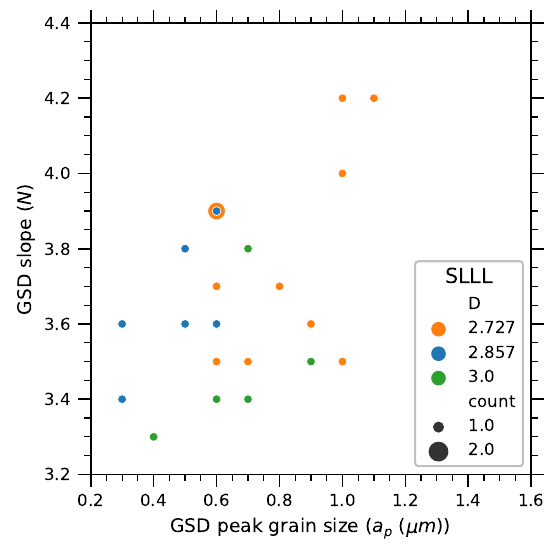}{0.35\textwidth}{(a)}
}
\gridline{
\hspace{-0.05cm}\rotatefig{0}{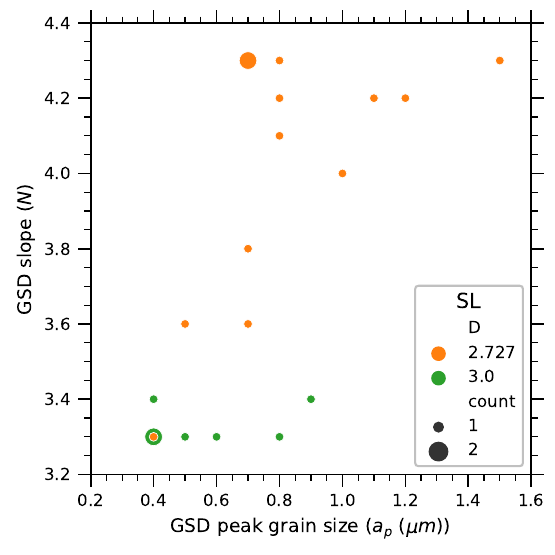}{0.35\textwidth}{(b)}
}
\caption{Grain size distribution slope ($N$) versus peak grain size, $a_{p}$(\micron), colored by 
porosity parameter $D$ that characterizes the amorphous materials. (a) SLLL models. 
(b) SL models. }
\label{fig:ap_vs_N_colored_by_D} 
\end{center}
\end{figure}

For the Spitzer survey models, best-fit HGSD parameters ($a_{p}$, $D$, $N$) are 
distinguishable with 95\% or greater confidence from non-best-fit choices of HGSD parameters.
A positive correlation exists between higher porosity (lower $D$) and larger $a_{p}$ and 
larger $N$ because increase in porosity causes increased particle temperatures whereas 
an increase in $a_{p}$ results in lower particle temperatures for the grain size distribution 
and a steepening of the slope lessons the emphasis on the largest and coolest particles.
Figure~\ref{fig:ap_vs_N_colored_by_D} shows the HGSD parameters of slope $N$ 
versus peak grain size $a_{p}$ (in \micron) and colored by $D$ for best-fit SLLL models and SL 
models to demonstrate their relationship with $D$.  

\subsection{Sensitivity to fitted wavelength ranges}
\label{sec:waves_SLLL_vs_SLLLdashSL_vs_SL}

The Akaike Information Criterion (AICc; Appendix~\ref{sec:AIC-Appendix}) may be used to
compare the relative suitability of two different model analyses of the same spectrum, 
or model analyses of two renditions of the spectra. In our analysis, we have used the AICc
to demonstrate that even though models fitted to fewer data points in SLLL--SL may have 
lower \chisqnu{}, these models have less information than models fitted to SLLL.

The relative likelihood of SLLL--SL referenced to SLLL, i.e., 
the probability that the SLLL--SL model minimizes the (estimated) information loss is 
exp$((AICc_{SLLL} - AICc_{SLLL-SL})/2) =~$exp$(\Delta AICc/2)$. 
On average, for a  subset of 20 of the 22 comet spectra, $\Delta AICc  = -514 \pm 186$, so the 
probability that the SLLL--SL minimizes the 
information loss is an extremely small probability. However, comet 
9P/Tempel~1 has a $\Delta AICc = 4.5$, so the SLLL--SL model fits the data better than the SLLL model. 
The SLLL model has insufficient crystalline silicate flux in the 10~\micron{} feature, or alternatively,
the model predictions of the far-infrared crystalline features are in excess of the observed flux.
Thus, a larger mass of crystals is excluded from the SLLL model fit. An extreme exception is comet 
17P/Holmes (Section~\ref{sec:special_17p}), which is discussed later, 
where $\Delta AICc = 468$ so the SLLL--SL model fit is superior to SLLL model. 

In some comets, there appears to be some emission in the 13.5-15.7~\micron{} region, 
which is not attributable to the Si-O vibration modes in Mg-crystalline and amorphous 
Mg:Fe silicates and therefore not predicted by the thermal model. For only a handful of 
comets is the `14.5~\micron{}' emission strong enough to affect the thermal model fitting 
of the mid-infrared 10~\micron{} silicate features, which typically extend to about 12.8–13.2~\micron , 
and the far-infrared silicate features, which include the 16.5~\micron{} crystalline olivine 
peak. To assess whether the model fitted to SLLL without the 13.5-15.7~\micron{} data points 
minimizes the information loss, we assess whether $exp(\Delta AICc/2) \ge 0.1$ or $\Delta AICc \ge -4.6$, 
i.e., if the relative probability of the model with fewer data points is greater than 0.1 then that 
model cannot be ruled out. The comets with $\Delta AICc \ge -4.6$ include: 78P, 88P, 
C2003K4-20050913, C2003T4-20051122, and C2006P1-20070802. Vibration modes of 
minerals with Ti-O and Al-O bonds occur near these wavelengths but modeling this $\sim14.5$~\micron{} 
emission is beyond the scope of this paper. 

\begin{figure}[!h]
\figurenum{6}
\begin{center}
\includegraphics[width=0.45\textwidth]{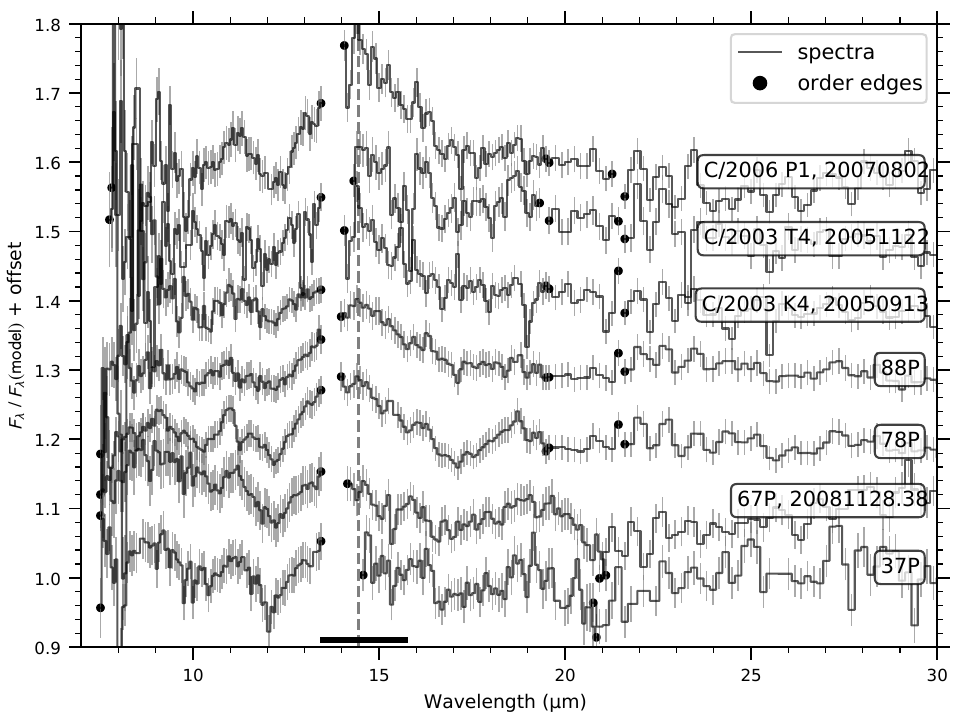}
\caption{Unidentified 14-\micron{} feature.  Seven spectra have been normalized by their 
best-fit thermal models to emphasize deviations from the model.  Five of the spectra 
(C/2006 P1, C/2003 T4, C/2003 K4, 88P, and 78P) were identified with our Akaike 
Information Criterion (AIC) tests, which compared model fits with and without the spectral data 
from 13.5--15.7~\micron{} (marked with a thick horizontal bar).  A vertical dashed line 
marks a proposed feature peak at 14.45~\micron.  Two additional high-quality spectra 
(67P and 37P) are shown for comparison.  These additional comets may also have 
evidence for a weak feature. }
\label{fig:fourteen_spec_roi}
\end{center}
\end{figure}

In Figure~\ref{fig:fourteen_spec_roi}, we present these five spectra, normalized by their 
best-fit thermal models to emphasize the unidentified spectral feature.  The feature spans 
two of the spectral settings (SL1 and LL2), from 12 to 17~\micron, with a peak around 
14.45~\micron.  Two additional spectra are shown, 67P-20081128.38 and 37P, for comparison. 
The feature may also be present in these additional data, but at a weaker level. 

\subsubsection{Model parameters and wavelength coverage} 

Roughly half of the IRS spectra in this survey are observed with the SL module 
only, and lack the far-infrared spectral coverage that LL observations would have provided.
To increase the sample size, the thermal models for the SL need to be placed in context 
with the thermal model parameters for SLLL.  Furthermore, the systematic differences 
in the model parameters between full wavelength coverage and only mid-infrared need to 
be evaluated and discussed. In many circumstances, especially 
non-space based observatories, spectra of comets are obtained only in the mid-infrared.

Table~\ref{tab:KStest_all_models} gives the median, mean, and standard deviation 
for the sets of models designated by SLLL, SLLL--SL, SL, and SLLL--SL \& SL (combined), 
employing the thermal model mean parameters for each model for a designated comet and 
epoch. In principle, compositional variant populations may be revealed from modeling 
studies of a large sample of comets. The standard deviations of the model population's 
composition parameters are large, so we investigate the cumulative distribution functions 
(CDF) for each parameter and apply the Kolmogorov-Smirnov (KS) test to compare the 
distributions of comae dust compositions revealed by the thermal models fitted to 
SLLL, SLLL--SL, SL, and SLLL--SL \& SL (combined) data sets, where SLLL and 
SLLL--SL are for the same set of spectra and SL is a distinct set of comets.

Taking crystalline pyroxene f(cp) for example, the median and mean of SLLL shows 
SLLL--SL and SL have greater mass fractions than for SLLL. 
The KS test passes for f(ao50) and f(ac) while the KS tests fails in other model parameters 
so when considering all the parameters the SLLL and SL are not drawn from the same 
distribution of models (Table~\ref{tab:KStest_all_models}). For model pairs of SLLL--SL and SLLL, 
forming the differences between model pair parameters, namely SLLL--SL minus SLLL, shows 
that there is a decrease in amorphous carbon, together with changes in the crystalline silicates 
that are anti-correlated with changes in amorphous silicates. The response of the thermal 
models to fitting a wavelength range limited to SL reveals the interplay between the changes in 
the grain size distribution parameters $a_{p}$ and slope $N$, the changes of which are 
positively correlated (if $a_{p}$ is larger, then the slope $N$ is steeper), as well as changes 
in compositional mass fractions.

We consider the possibility that the model’s predictions of 
the spectral behavior of the thermal emission from crystalline pyroxene in cometary comae 
may be better predicted in the SLLL--SL spectral range compared to the SLLL full spectral range 
as a component to understanding the model's response of fitting greater mass fractions of crystalline 
pyroxene to SLLL--SL models compared to the SLLL models. Our choice of optical constants for 
crystalline ortho-pyroxene is consistent with other modelers of cometary SEDs. However, optical 
constants for crystalline pyroxene do show variation
\citep{1998A&A...339..904J, 2001PASJ...53..243C, 2009ApJ...695.1024J, 2023arXiv230503417E}
with significant variations depending on ortho-pyroxene versus clino-pyroxene and depending 
on Fe-content. Given these caveats about crystalline pyroxene and the co-dependence of grain 
size distribution parameters and compositional mass fractions, the population of models for 
SL-only are combined into SLLL--SL \& SL and presented here for consideration. 

\input{table_5}

\subsubsection{Fitting outcomes as represented by six comets } 
\label{subsubsec:sixcomets-analysis}

\begin{figure*}[!ht]
\figurenum{7}
\begin{center}
\centering
\gridline{
\hspace{-0.05cm}\rotatefig{0}{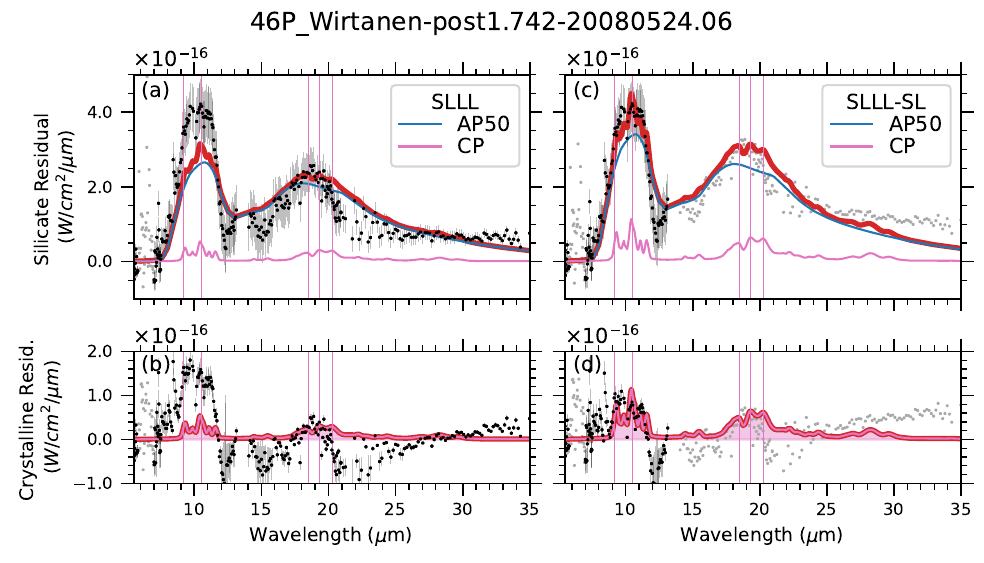}{0.45\textwidth}{(left)}
\hspace{-0.05cm}\rotatefig{0}{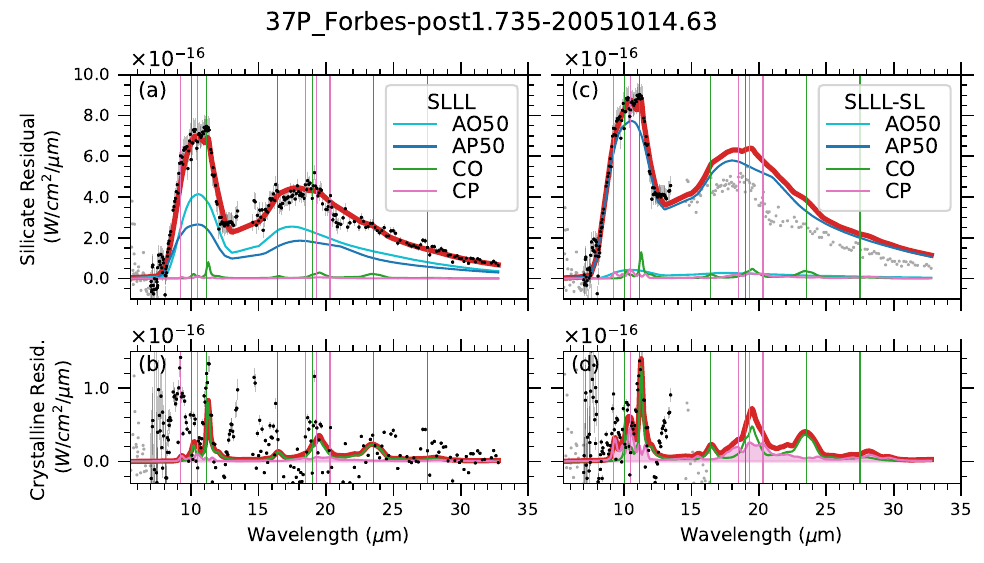}{0.45\textwidth}{(right)}
}
\caption{Crystalline features constrained by thermal modeling fitting for comets 46P/Wirtanen (right)
and 37P/Forbes (left). (a, b) SLLL. 
Models using the mid- and far-infrared SEDs. (c, d) SLLL--SL. Models using the mid-infrared 
SEDs. Upper panels (a, c) show the silicate residuals for the comet spectra 
(black points with error bars) and for the thermal model silicate residual and its 
decomposition.  Points without error bars (gray) are not used in the fit, e.g., the $\nu_{2}$ 
bands of water and restricted wavelengths for SLLL--SL. The lower panels (b, d) 
present the silicate crystalline residuals. The vertical colored lines are at positions 
of mineral resonances. The solid lines depict the contribution to the total residual 
(solid red line) of  the silicate dust component composition comprising submicron 
crystalline olivine (green) and crystalline pyroxene (pink), and amorphous olivine 
(cyan), and amorphous pyroxene (blue). Model fitting SLLL--SL yields increased 
contrasts in the crystalline features and higher crystal mass fractions. 
A complete figure set for  six comets is available in the online journal.} 
\label{fig:sil_resid_6comets}
\end{center}
\end{figure*}

Selecting six comet spectra from the survey ensemble, the differences between models fitted to SLLL 
and SLLL--SL can be highlighted. These six comet spectra include crystalline olivine rich f(co)$>$0.1, 
crystalline pyroxene rich f(cp)$>$0.1, and crystal poor f(co)$<$0.1 and f(cp)$<$0.1.  
Figure~\ref{fig:sil_resid_6comets} shows in the left panel the comet spectral silicate residual 
and model silicate residual and in the right panel the comet spectral crystalline silicate 
residual and the model crystalline silicate residual (Figure~\ref{fig:sil_resid_6comets}(a) is SLLL; 
Figure~\ref{fig:sil_resid_6comets}(b) is SLLL-SL). The silicate residual is defined as 
the modeled amorphous carbon subtracted from the observed flux resid$_{sil} = $ [flux - flux(ac)], 
and the crystalline silicate residual, is defined as the modeled amorphous carbon plus 
modeled amorphous silicates subtracted from the observed flux resid$_{cryst} =$
[flux - (flux(ac)+flux(ao50)+flux(ap50))].

The 11.15-11.2~\micron{} peak is well fit by a crystalline olivine component, which also produces
two of the three resonances at $\lambda \gtsimeq 15$~\micron{} (16.5~\micron{}, 
19.5~\micron{}, 23.5~\micron{}, 27.5~\micron{}, and 33.5~\micron{}). 
The three crystalline olivine peaks span the width of the far-infrared amorphous olivine 
feature.The shorter wavelength far-infrared resonance (16.5~\micron{}, 19.5~\micron{}) can 
be affected by the emission features from the far-infrared amorphous silicate bands.  The two 
or three far-infrared crystalline pyroxene peaks occur nearly spectrally coincident with the 
far-infrared amorphous pyroxene feature and at slightly longer wavelengths than the far-infrared 
amorphous olivine feature. Even if a weak contribution, the relative mass fraction of crystalline 
pyroxene is significant if the far-infrared crystalline pyroxene  features successfully can be 
fitted by the model.

\begin{figure*}[!ht]
\figurenum{8}
\begin{center}
\gridline{
\hspace{-0.05cm}\rotatefig{0}{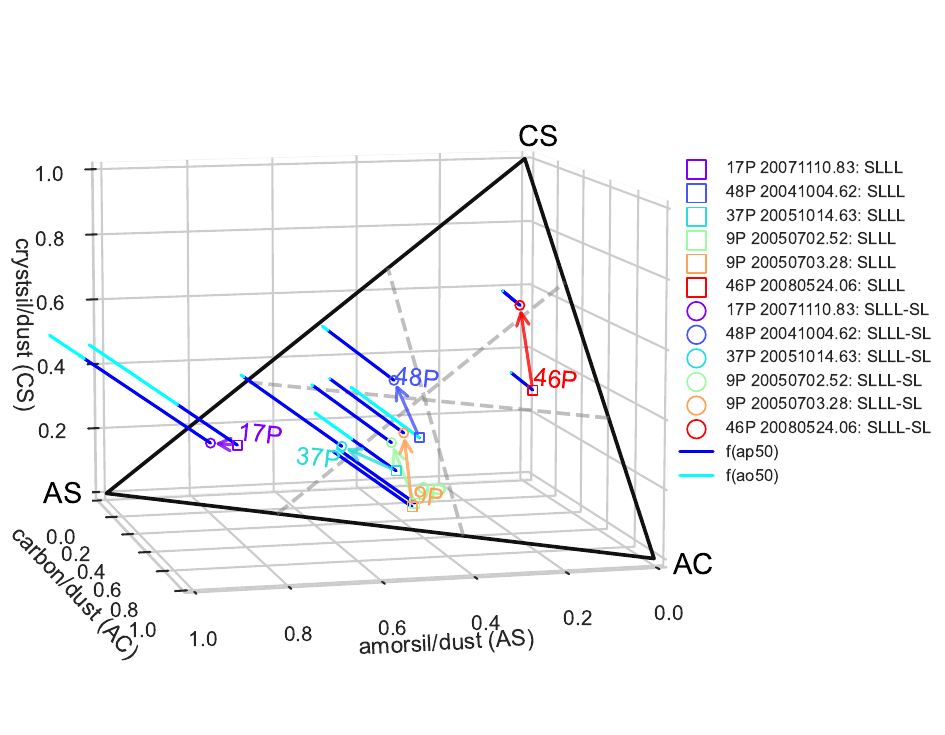}{0.48\textwidth}{(a) SLLL$\longrightarrow$SLLL--SL, amor.\ sil. rods}
\hspace{-0.05cm}\rotatefig{0}{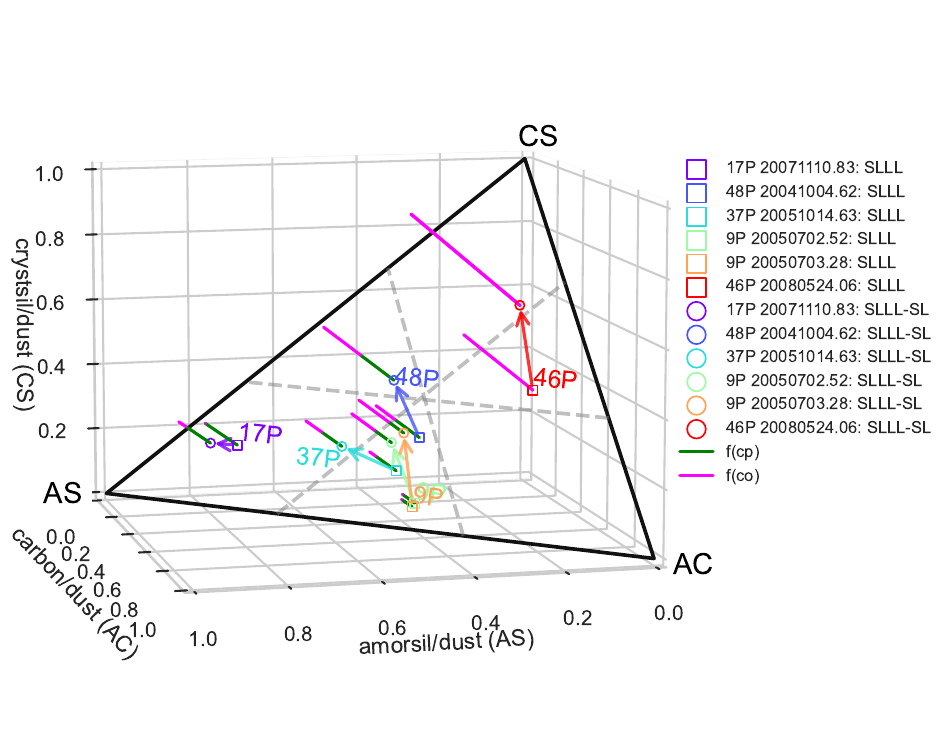}{0.48\textwidth}{(b) SLLL$\longrightarrow$SLLL--SL, cryst.\ sil. rods}
}
\gridline{
\hspace{-0.05cm}\rotatefig{0}{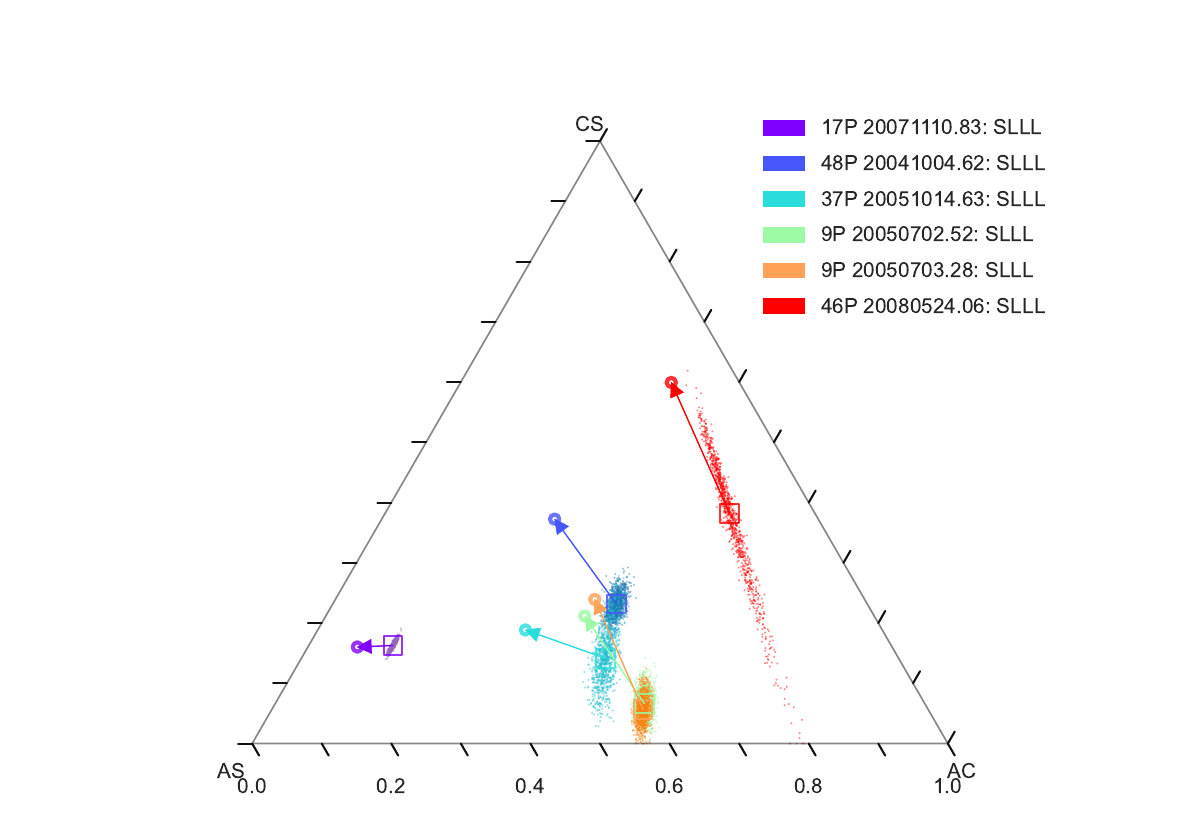}{0.48\textwidth}{(c) SLLL (1K MC trials)}
\hspace{-0.05cm}\rotatefig{0}{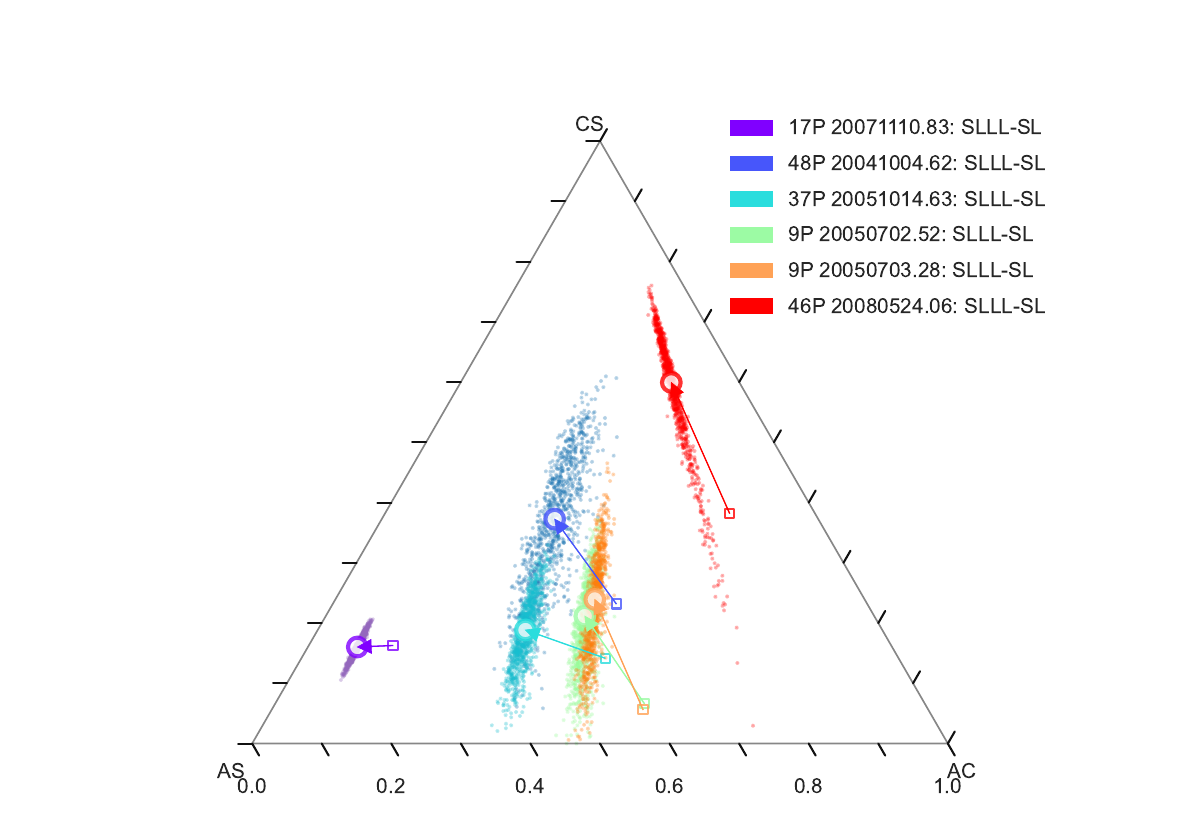}{0.48\textwidth}{(d) SLLL--SL (1K MC trials)}
}
\caption{Projection into 3D and 2D ternary diagram of five model parameters for six comets 
showing changes in fitted parameters from SLLL to SLLL--SL (Figure~\ref{fig:sil_resid_6comets}). 
From SLLL to SLLL--SL ({\it shown by arrows}), 
crystalline silicates (CS) increase and amorphous carbon (AC) decreases, where ($a$, $b$) 
respectively show the changes in amorphous silicates and crystalline silicates by rods 
perpendicular to the \{AC, AS, CS\} plane in 3D. ($c$, $d$) Ternary diagrams showing 
changes from SLLL $\longrightarrow$ SLLL--SL, with mean model parameters as 
points and extent of 1000 (1K) subset of 10,000 (10K) Monte Carlo (MC) trials as dots. SLLL parameter space is 
less tightly constrained compared to SLLL--SL (see Section~\ref{sec:waves_SLLL_vs_SLLLdashSL_vs_SL}).
} 
\label{fig:ternary_SLLL_SLLLdashSL_6comets}
\end{center}
\end{figure*}

To show the model parameters and how they change depending on the wavelength range 
for the six comet spectra in Figure~\ref{fig:sil_resid_6comets}, the ternary diagram, first in 3-D and 
then in 2-D, is introduced. Projecting the 5-dimensional parameter space onto 3-dimensions 
enables visualization of the shifts in relative mass fractions between models fitted to SLLL 
and SLLL--SL. Using the definitions CS=f(co)+f(cp) and AS=f(ao50)+f(ap50), and simply 
AC=f(ac), the model parameters are shown in Figure~\ref{fig:ternary_SLLL_SLLLdashSL_6comets} 
as coordinates a 3-dimensional space to demonstrate that they lay in a 2-dimensional 
plane because AC+AS+CS=1 (by definition of relative mass fractions). Model 
parameters for SLLL are squares and SLLL--SL are circles 
with arrows showing the shifts between SLLL and SLLL--SL. 

Rods perpendicular to the 2-d plane in 3D-space show the summands of amorphous silicates
and the summands of crystalline silicates (Figure~\ref{fig:ternary_SLLL_SLLLdashSL_6comets}(a), 
and f(co) and f(cp) in Figure~\ref{fig:ternary_SLLL_SLLLdashSL_6comets}(b)). Between SLLL 
and SLLL--SL, crystalline olivine may increase and often 
crystalline pryoxene increases; amorphous olivine decreases and amorphous pyroxene 
increases because the short wavelength shoulder of the Spitzer IRS SL mode is systematically 
better fitted by amorphous pyroxene. 

For each of the six comets, shifts in model parameters between for SLLL and for SLLL--SL 
are shown by squares for SLLL and circles for SLLL--SL as well as shown by dots are a 
randomly selected 1K subset of the 10K MC trials, 
Figure~\ref{fig:ternary_SLLL_SLLLdashSL_6comets}(c) and (d). For a 
given comet, the shape of parameter correlations are preserved but the domains in parameter 
space defined by models for SLLL are more compact, reflecting the smaller 2$\sigma$ 
uncertainties (95\% extrema of the trials) for SLLL models for these comets 
(Table~\ref{tab:KStest_all_models}). The SLLL models, which have tighter parameter 
confidence intervals, are proven to have higher information content than their SLLL-SL counterpart 
models by application of the AICc. Analysis of the SLLL models, which have tighter parameter 
confidence intervals, are the focus of this paper.

\subsubsection{Model correlation analyses for SLLL}
\label{subsubsec:SLLL_correlations}

\begin{figure}[!ht]
\figurenum{9}
\begin{center}
\gridline{
\hspace{-0.05cm}\rotatefig{0}{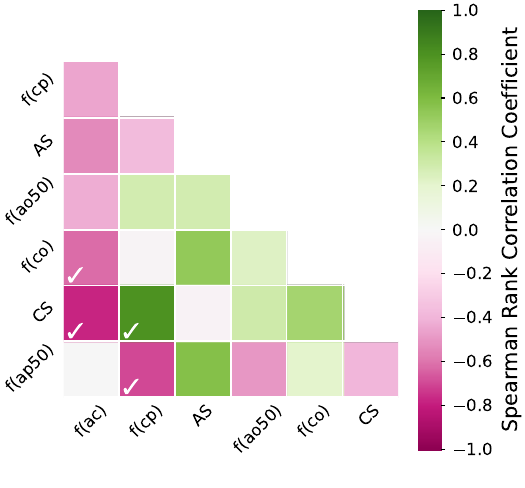}{0.35\textwidth}{(a)}
}
\gridline{
\hspace{-0.05cm}\rotatefig{0}{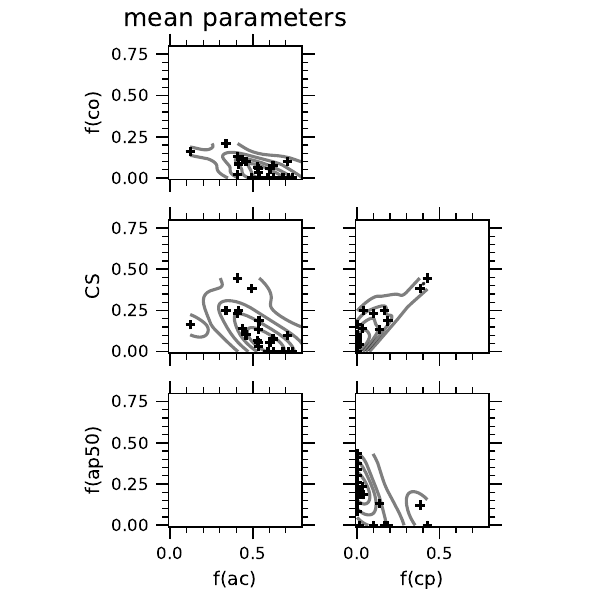}{0.35\textwidth}{(b)}
}
\gridline{
\hspace{-0.05cm}\rotatefig{0}{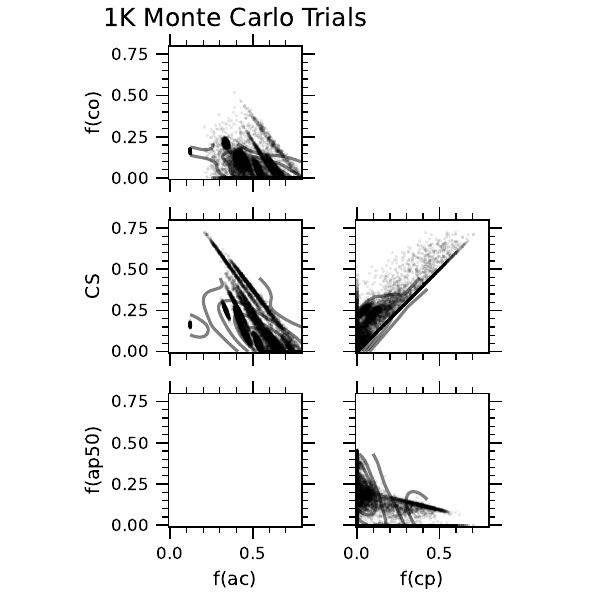}{0.35\textwidth}{(c)}
}
\caption{Spearman Rank Tests (for Correlations). (a)~The Spearman rank test correlation matrix for each of 
SLLL, with $\checkmark$ marks where pvalue (Permutation Test) is $<$0.003 (3$\sigma$). (b) The kernel density 
distribution (KDE) plots for mean parameters for those parameter pairs where the `$+$' marks indicate
those with permutation test pvalue$<$0.003 (3$\sigma$).
(c)~The KDE for 1000 Monte Carlo trails. 
}
\label{fig:SLLL_spearmanr} 
\end{center}
\end{figure}

\vspace{-0.2cm}
There are no positive correlation values greater than 0.8 between model mass fraction parameters for SLLL. 
Spearman Rank tests provide the correlation matrix values and permutation tests yield the 
correlation p-values. Figure~\ref{fig:SLLL_spearmanr} shows the correlation matrix with check 
marks for parameter combinations with p-values lower than 0.003 (97.7\%) used to reject the null hypothesis.
When the probabilities are too low for a random chance for there to be a correlation, the values of the 
correlation for the subset of parameters is in Table~\ref{tab:SLLL_spearmanr}. 

\input{table_6}


There is a moderate positive correlation between f(cp) and derived model parameter CS. 
Specifically, by doing Spearman Rank tests 
(computing the correlation matrix), we find f(cp) is anti-correlated with f(ap50) with a moderate 
value for the correlation parameter of $-0.70.$ There is an anti-correlation between f(co) and 
f(ac): modeled comets with higher amorphous carbon have less crystalline olivine. This
is commensurate with a bulk compositional study of anhydrous IDPs that shows higher mineral 
modality of crystalline olivine is correlated with lower wt\% of carbon (\ltsimeq 10~wt\%) 
\citep{1993GeCoA..57.1551T}. Other plots show f(co) does not change its KDE (kernel density estimation) 
from SLLL to SLLL--SL and  SL; f(cp) is not correlated with f(co) in SLLL.
Otherwise, correlations and anti-correlations are as expected from correlations of model 
parameters being mass fractions. This is shown by the full range of parameters 
spanned by the Monte Carlo trials in Figure~\ref{fig:SLLL_spearmanr}(c). \vspace{0.1cm}


\subsection{Dynamical Taxonomy} 
\label{sec:taxonomy_discuss}

The large sample of comets modeled herein enables a robust assessment of whether comets 
have a taxonomy based on refractory species composition and their relative abundances
factions as measured in the coma. This proposition can be explored through
a variety of statistical tests. An under lying key question to address with this survey of Spitzer 
comets is whether Jupiter-family (JFCs, Ecliptic comets) and Oort cloud comets (OCCs, 
Isotropic comets), two different dynamical families, have similar, or dissimilar, mineralogies 
and grain properties.  Such a comparison can give insight to 
conditions in the early protosolar nebula, including the mixing of inner disk material 
into outer comet forming zones. JFCs and OCCs occupy similar regions of model 
parameter space in terms of relative mass fractions.  A KS test comparing JFCs and OCCs 
does not identify any significant difference, Table~\ref{tab:KStest_all_models}.

\subsubsection{Dynamical populations: KS test of model sets}
\label{subsubsec:ksmodel_test}

The Kolmogorov–Smirnov (KS) test can be used to assess if two collections of models 
are drawn from different populations.  If the probability or pvalue of the KS test is less 
than 0.01 (1\%) then the null hypothesis (a claim that the models are not drawn from 
different populations) can be rejected, and the conclusion that the two model collections 
are statistically different may be reached. Sometimes 5\% is used. When the pvalue is 
high then we cannot say that they are not drawn from the same distribution but rather say 
that the \pagebreak

\begin{figure}[!ht]
\figurenum{10}
\begin{center}
\includegraphics[trim=0.0cm 0.0cm 0.0cm 1.0cm, clip, width=0.48\textwidth]{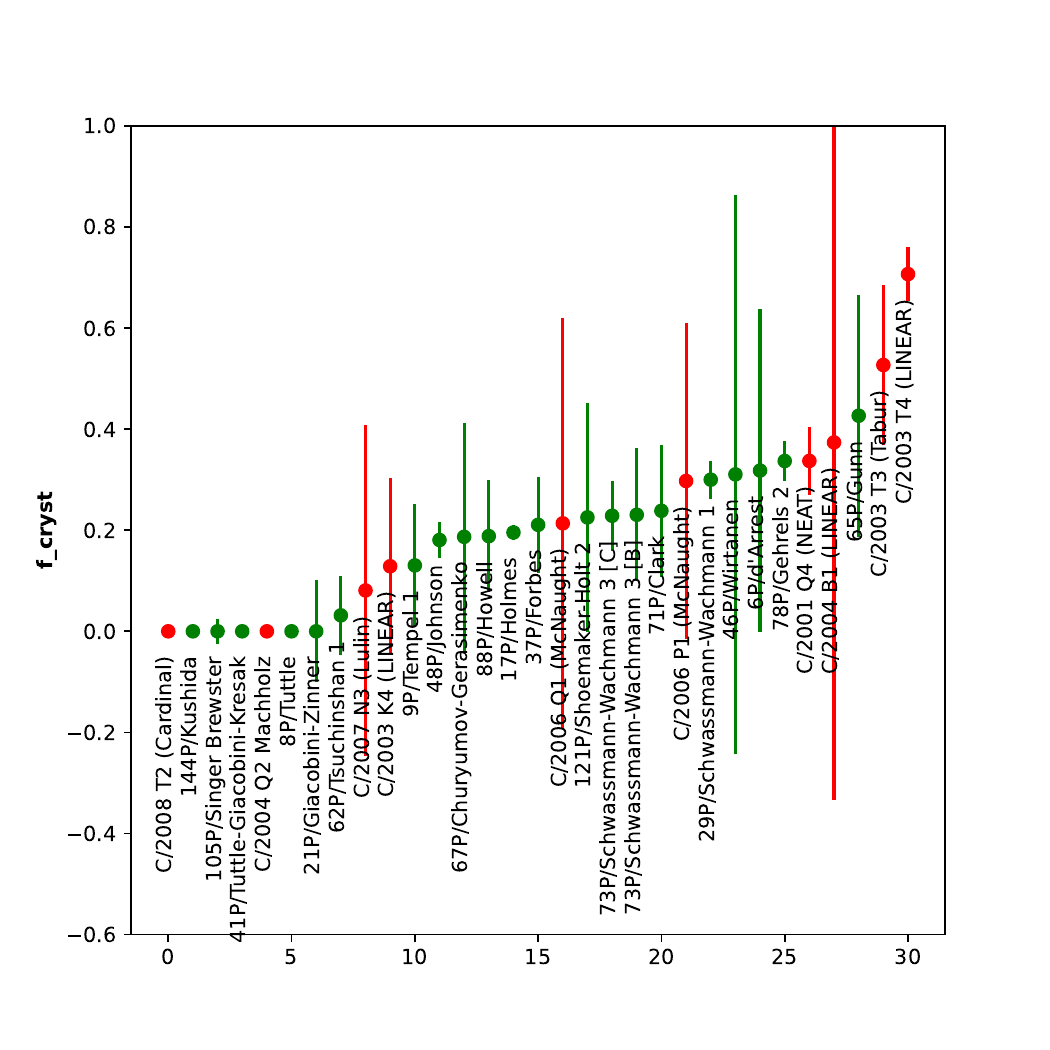}
\caption{The derived silicate crystalline mass fraction, \fcryst{}. Filled red circles are 
Oort cloud comets, filled green circles are Jupiter-family comets, where each
point represent the average values over all observations for the parameter $f_{cryst}$
if multiple observations of the same comet were obtained. The vertical bars represent
the range of $f_{cryst}$ within the 95\% confident interval.}
\label{fig:relative_massfraction_fig}
\end{center}
\end{figure}

\begin{figure}[ht!]
\figurenum{11}
\begin{center}
\includegraphics[width=0.50\textwidth]{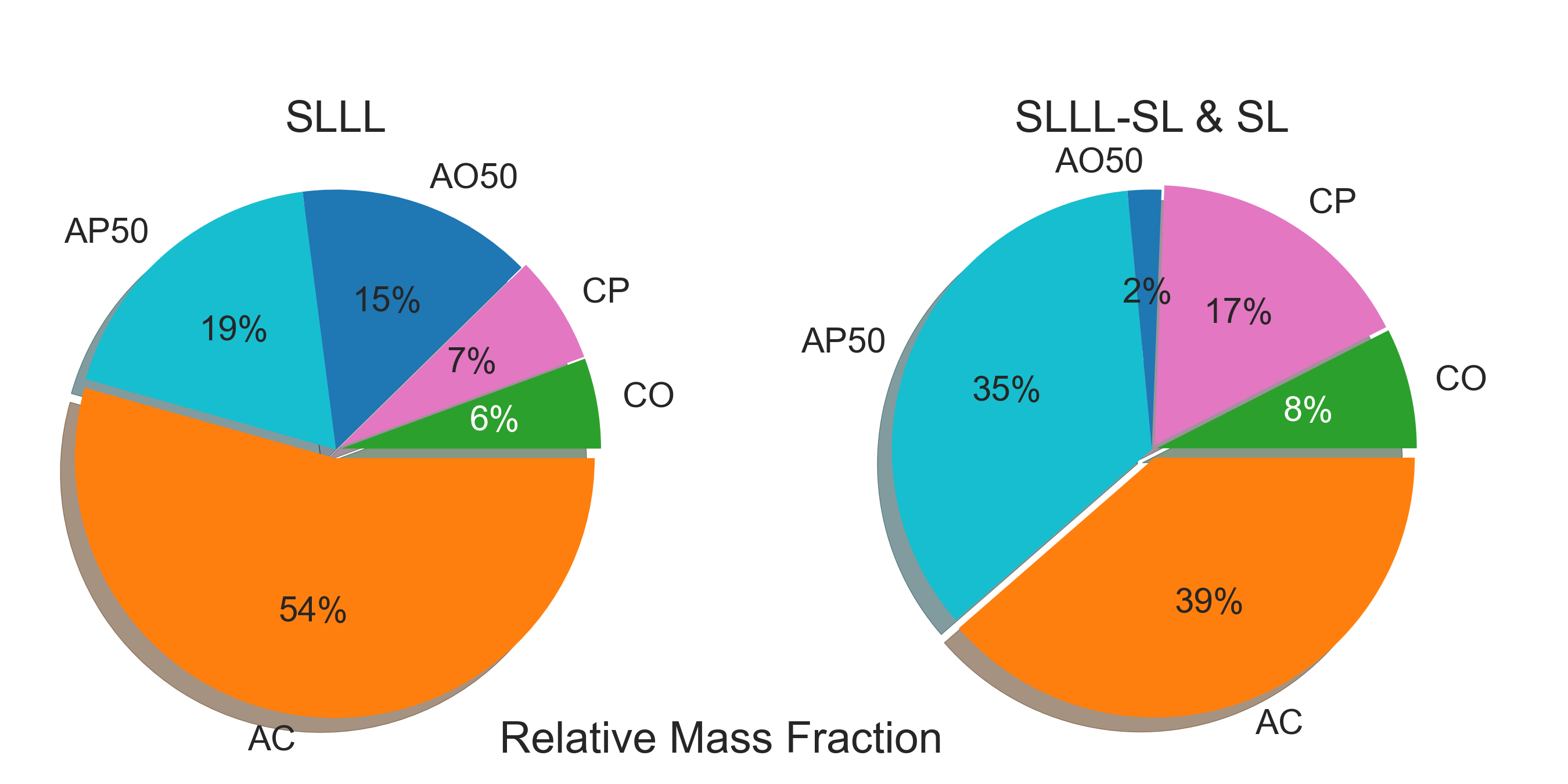}
\caption{Relative mass fraction of five primary materials in cometary comae derived from the
Spitzer comet survey. Left. SLLL.  Amorphous carbon (AC) is $>50$\%, dominating the mass fraction. 
Right. Combined (SLLL--SL and SL). AC is greater than one-third
and is comparable to the amorphous silicate mass fraction, and 
the crystalline silicates (CS) (sum of the crystalline pyroxenes [CP] and crystalline olivine [CO]
components) comprises 25\%. The increase in CS from SLLL to (SLLL--SL \& SL) 
is 12\% to 25\%, or an increase by a factor of 2. AC is a highly absorbing carbonaceous species. There could 
be more carbon in organics that are not as highly absorbing as AC. However, spectroscopic
observations that take full advantage of JWST's S/N ratio and superior spectral resolution are necessary
in order to potentially detect weaker features from organics. A complete figure set of the relative mass faction of
the five primary materials for each individual comet (some observed at multiple epochs) is available in the online journal. }
\label{fig:pie_SLLL_vs_SLLLdashSL_and_SL}
\end{center}
\end{figure}

\clearpage

\noindent  distributions are not distinguishable. Table~\ref{tab:KStest_all_models} shows the KS test for 
pairs of models (set 1 and set 2) for the five model mean parameters or mean relative mass 
fractions \{f(ap50), f(ao50), f(co), f(cp), f(ac)\}, and for the sum of amorphous silicates 
AS $\equiv$ f(ap50)+f(ao50) and of crystalline silicates CS $\equiv$ f(co)+f(cp). 

Although we focus primarily on interpreting the model outcomes from SLLL alone
 (section 4.5.3) for completeness we briefly comment on interpretative 
outcomes from the other wavelength fitted ranges contained in Table~\ref{tab:KStest_all_models}. 
For the set 1 and set 2 pair being (SLLL, SLLL--SL), the parameters that pass the KS test 
and thereby show significant differences include \{f(ao50), f(cp), f(ac)\}. 
For the set 1 and set 2 pair being SLLL and SL, the parameters that show significant
differences include include \{f(ao50), f(ac)\}. Only for the set 1 and set 2 pair of 
SLLL--SL and SL is no difference found in any parameter, and therefore we proceed 
with the analysis forming the SL-only data set from SLLL--SL \& SL.

The contrast of the silicate spectral features depends on the particle size and the 
silicate-to-amorphous carbon ratio. Weak spectral features from crystalline silicates contribute 
at multiple wavelengths. With the S/N presented in the Spitzer spectra, even silicate features 
of low spectral contrast can reveal the mineralogy of amorphous and crystalline components. 
The thermal model fitting accounts for the grain size distribution dependence and 
radial-heliocentric dependence when deriving the relative mineral mass 
fractions. Thus, we discuss taxonomy based on the the model parameters of 
relative mass fractions. 



\subsection{Silicate crystalline mass fraction}
\label{sec:cs_mass_frac}
Crystalline features of low spectral contrast, when fitted by the thermal models, contribute 
significantly to the crystal mass fraction ($f_{cryst}$) because (a) Mg-crystals are cooler and therefore more 
mass is required to account for their observed flux and (b) crystals are solid whereas the 
amorphous materials are considered to range from solid to porous and porous particles 
produce greater flux per mass (greater surface area and higher temperatures). The assessment of the 
crystal mass fractions may be one of the important contributions of JWST MIRI IFU to quantifying the 
comae dust properties and impacting our understanding of cometary materials and their origins.
In Figure~\ref{fig:relative_massfraction_fig} we present all of the comets in our analysis, constrained 
to a 95\% confidence level.  This includes comets for which \fcryst{}$= 0$.  There is no clear 
separation between comet families.

Generally, there is an increase in the relative mass of crystalline silicates when only modeling 
the SL modules, resulting in a higher \fcryst{}  on average by a factor of about 60\%.
This is illustrated in Figure~\ref{fig:pie_SLLL_vs_SLLLdashSL_and_SL} which
shows on the left the mass fraction for mean properties derived from SLLL models of 
22 comet spectra in the Spitzer survey for the 5 mineral components in the thermal model  
designated in the figure as: amorphous carbon (AC), amorphous 
pyroxene (AP50 for Mg:Fe=50:50), amorphous olivine (AO50), crystalline olivine (CO) and 
crystalline pyroxene (CP) (see Section~\ref{sec:dust_comp}).
The right is the mass fraction of the entirety of the sample (combined SLLL--SL and
SL). For the population of models of SLLL versus SLLL--SL (for the same 22 comet spectra), 
the increase in crystalline pyroxene is what drives the increase in \fcryst.

\vspace{0.4cm}
\subsection{Amorphous carbon}
\label{sec:ac_discuss}

For the distribution of SLLL models, the full range of amorphous carbon 
is large, spanning 0.1--0.9, with the distribution of models having a mean and standard deviation
of $0.54 \pm 0.16$, and 16 of 22 models or 73\% lie within the 
1$\sigma$ range of the mean as expected for a gaussian distribution. 
 
\citet{2021PSJ.....2...25W} discuss the the importance of amorphous carbon as a significant
component of coma refractories derived from thermal modeling and the association of these
result with cometary materials observed {\it in situ} measurements (Rosetta on 67P/Churyumov-Gersimenko, Vega 1 
and Vega 2 on 1P/Halley) and laboratory investigations on cometary samples and cometary IDPs.
Carbon Xray Absorption Near Edge Spectroscopy (C-XANES) has identified carbon in the 
phase of amorphous carbon in some cometary samples \citep{2009M&PS...44.1611W} but 
many cometary samples have other forms of carbon including aromatic and aliphatic carbon 
bonds and rarely graphite \citep{2005A&A...433..979M, 2017ApJ...848..113D, 2023arXiv230503417E}.
Optical constants for solid state organics with suitably highly absorbing properties, similar 
to amorphous carbon, are lacking and the available optical constants for organic solid 
state residues such as tholin \citep{2015P&SS..109..159B} are significantly too transparent 
to account for the observed near-infrared cometary flux densities and high particle temperatures of the 
carbonaceous dust component in cometary comae that is well modeled by amorphous 
carbon \citep{2017MNRAS.469S.443B, 2017MNRAS.469S.842B, 2023arXiv230503417E}.

Cometary IDPs are visibly very dark but when sliced or when crushed these same particles 
become more transparent \citep{2000ASPC..213..191F} so the optical properties of cometary IDPs may
not be identical throughout the particle. Thermal models presume that the optical properties 
are the same throughout each particle and use of mixing theories to compute the absorptivities 
and emission spectra of porous amorphous particles. In support of these modeling methods, 
computations of porous aggregates with amorphous silicate and amorphous carbon monomers 
distributed though out the particle and with amorphous carbon as a fluffy mantle do not yield 
significantly different predicted optical properties \citep[e.g., Figure~8 in][]{2016ApJ...818..133S}. The 
interstitial porous material that appears to hold together the crystalline silicates and amorphous 
silicates in cometary IDPs has long been described as the carbonaceous matrix or glue that 
holds the particles together \citep{1993GeCoA..57.1551T, 2011Icar..212..896B, 2012GeCoA..76...68D}. 
In Stardust return samples and in IDPs, carbonaceous matter appears as aromatic-rich 
nanoglobules \citep{2006Sci...314.1439N, 2009E&PSL.288...44B, 2011M&PS...46.1376D}. 
The carbonaceous matter is where D-enrichments reside \citep{, 2009E&PSL.288...44B, 2023arXiv230503417E}  
as well as rarely revealing higher N/C ratios \citep{2017MNRAS.469S.506F}, giving 
clues to some of its origin in cold molecular cloud environments via ion-molecule reactions 
that enhance D \citep{2018PNAS..115.6608I}. 

Bulk X-ray measurements of carbon in eleven anhydrous IDPs were assessed in the 1990s 
and shown to be 5--25~wt\% of elemental carbon with pyroxene-dominated crystalline silicates 
having the higher wt\% carbon \citep[Figure~3;][]{1993Metic..28R.448T}. By comparison with bulk S/Si and 
Fe/Si, the higher carbon, higher crystalline IDPs were suggested to be the best candidates 
for cometary IDPs collected in the stratosphere. From these bulk IDPs studies, the highest 
reported bulk C of $\sim$45~wt\% was reported for the pyroxene-rich IDP L2006B23, which 
also was characterized by $\sim$50\% porosity that yielded an estimated 90~vol\% for its 
carbonaceous matter \citep{1994Metic..29R.480K}. 

In comparison to these prior studies, Rosetta' s COSIMA {\it in situ} mass spectrometry of 
30 particles from comet 67P/Churyumov-Gersimenko, which are representative of 
$>$250 particles, demonstrated that  carbonaceous matter comprises 45wt\% of the dust 
\citep{2017MNRAS.469S.712B}.   From thermal models, comet 67P/Churyumov-Gersimenko (4 models) 
has an amorphous carbon (where amorphous carbon has optical constants that are
 taken as proxy for the carbonaceous matter that was measured by COSIMA) mass 
 fraction of 0.46$\pm$0.13. \citet{2021PSJ.....2...25W} showed that 
comet C/2013 US10 (Catalina) has a mass fraction of $0.473^{+0.015}_{-0.017}$ for amorphous 
carbon, similar to 67P/Churyumov-Gersimenko's {\it in situ} determination. 

This Spitzer survey finds that the comet model population has a mass fraction of $0.54 \pm 0.16$ 
of dark and highly absorbing carbonaceous matter that is well modeled by amorphous carbon. 
The emerging paradigm is that most comets have a carbonaceous content of $\sim$50~wt\%. 
Furthermore, comets present an important metric of 50~wt\% carbon in the solid state dust 
that is key to our understanding of the reservoir of the protoplanetary disk out of which 
comets formed and how it differed from the reservoir out of which asteroids like Ryugu 
\citep{2023Sci...379.8671N} and other carbonaceous chondrites were derived 
(possibly by collisions with large bodies).

\vspace{0.4cm}
\subsection{Amorphous silicates}
\label{sec:as_discuss}

The fits to the mid-infrared preferentially select amorphous pyroxene over amorphous olivine. We 
attribute this preferential selection of amorphous pyroxene in the mid-infrared to the 
prevalence of a short wavelength shoulder on the 10~\micron{} feature of comets in this 
Spitzer survey. However, this shoulder is not ubiquitous amongst cometary 8--13~\micron{} 
spectra. \citet[][Table~2 and Figure~5]{1994ApJ...425..274H} called attention to the mid-infrared spectrum of
Oort comet C/1990~K1~(Levy) \citep[prior designation 1990~XX,][]{1995JBAA..105..295S} 
 because of its a narrower 10~\micron \ feature with short wavelength rise at longer wavelengths.
 This was attributed to amorphous olivine by comparison with laboratory absorption spectra of 
 laser-vaporized (amorphous) forsterite. Thermal modeling of 10~\micron{} emission features 
 using optical constants of Mg:Fe amorphous olivine (`bronzite') and amorphous pyroxene 
 provided by \citet{1995A&A...300..503D} demonstrated the need for Fe in the amorphous 
 silicates in order to derive the required radiative equilibrium temperatures to 
 predict and fit the spectral contrast and shape of the 
 features \citep{, 1996Icar..124..344H, 2002ApJ...580..579H}. 
 
There is some concurrence that amorphous silicates in the ISM are Mg-rich
\citep{2008A&A...486..779M, 2021ApJ...906...73H}.   However, factors of 
composition, shape, porosity affect 10~\micron{} band shape \citep{2010ARA&A..48...21H}
and the 10~\micron{} to 20~\micron{} band ratio for lines-of-sight through the ISM.  
Determining the composition of ISM amorphous silicates is thereby challenging.  
In contrast, the composition of the amorphous silicates in comets is known because, 
given that comet comae particles are emitting in radiative equilibrium at a given 
distance from the sun, then the size distribution, porosity and 
composition of amorphous silicates can be constrained by comet thermal models. Although 
heritage of ISM materials into the outer protoplanetary disk is often discussed, the 
uncertainties in the composition of the ISM amorphous silicates then makes it difficult
to presume a heritage (link) between cometary amorphous silicates and ISM amorphous 
silicates.  

GEMS (glasses embedded in metals and sulfides) are amorphous silicates that are found 
in most cometary IDPs and in Ultra-Carbonaceous Antarctic Micro Meteorites (UCAMMs). 
GEMS are porous and have (Mg+Fe)/Si  $\le 1$ \citep{2022GeCoA.335..323B} and 
therefore are not stoichiometric. In contrast, thermal models for comets employ optical 
constants of stoichiometric amorphous pyroxene and olivine so one cannot compare 
the cometary model (Mg+Fe)/Si ratio to that of GEMs because of the constraints of the materials 
in the models. There only are a few infrared  spectra of GEMS-rich regions of IDPs 
\citep{1994Sci...265..925B, 2018PNAS..115.6608I}. Therefore, there are insufficient spectra 
of GEMS for a robust comparison between cometary comae infrared spectra and GEMS. 
All cometary infrared spectra have amorphous silicate emission features and studies 
of cometary IDPs ubiquitously discuss GEMS. This fact does not justify nor deny a 
potential link between the GEMS and cometary spectral thermal models.   

Amorphous pyroxene to amorphous silicates ratios in the diffuse ISM and towards 
protostars are based on 10~\micron{} features only and our analyses of 
SLLL versus SLLL--SL indicates that the composition deduced 
from the 10~\micron{} region alone may be biased towards pyroxene identifications. 
Thus, determining the origin and heritage of silicates in comets is challenging. The derived 
amorphous olivine to amorphous pyroxene compositional ratios in comets may provide clues.

\vspace{-0.15cm}
\subsubsection{Silicate irradiation and amorphization}
\label{sec:irradiation_discuss}

The “short wavelength shoulder” on the mid-infrared 10~\micron{} feature fitted by the
amorphous pyroxene component in the model is a potential signature of amorphization by 
ion irradiation (cosmic rays). For irradiated amorphous Mg:Fe olivine measured in the laboratory
only relative wavelengths shifts of the shoulder \citep{2016ApJ...831...66J} are reported 
and no optical constants are provided. Hence, assessing the amount irradiation that 
the amorphous silicates present in the refractory component of comet comae dust experienced
is beyond the scope of this work. 

\citet{2020MNRAS.493.4463D} model the 10~\micron{} feature along lines-of-sight through 
the galactic diffuse ISM and quote the ratio of amorphous pyroxene to amorphous silicates 
is ~17\% towards the Galactic Center.  In the circumstellar environments of O-rich Asymptotic 
Giant Branch (AGB) stars, where olivine condenses prior to pyroxene, the amorphous 
pyroxene to amorphous silicate ratio is low.  Amorphous olivine dominates the 
line-of-sight towards $\eta$~Ceph \citep{2006ApJ...645..395S}. On the other hand, amorphous 
olivine and amorphous pyroxene are seen towards molecular clouds \citep{2011A&A...526A.152V} 
and embedded protostars \citep{2020MNRAS.493.4463D}. Irradiation by cosmic rays has 
been long postulated as the cause of the trend between amorphous olivine-dominated 
AGB circumstellar medium versus the diffuse interstellar medium and protoplanetary 
disks with higher amorphous pyroxene to amorphous silicate ratios. Only one 
laboratory experiment provides spectral measurements of both the 10~\micron{} and 
18~\micron{} silicate features before and after irradiation of amorphous olivine. 
\citet{2016ApJ...831...66J} shows a diminishment in the 18~\micron/10~\micron{} 
feature ratio after H$^{+}$ irradiation, as well as shifts to longer wavelengths due to strong 
interactions with the experimental substrate that they do not consider relevant.

Recent JWST MIRI spectra of Class 0 protostar provides exquisite S/N spectra and modeling 
of these SEDs, specifically the 10~\micron{} absorption feature,
demonstrates the dominance of amorphous pyroxene over amorphous olivine \citep{2023ApJ...945L...7K}. 
The JWST modeling uses the same optical constants used in our comet thermal model. 
The Spitzer survey shows that there is a wide range of ratios of amorphous pyroxene 
to amorphous silicate for the SLLL models. However, additional laboratory measurements 
are clearly required at 10~\micron{} and 18~\micron{}, which is critical to breaking the degeneracy 
between olivine and pyroxene compositions. Measurement of the far-infrared resonances that are 
less overlapping that those in the mid-infrared, will enable an analyses of 
the potential signatures of irradiation of silicates as probed by cometary infrared SEDs and 
thermal models.

\subsubsection{Feature Ratios: 19~\micron{} and 10~\micron}
\label{sec:ratio1020_discuss}

The analysis of the Spitzer comet set may provide some clues by careful assessment
and comparison of the observed and model amorphous silicate feature 
19~\micron-to-10~\micron{} ratios and the wavelengths of the rise of the 10~\micron{} 
silicate feature. These data and the set of thermal models for this survey, which is uniquely suited 
because of its uniformity of data reduction and modeling methods, reveals an intriguing 
correlation (albeit at the 95\% confidence level) between the observed and model 
19~\micron-to-10~\micron{} feature ratios. The observed ratios may have a lower value than 
the model as well as the observed wavelength of the rise of the 10~\micron{} feature occurs at 
shorter wavelengths than the model. 

These differences between the data and the model are subtle: the optical constants of the 
amorphous Mg:Fe olivine and amorphous Mg:Fe pyroxene well represent the spectral shapes 
of the SLLL data (see 73P(B)-20060806.18 for amorphous pyroxene, 
48P/Johnson for amorphous olivine, and 46P-20080424 for a combination of 
amorphous pyroxene and amorphous olivine). Also, these same optical constants are 
used to model amorphous silicates in many astrophysical 
environs \citep{2010ARA&A..48...21H, 2020MNRAS.493.4463D, 2023NatAs...7..431M}. 
Optical constants for amorphous silicates made by the sol-gel method are available
 for compositions similar to Mg-rich pyroxene, Mg-rich olivine, and Mg-,Fe-pyroxene 
 \citep{2003A&A...401...57J, 2022A&A...666A.192D}  but thermal models require 
 amorphous Mg:Fe olivines and amorphous Mg:Fe pyroxenes.

\begin{figure*}[!ht]
\figurenum{12}
\begin{center}
\gridline{ 
\hspace{-0.15cm}\rotatefig{0}{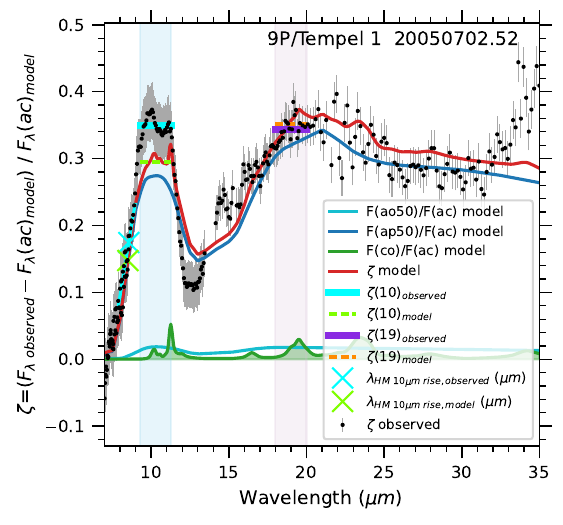}{0.4\textwidth}{(a)}
 }
\gridline{
\hspace{-0.15cm}\rotatefig{0}{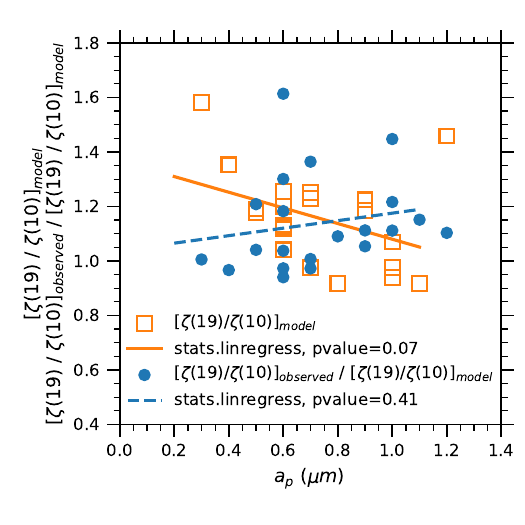}{0.4\textwidth}{(b)}
\hspace{-0.15cm}\rotatefig{0}{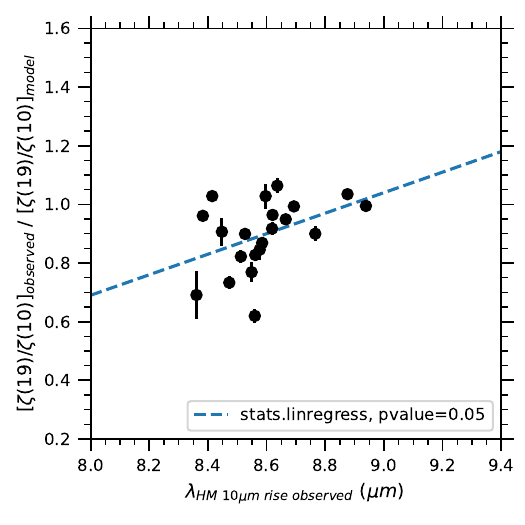}{0.4\textwidth}{(c)}
}
\caption{Analyses of the silicate feature ratios of 10~\micron-to-19~\micron. 
(a)~$\zeta$ versus $\lambda$~9P/Tempel 1 (20050702.52). For 9P/Tempel~1, we define 
$\zeta_{model} = (F_{\lambda , total} - F(ac))/F(ac)$ and $\zeta_{observed} = (F_{\lambda , observed} - F(ac))/F(ac)$. 
$\zeta(10)$ is the flux averaged over $\Delta \lambda$ = 2~\micron{} centered at 10.3\micron, and 
$\zeta(19)$ uses $\Delta \lambda$ = 2~\micron{} centered at 19.02~\micron.
(b)~Silicate feature ratios of 19~\micron-to-10~\micron{}  versus $a_{p}$. The models' silicate 
feature ratio 19~\micron-to-10~\micron{} ($[\zeta(19)/\zeta(10)]_{model}$) is dependent upon the grain 
size distribution peak grain size $a_{p}$. However, $[\zeta(19)/\zeta(10)]_{observed}/[\zeta(19)/\zeta(10)]_{model}$ 
is independent of $a_{p}$. 
(c)~$[\zeta(19)/\zeta(10)]_{observed}~/~[\zeta(19)/\zeta(10)]_{model}$ versus $\lambda_{HM~10\mu m ~rise~observed}$. 
A linear regression using the python statsmodels linregress \citep{seabold2010statsmodels} yields 
a slope of 0.350$\pm$0.170  and intercept of -2.109$\pm$1.462, and pvalue of 0.05. 
A firm correlation requires a pvalue of $\leq$ 0.007 (97.3\% probability of the failure of the 
null hypothesis of a zero-valued slope) so the pvalue of 5\% indicates there is only 95\% confidence
that this non-zero-valued slope represents a correlation.  }
\label{fig:zeta_3panels}
\end{center}
\end{figure*}

For the populations of models SLLL versus SLLL--SL, systematically, the crystalline fraction 
assessed from only the mid-infrared compared to the full wavelength data is higher and the wavelength
of onset of the 10~\micron{} feature is better fitted by amorphous Mg:Fe pyroxene, which has an 
onset at shorter wavelengths compared to the amorphous Mg:Fe olivine feature \citep{2020ApJ...897L..37M}. 
The far-infrared crystalline features of Mg-olivine and Mg-pyroxene, which are weak in comparison to the 
broad far-infrared amorphous silicate features, appear in some thermal model fits to not 
be allowed to be `squeezed' into the observed far-infrared flux level, simultaneous
with the thermal model fit having a notable dearth of emission in the mid-infrared, Figure~\ref{fig:zeta_3panels}(a). 

Aspects of the observed cometary amorphous silicate emissions may not be well-accommodated by the 
model and/or alternatively proffer intriguing support of a commonly held view that in the ISM, the 
amorphous silicates trend from olivine to pyroxene compositions due to ion 
bombardment \citep{2001A&A...368L..38D, 2022ApJ...941...50T}. Further, the laboratory experiments
exploring H$^{+}$ irradiation of amorphous Mg:Fe olivine showed the 19~\micron-to-10~\micron{} 
feature ratio is diminished compared to the un-irradiated samples by about a factor of 0.6 
\citep{2016ApJ...831...66J}. Comet 8P/Tuttle is excluded from this exercise because its silicate-to-carbon ratio
is 0.1 and for the purpose of the exercise its silicate feature is too weak. 

Let us define $\zeta \equiv (F_{\lambda}-F(ac))/F(ac)$ with $F_{\lambda}$ being observed or 
model flux density and $F(ac)$ being the model flux density of amorphous carbon; i.e., using 
the prior notation $\zeta$ is the silicate residual flux divided by the amorphous carbon (AC) flux. 
We characterize the 10~\micron{} and 19~\micron{} silicate features by assessing, $\zeta(10)_{observed}$, 
$\zeta(19)_{observed}$, $\zeta(10)_{model}$, $\zeta(19)_{model}$, 
respectively, from the mean data (with uncertainties) and mean model centered at 10.3~\micron{} 
spanning the peak of the 10~\micron{} feature and at 19.0~\micron{} in the range where there 
are contributions from amorphous pyroxene, amorphous olivine, the 19.5~\micron{} 
crystalline olivine and the 19--20~\micron{} crystalline pyroxene features. A feature sampling
bandwidth of $\Delta \lambda$ = 2~\micron{} is used.

We assess the wavelength of the rise of the 10~\micron{}  feature at the half maximum of 
the feature $\lambda_{HM~10\mu m~ rise~observed}$ by a linear regression to point 
pairs ($\lambda, \zeta$) such that $7.8\mu m \leq \lambda \leq 9.3\mu m$ and 
$0.25\zeta(10)_{observed}\leq \zeta \leq 0.75\zeta(10)_{observed}$. Thereafter, we can 
define $\lambda_{HM~10\mu m~ rise~observed}$ by the wavelength at which the line fit 
has value $\zeta(10)_{observed}/2$. For the model, the wavelength of the rise of the 10~\micron{} 
feature ($\lambda_{HM~10\mu m~ rise~ model}$) is similarly defined.  
$[\zeta(10)_{model}$/$\zeta(19)_{model}]$ is not correlated with heliocentric distance ($r_{h}$ (au)) 
and grain size distribution slope $N$ because dividing the silicate residual flux by the flux of 
amorphous carbon model is effectively normalizing by a flux that appropriately represents 
the grain size distribution-weighted particle fluxes with their $r_{h}$- and size-dependent 
radiative equilibrium temperatures. There is no apparent correlation with the shift of the 
wavelength of the rise of the 10~\micron{} feature with respect to the model. 

The model feature ratio $[\zeta(10)_{model}$/$\zeta(19)_{model}]$ is correlated with 
$a_{p},$ Figure~\ref{fig:zeta_3panels}(b), -- orange line. In contrast, 
$[\zeta(10)_{observed}$/$\zeta(19)_{observed}]/[\zeta(10)_{model}$/$\zeta(19)_{model}]$, 
is independent of $a_{p}$, Figure~\ref{fig:zeta_3panels}(b), -- blue line,  because to first order 
the model fits the observed 19~\micron-to-10~\micron{} feature ratio. 
The $[\zeta(10)_{observed}$/$\zeta(19)_{observed}]/[\zeta(10)_{model}$/$\zeta(19)_{model}]$
is independent of f(ap50)/AS, that is, it is not correlated with the amorphous silicate compositions.
In this exercise, we are pursuing the subtle differences between the observed and model 
amorphous silicate features by utilizing $\zeta$. 

With these parameters established, Figure~\ref{fig:zeta_3panels}(c) shows that there is a relationship between 
$[\zeta(10)_{observed}$/$\zeta(19)_{observed}]/[\zeta(10)_{model}$/$\zeta(19)_{model}]$ 
and $\lambda_{HM~10\mu m~rise~observed}$, given by the linear regression \citep{seabold2010statsmodels}.  
We interpret the trend-line in Figure~\ref{fig:zeta_3panels}(c) to mean that when the observed 
19~\micron-to-10~\micron{} ratio is smaller than the model 19~\micron-to-10~\micron{} ratio, the 
short-wavelength-rise of the 10~\micron{} feature will also occur at shorter wavelengths. 

There are sparse laboratory IR spectroscopic studies that measure the changes in the mid-infrared
and far-infrared spectral features from ion irradiation of Mg:Fe silicates and in particular of 
Mg:Fe amorphous silicates. \citet{2016ApJ...831...66J} shows that the 19~\micron-to-10~\micron{} 
ratio is lowered by a factor of about 0.6 by H$^{+}$ irradiation of amorphous Mg:Fe olivine. The 
decrease in the strength of the bending mode (far-infrared) relative the Si–O stretching mode (mid-infrared) 
is a consequence of the restructuring of the amorphous silicate matrix due to the destruction of 
bridging oxygen atoms. Nano-phase Fe also forms. In this experiment, the wavelength of the peak 
of the 10~\micron{} feature shifts to longer wavelengths by 0.46~\micron{} but the far-infrared absorption
feature is only slightly shifted. This can be explained via a reduction of bridging oxygen 
by sputtering since ``...10~\micron{} band of silicates is usually a measure of the polymerization of 
SiO$_{4}$ tetrahedrons in the amorphous network...'' \citep{2016ApJ...831...66J}. Their 
ion irradiation of the amorphous Mg olivine resulted in nanometer-sized silicon particles and less 
modification to the surrounding silicate matrix as well as significantly less formation of nano-phase Fe.
However, no IR spectra are shown. 

For the population of SLLL models in this Spitzer survey, the 
minimum of $[\zeta(19)/\zeta(10)_{observed}]/[\zeta(19)/\zeta(10)_{model}]$ is about 0.6, which may 
be accompanied by a systematic shift towards shorter wavelengths (Figure~\ref{fig:zeta_3panels}(c)). 
The decline in the 19~\micron-to-10~\micron{} feature presented here is in agreement with the 
experiment but not the wavelength shift of the rise 
of the 10~\micron{} feature that may be attributable to 
particle-irradiative processing of amorphous silicates prior to their incorporation into comets 
compared to the laboratory-prepared glasses from which the employed optical constants are derived.
In this context, studies indicate loss of Mg and O with ion irradiation and discuss trends towards
pyroxene compositions upon irradiation \citep{2002M&PS...37.1599C, 2001A&A...368L..38D, 2001M&PSA..36Q..37C}. 

Various laboratory irradiation experiments 
\citep{2002M&PS...37.1615C, 2003A&A...401...57J, 2004A&A...420..233D, 2004A&A...413..395B, 2010ApJ...708..288S, 2011M&PS...46..950C,  2009ApJ...705..791R, 2016ApJ...831...66J}
have interesting implications  for ISM silicate evolution and studies of cometary 
refractories. The origins the amorphous silicates in cometary 

\clearpage

\input{table_7}

\noindent  samples, the GEMS, is highly debated between potential solar system, cold cloud, and ISM origins 
\citep{2008A&A...486..781B, 2008A&A...486..779M,  2011GeCoA..75.5336K, 2013GeCoA.107..341K, 2022GeCoA.335..323B}.
Regardless of whether the source of ion irradiation is local or ISM, this Spitzer comet survey 
is a unique set of data that suggests the complexity of cometary materials as probed by thermal 
modeling of IR SEDs may be able to be enhanced by additional laboratory IR spectroscopic studies 
of ion bombardment of amorphous Mg:Fe silicates to assess the shifts in 
wavelengths of the rise of the 10~\micron{} spectral features and the 19~\micron-to-10~\micron{} 
feature ratios. 

\subsection{Carbon/Silicon elemental ratios}
\label{sec:c_si_ratio}

The atomic C/Si ratio for each of our spectra is based on the relative mass fraction of 
submicron grains for each material used in the best-fit thermal model 
(Section~\ref{sec:models}; Table~\ref{tab:modparams}).  
The method for calculating the atomic C/Si 
ratio is based on the methodology presented in \citet{2021PSJ.....2...25W}.  
They make some suppositions related to their calculation of 
atomic C/Si ratio. One, is that the amorphous carbon used in the  thermal dust model
is considered to be a good representation of the highly absorbing carbonaceous
based material in comet comae.  And two, that the relative mass fraction of
carbon calculated from the model fits is a good representation of the bulk 
of the solid state carbon (not the carbon in  the gases of ices)
material within the comet comae.  When comparing
our results with \textit{in situ} and laboratory studies, we are comparing our 
spectroscopic observations of a mix of thermally radiating grains of various 
radii to studies of single or isolated domains.

As in \citet{2021PSJ.....2...25W}, to allow for error propagation in our calculation
of C/Si, we ``symmetrize'' the relative mass fractions in Table~\ref{tab:modparams} using 
Method~\#2 of \citet{2017ChPhC..41c0001A}, being aware of the limitations of using this method 
\citep{2019Metro..56d5009P}. The C/Si ratio is calculated by assuming the 
mass fraction in the numerator is 100\% carbon, and the denominator is the sum of the 
mass fractions of the silicate bearing species with a atomic Si mass fraction of 
24.2\%, 16.3\%, 20.0\%, and 28.8\% for amorphous pyroxene, amorphous olivine, 
crystalline olivine,  and orthopyroxene, respectively.  Table~\ref{tab:c_si_tab} lists the values 
of C/Si for each spectrum in our sample.

The C/Si elemental ratio derived for comets (Table~\ref{tab:c_si_tab}) in the Spitzer survey 
are near the ISM values and higher than those of carbonaceous chondrites.  In the Spitzer survey, 
comets with higher amorphous carbon relative mass fractions are more numerous than 
comets with lower amorphous  carbon, Figure~\ref{fig:pie_SLLL_vs_SLLLdashSL_and_SL} (left).
That is, comets with AC $\ltsimeq 40$\%, like 17P/Holmes 
in outburst and Hale-Bopp, appear to be uncommon. Comet 67P, which can be seen 
in the ternary diagram to have thermal model parameter (mass fraction) AC $\gtsimeq 50$\%, was 
assessed by Rosetta COSMIA mass spectrometry to be 45~wt\% organics. Thermal models derive a 
similar C/Si ratio from the mass fraction of amorphous carbon \citep{2021PSJ.....2...25W}. 
The C/Si elemental ratios that are near the ISM value suggest comets either efficiently 
sequestered carbonaceous materials from the ISM or the outer disk produced refractory 
organics as recorded by comets. The high C/Si in comets is not seen in the carbonaceous 
chondrites, which are considered to be the most carbon-rich of the asteroid populations. 

\begin{figure}[!ht]
\figurenum{13}
\begin{center}
\gridline{ 
\hspace{-0.15cm}\rotatefig{0}{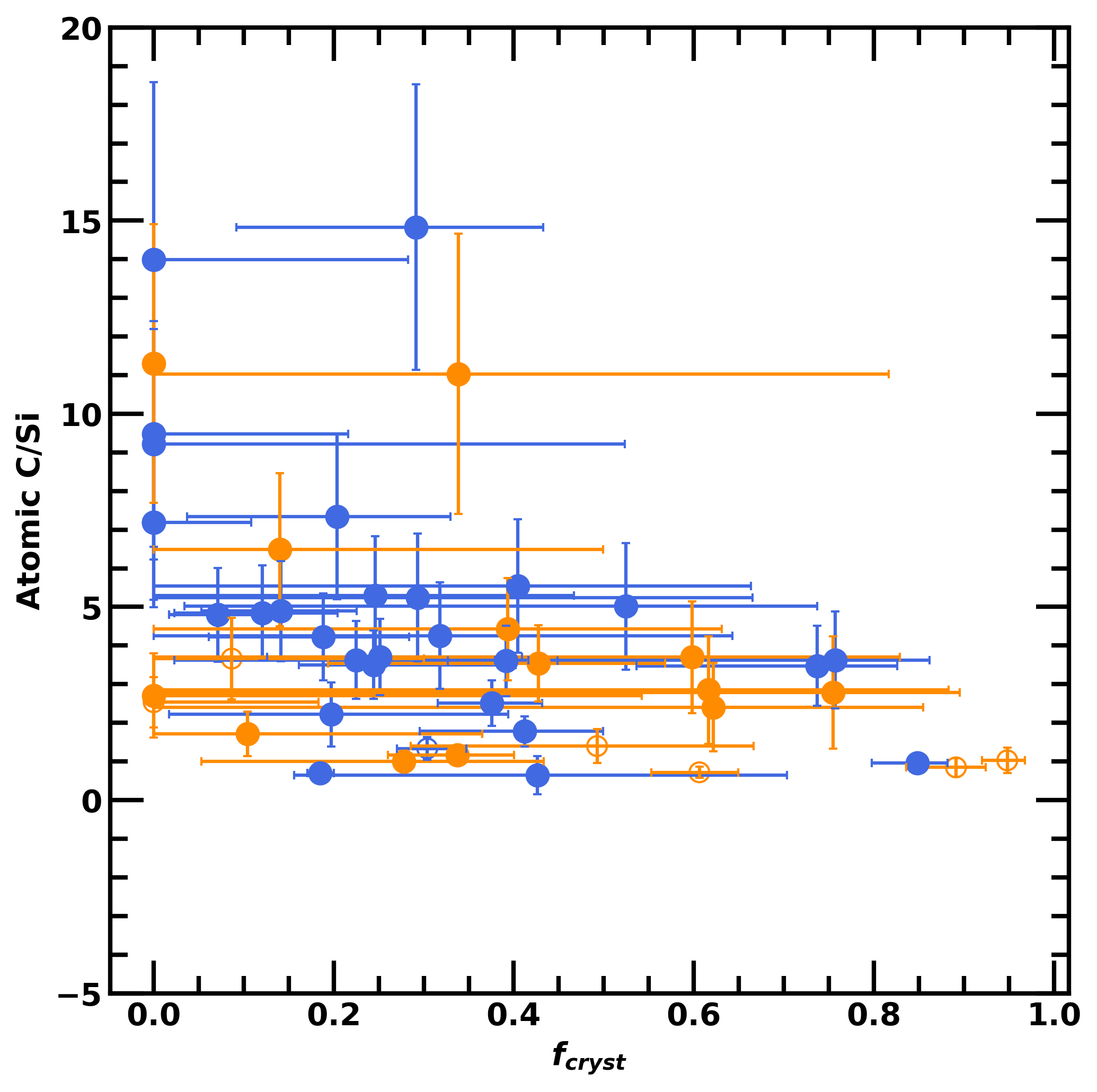}{0.40\textwidth}{(a)}
}
\gridline{
\hspace{-0.15cm}\rotatefig{0}{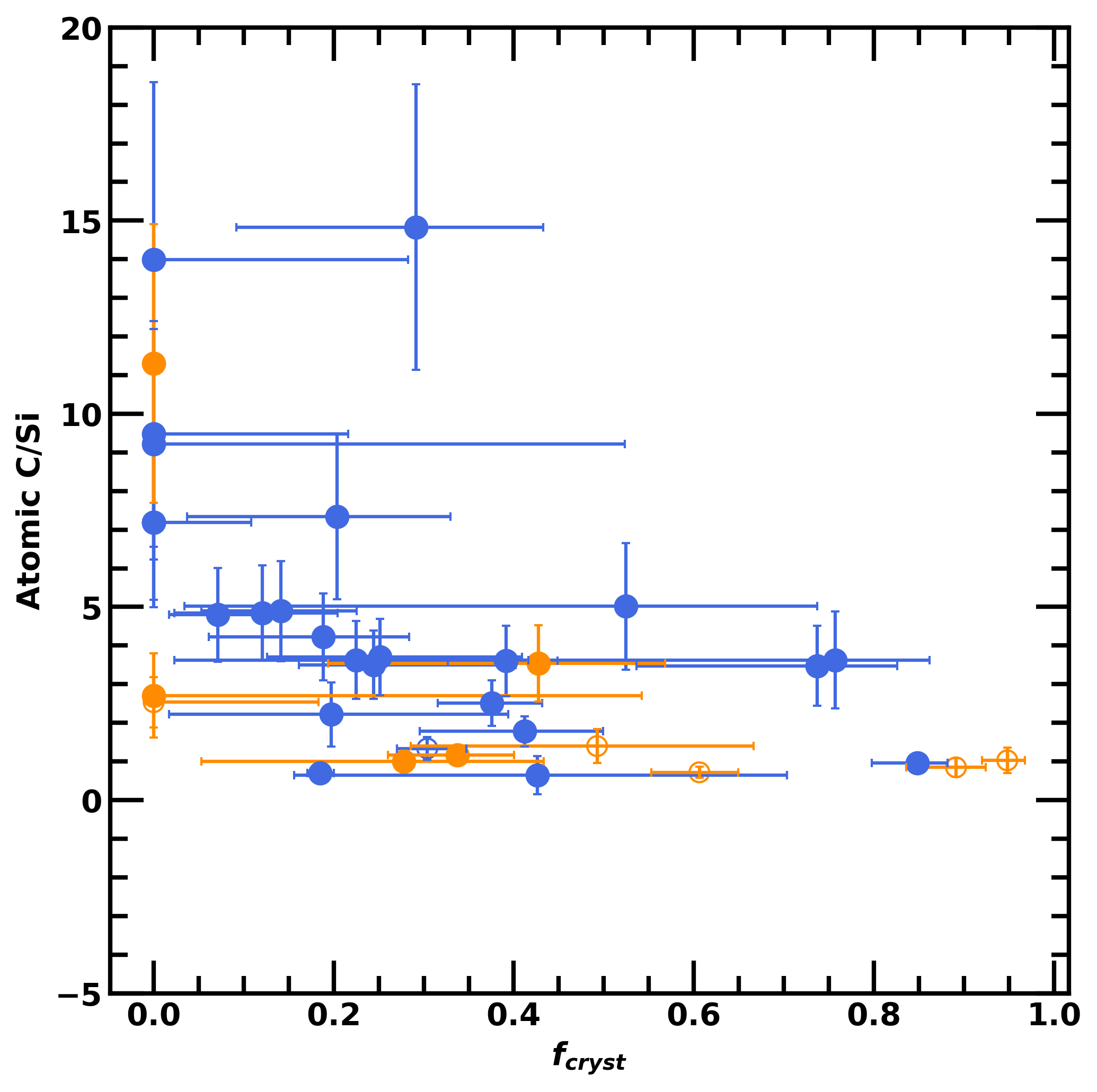}{0.40\textwidth}{(b)}
}
\caption{Atomic carbon-to-silicate ratio versus $f_{cryst}$ for the comet spectra in our sample.  
JFC (short period) comets are shown in blue, and OC (long period) comets in orange. (a) All comet spectra in the survey.
(b) Comet spectra where $f_{cryst}$ is constrained to the 95\% confidence level.}
\label{fig:c2si_plt}
\end{center}
\end{figure}

To look for any trends or differences between comet families, Figure~\ref{fig:c2si_plt} shows the 
atomic C/Si ratio  versus \fcryst{} for all of the spectra in our sample.  This includes comets for 
which \fcryst{}$= 0$. In Figure~\ref{fig:c2si_plt}(a), we present all of the spectra in our analysis, 
and in Figure~\ref{fig:c2si_plt}(b), we present those spectra for which \fcryst{} has been 
constrained to a 95\% confidence level.  In neither plot is there a clear separation between
comet families. 

\subsection{Comets spectra of special interest}
\subsubsection{17P/Holmes} \label{sec:special_17p}

The spectrum of comet 17P was obtained less than about 20 days after a mega-outburst, 
and soon after a possible mini-outburst \citep{2010Icar..208..276R}.  During the Spitzer 
observations 17P was  centered in each of the modules using the peak-up array. 
Subsequent background observations were taken offset from the comet center.
Details of the observations are presented in \citet{kelley-pds2021}.


The resulting spectrum of 17P exhibits a large silicate feature in the 
10~\micron{} region as well as smaller, but distinctive crystalline
silicate peaks at longer wavelengths.  The initial attempt to model
the spectrum of 17P did not produce a particularly good fit to the spectrum.
The 10~\micron{} silicate feature was fit poorly as well as the continuum
less than about 7.8~\micron.  However, the fit revealed what is possibly
an absorption feature between 6.2~\micron{} and 7.8~\micron. 
Careful examination of the observing sequence shows that the sky background was 
measured in a diffuse region of the coma prior to the primary science coma-centered 
observation. In the non-background subtracted image cubes, there appears excess 
emission in this offset coma position but only at these wavelengths and not at $\gtsimeq 8$~\micron. 
We speculate that a potential source of emission at these extended coma distances could 
be from PAHs but the S/N ratio is poor as well as the spectral resolution insufficient to 
model and subtract any cometary water vapor emission. Alternatively, the outburst may have 
released grains of unknown composition that is generating an absorption feature. 

In summary, the ambiguity of the identification of the `dust continuum' shortward of 
7.5~\micron{} led us to model the SED longward of 7.5~\micron, which results in the
appearance of an absorption feature in Figure~7. Future observations of comets with
JWST in this wavelength regime will provide SEDs with superior spectral resolution
and sensitivity to explore whether such absorption features are extant.

As for possible absorption sources in the coma, since the comet was observed after a 
major outburst on 2007 October 23 UT \citep{2010Icar..208..276R}, less than a month before it 
was observed by Spitzer.  This outburst could have released grains of some unknown 
composition that lingered in the coma for an extended period of time.  JWST 
has superior sensitive and spectral resolution in the 6.2--7.8~\micron{} region, 
enabling not only detect of water lines \citep{2023AAS...24113606W}, but when subtracted 
from the spectrum, residuals that may show potential spectral feature of grains of unknown 
composition that might be contributing to emission or absorption features in comets. Accounting for, 
and modeling the nature of, this possible absorption feature is out side the scope of this paper. 
We are primarily concerned with the refractory materials in the coma, so for our purposes, we 
modeled our SEDs longward of 7.5~\micron{} and will explore the origins of this absorption in
 future analyses. 

\subsubsection{73P/SW} \label{sec:special_73p}

Comet 73P(C)-20060806.05 also has a unique model fit. The grains were very fractally porous
with $D = 2.607$ and with a large peak to the HGSD of $a_p = 2.3$~\micron{}.  The
resulting \fcryst $= 99.4\%$, which is extremely large and we think is not a robust determination.  
There is nothing strange with the observation or the spectral extraction.  It is possible for the 
comet to have changed its grain properties post-perihelion, but and if so, then it is likely the model 
is challenged to adequately represent the properties of large, possible crystalline grains.
To match the rounded feature shape strength, the amorphous grains had to be fairly 
porous with a relative large peak to the grain size distribution. Given that we model crystalline grains 
as solids, to match the feature shape contrast, a significantly more crystalline material have to
be compacted into an equivalent grain of radius of $a$, resulting in such a large $f_{cryst}$.  
Hence, based the shape of spectral feature observed in the SED, the model approach is
not applicable and results in a non-physical outlier. 

\subsection{Implications for protoplanetary disk processes}
\label{sec: ppdp_discuss}

Cometary dust compositions probe the conditions in the protoplanetary disk in the realms 
of icy small body accumulation because the temperatures experienced by refractory dust 
particles in cometary nuclei are too low to change their composition and structure and 
there is no significant evidence of aqueous alteration of minerals in the Stardust 
return samples from comet 81P/Wild~2. Cometary dust particles are probed by laboratory 
examinations of chondritic anhydrous IDPs \citep[CA IDPs,][]{2023arXiv230503417E}, which vary in 
porosity, organic matter content and composition, and  the Mg:Fe-content of their crystalline silicates. 

Stardust return samples and giant IDPs contain crystalline olivine with a wide range of Mg:Fe-contents 
from Mg:Fe=100:0--0:100. Studies of their minor element abundances confirm their 
compositional links to Type II (Fe-rich) micro-chondrules 
\citep{2009M&PS...44.1675O, 2014M&PS...49.1522F, 2017RSPTA.37560260W, 2017M&PS...52..471B}, 
with one age-dated at $\ge 3$~My after CAI formation \citep{2012ApJ...745L..19O}. Oxygen isotope 
studies of crystalline olivine in Stardust return samples and CA IDPs shows that the highest 
Mg-content crystalline olivines (Mg:Fe=100:0--80:20) are $^{16}$O-rich with 
$\Delta^{17}$O~$\approx$~-23 \citep{2016LPI....47.1584D, 2017E&PSL.465..145D}.
This finding suggests the formation (condensation) of Mg-rich crystalline olivine occurs 
in early $^{16}$O-rich reservoir in the protoplanetary disk. Minor element studies also 
reveal Low-Iron-Manganese Enriched (LIME) olivine crystals in Stardust return samples
 (Mg:Fe=100:0--90:10), which characterize them as 
 condensates \citep{1994AIPC..310...51K, 2017M&PS...52.1612J}. Thermal models 
 presented here of 42 Spitzer IR SEDs successfully fit crystalline olivine with 
 Mg:Fe=100:0--80:20 and do not find features from higher Fe-contents crystalline 
 olivine \citep{2003A&A...399.1101K}. By comparison to these laboratory studies, the 
 crystalline olivine fitted in thermal models in these 42 comets are early-forming condensates 
 and not later-forming micro-chondrules. 
 
Disk models for radial transport via diffusion have achieved \fcryst{} of 10\%--20\% 
\citep{2002A&A...384.1107B,  2004GeCoA..68.2577K, 2006ApJ...640L..67D, 2010Icar..208..455C}.  
Disk models with annealing during infall, crystal condensation in the inner disk, and radial transport can 
produce \fcryst{} of 40\% but the refractory materials that formed after infall are abundant
in the 2--5~au region \citep{2012M&PS...47...99Y}, which is interior to the comet-forming zone. Magnetically 
driven disk wind transport may produce \fcryst{} of 1 out to 20~au \citep{2021ApJ...920...27A}. 
Disk models may consider annealing of amorphous silicates as a mechanism to explain 
enhanced \fcryst{} because annealing happens at lower temperatures and thus over larger 
disk volumes. However, the amorphous silicates in cometary thermal models must have 
Mg:Fe$\approx$50:50 to explain their radiative equilibrium temperatures \citep{2002ApJ...580..579H}. 
Highly specialized circumstances such as annealing of only micron-sized and smaller 
Fe-bearing glasses under low oxygen fugacity conditions that allows for inter-diffusion 
and loss of Fe metal could explain Mg-rich cometary crystals \citep{2007prpl.conf..815W}. The 
crystal mass fraction for the population of models \fcryst{} is 35--39\% and ubiquitously 
the crystals are Mg-rich so this study implies there was was significant and efficient 
radial transport of Mg-rich crystals, which likely are condensates, from the hot inner disk 
out to the comet-forming regime.  

Mg:Fe amorphous silicates are modeled in all cometary IR spectra. Based on laboratory 
studies of cometary CA IDPs, the amorphous silicates in cometary samples are the 
so-called GEMS, with FeS inclusions, non-stoichiometric compositions, substantial porosities, as well as 
some evidence for ion irradiation. The subtle spectral behaviors of the amorphous silicate 
18~\micron{}-to-10~\micron{} ratios in 21 SLLL models (Section~\ref{sec:ratio1020_discuss}) 
together with the wavelength of the rise of the 10~\micron{} feature suggests the 
Mg:Fe amorphous silicates may have experienced differing degrees of ion irradiation. 
Optical constants of irradiated amorphous Mg:Fe silicates are a future necessity.

The crystalline pyroxene to crystalline olivine ratios (f(cp)/f(co)) assessed from thermal models 
for individual comets have a large range of values, spanning 0 to $\infty$. The mean and 
standard deviation of the ratios of f(cp) to f(co) per comet model that are not at these two 
extremes is approximately equal to the ratio and propagated uncertainties of the model 
population Mean and Std of f(cp) to f(co) (Table~\ref{tab:KStest_all_models}). These values 
are for f(cp)/f(co): 1.9$\pm$2.6 (SL), 1.2$\pm$2.6 (SLLL), 2.5$\pm$2.6 (SLL--SL),
2.2$\pm$2.7 (SLLL--SL \& SL), which are greater than 1. Crystalline Mg-pyroxene has a 
greater mass fraction in models fitted only to the short wavelength mid-infrared data 
(SLLL--SL and SL models). Comparison of model-derived crystalline pyroxene-to-crystalline 
olivine ratios to those assessed for carbonaceous chondrites ($<1$), UCAMMs ($>1$), 
and for Stardust return samples ($\sim$1) \citep{2012GeCoA..76...68D} are affected by the 
bias towards higher ratio values for comet spectra models fitted only to the 
mid-infrared data.  \citet{2012GeCoA..76...68D} discuss the 
evidence of higher crystalline pyroxene to crystalline olivine ratios with increased disk radii from 
observations and from UCAMM studies. They suggest high pyroxene/olivine ratios 
may trace small body accretion at the greater disk radii.  The population of comet models 
presented here and UCAMMs share similarly high values of crystalline 
pyroxene-to-crystalline olivine ratios. 

The cometary amorphous carbon mass fraction (Sections~\ref{sec:ac_discuss}, \ref{sec:c_si_ratio})
is high compared to carbonaceous chondrites, which supports the concept of a carbon 
gradient in the solar system as well as offering support to the growing evidence of the 
effects of a gap in the disk created by Jupiter formation \citep{2021PSJ.....2...25W}. 
In UCAMMs, which are compared to CA IDPs yet may be particles from a N-rich parent 
body, crystalline Mg-pyroxene is correlated with higher wt\% of carbonaceous 
matter \citep{2012GeCoA..76...68D}. In this study, crystalline Mg-olivine is anti-correlated 
with amorphous carbon, which is similar to bulk studies of limited numbers of 
CA IDPs \citep{1994AIPC..310..165T}. Without this anti-correlation between 
Mg-rich crystalline olivine and amorphous carbon in the thermal models, the reservoirs of 
Mg:Fe amorphous silicates, Mg-crystalline silicates and dark carbonaceous matter might appear 
to be separate and unrelated reservoirs in the disk. Taking together the anti-correlation 
of amorphous carbon with crystalline olivine and the model populations' having crystalline 
pyroxene-to-crystalline olivine $\ge 1$ suggests that amorphous carbon and 
crystalline pyroxene may have been together more abundant in the outer disk. 

Each comet offers a probe of disk conditions. The populations of thermal models of the 
Spitzer survey reveal a significant range in compositions between individual comets 
as well as these intriguing trends. The JWST era of comet observations will compliment 
this extensive data set and offer more information and constraints on protoplanetary 
disk conditions under which comet nuclei agglomerated. 

\section{Conclusion}
\label{sec:concl}
Thermal dust models of 57 spectral energy distributions from 
5--35~\micron{} covering 33 individual comets (some observed at multiple epochs)
obtained from the Spitzer Heritage Archive uniquely allow the assessment of the 
compositional properties and trends for a significant number of comets from the application 
of thermal models to data reduced by consistent methods from the same instrument. 

Analyses of the population of thermal model compositions utilize the relative mass 
fractions (defined as f(material)) of the five mineral components: amorphous carbon, 
amorphous Mg:Fe pyroxene, amorphous Mg:Fe olivine, crystalline Mg olivine, and 
crystalline Mg pyroxene (f(ac), f(ap50), f(ao50), f(co), f(cp)). The population of models 
for compositional analyses is restricted to $r_{h} < $3.5~au and utilizes only data with 
Spitzer IRS full wavelength coverage (SLLL) or mid-infrared data (SL), which yields 22 SLLL and 
20 SL infrared spectral energy distribution (SED) models for model compositional population studies. 
Detailed comparisons of thermal models fitted to SLLL and to only the mid-infrared  data points of the 
same data sets (SLLL--SL) elucidate aspects of the data and of the model that contribute to 
wavelength-dependent compositional differences, which is possible for the first time given this large set. 
Systematically, amorphous pyroxene is fitted to SL compared to SLLL without change in the 
amorphous silicate mass fraction.

Empirical analysis of the survey comet spectra reveals a range of 10~\micron{} silicate
feature shapes, generalized into three broad categories: nearly trapazoidal (indicating emission 
from amorphous silicates),  rounded/triangular with strong emission at 8.2~\micron{} (also indicative
of amorphous silicate emission with the possible inclusion of crystalline silicates), and strong narrow
features with a stronger 9~\micron{} shoulder (indicative of emission from crystalline silicates). 

The rise of the 10~\micron{} feature occurs at shorter wavelengths, better modeled using the 
optical properties for amorphous Mg:Fe pyroxene. Also, increased crystalline 
Mg-pyroxene mass fractions are assessed for SL or SLLL--SL with a significant 
change in the crystalline mass fraction \fcryst{}.  The average crystalline fraction of the 
submicron grain component, $f_{cryst}$, is 25\% for SLLL and $\simeq 35$--39\% for SL-only.  
Because crystalline silicates are hot nebular products formed in the inner regions of the 
protoplanetary disk, high values of $f_{cryst}$ demonstrate that efficient radial transport of 
these materials to the region where comet nuclei aggregated occurred. 

The survey comet spectra and models indicate the wavelength of the rise of the 10~\micron{} 
feature occurs at shorter wavelengths when the observed 18~\micron{}-to-10~\micron{} feature 
ratio appears weaker than the modeled feature ratio. This difference may result from 
variable ion irradiation exposure of the cometary amorphous Mg:Fe silicates. 

For the populations of models, the crystalline Mg-pyroxene to crystalline 
Mg-olivine ratio is $> 1,$ which agrees with laboratory studies of 
ultra-carbonaceous micro-meteorites (UCAMMs) and is larger than this ratio for 
carbonaceous chondrites. This outcome suggests that the population of comet models 
is revealing that the disk distances at which comets agglomerated are greater for comets than 
for carbonaceous chondrites. 

The crystalline pyroxene to crystalline olivine mass fraction ratios (f(cp)/f(co)) assessed from 
thermal models for individual comets have a large range of values, spanning 0 to $\infty$.
High pyroxene/olivine ratios may trace small body accretion at the greater disk radii. The 
population of comet models derived from the Spitzer survey analysis exhibit ratios
similar to those measured in laboratory UCAMM samples. Combined with the result that we find 
an anti-correlation of amorphous carbon with crystalline olivine and the model populations’ having 
crystalline pyroxene-to-crystalline olivine $\gtsimeq 1$. These findings suggest that amorphous carbon 
and crystalline pyroxene may have been more abundant in the outer disk.

The Spitzer comet model population has an average mass fraction of $0.54 \pm 0.16$ of dark
highly absorbing carbonaceous material that is well modeled by amorphous carbon. 
Jupiter-family comets (JFCs) and Oort cloud comets (OCCs) occupy similar regions of the 
amorphous carbon--amorphous silicate--crystalline silicate ternary diagram and
many comets have a carbonaceous content of order 45~wt\% or greater. 
We find that Jupiter-family and Oort cloud comets occupy similar regions of model parameter 
space in terms of relative mass fractions.  A Kolmogorov–Smirnov (KS) test comparing 
JFCs and OCCs does not identify any significant difference. Hence, we cannot definitively 
conclude that there is a distinct difference in the dust composition between OCC and JFC
population as a whole. Thus, the development of a robust dust taxonomy scheme based on 
dynamical family is not yet possible.

The analysis of comet dust characteristics from the Spitzer survey of objects makes 
it possible to investigate general characteristics of these solar system small body population. 
These analysis also advance insights into the origin of comets. In the context of the diversity 
demonstrated by this Spitzer survey, comets exhibit a silicate to carbon ratio and crystalline 
mass faction ($f_{cryst}$) that were more similar (to each other) than 
diverse. Each observation and model of a given comet is sampling and assessing the dust 
compositions of cometary materials in the coma. 

Remote sensing studies of heliocentric driven activity changes in an individual comet's 
coma dust characteristics at mid-infrared wavelengths (such as those observable with Spitzer) 
are few. The limited secular studies primarily are those of in-situ measurements of comet 67P/Churyumov-Gersimenko
by Rosetta, and those of comet C/1996 O1 (Hale-Bopp). The latter, which had
had moderate sampling epochs (i.e., heliocentric distance) cross its apparition showed 
differences in its mid-infrared spectral features and derived refractory 
composition \citep[e.g.,][]{2000ApJ...538..428H, 2002ApJ...580..579H,2003ApJ...595..522M} 

Without the broad IR spectral grasp of Spitzer, Rosetta’s VIRTIS-M and VIRTIS-H assessed 
the comae dust compositions and particle properties by modeling the scattered light color, 
albedo, and the color temperature of the thermal rise \citep{2017MNRAS.469S.598R, 2019A&A...630A..22B}.
Rosetta’s COSIMA mass spectrometer measured a diversity of dust particle compositions 
but reporting is limited to dozens of particles \citep{2022cosp...44..175S}.  The in-situ 
Rosetta studies show that within the coma and during the activity cycle of a single comet that 
there were variations in dust properties. Complimentary to Rosetta's in-situ studies are the 
dust properties assessed from thermal models fitted to Spitzer IR spectral measurements 
discussed herein that also show that comae reveal a diversity of dust compositions. 

In summary, every individual solar system comet observed offers a probe that reveals insight into
conditions extant in our early disk. The findings here will both guide future JWST 
investigation of comets and provide a link to the JWST study of the nascent evolutionary stages of 
protoplantery disk in the infrared. These studies will further advance our understanding of the 
evolution of planetary disks and the accretion process in the outer solar system.

\begin{acknowledgments}

The coauthors wish to thank the referees for insights and detailed critiques that improved the
manuscript and the clarity of the graphical presentations. This work is based in part 
on observations made with the Spitzer Space Telescope, 
obtained from the NASA/IPAC Infrared Science Archive (doi:~10.26131/IRSA433), both of 
which are operated by the Jet Propulsion Laboratory, California Institute of Technology 
under a contract with the National Aeronautics and Space Administration. Support for 
this work was, in part, provided by NASA through an award issued by JPL/Caltech.
CEW, DEH, and MSPK acknowledge support from NASA grant 80NSSC19K0868.
DHW acknowledges support from the STT Branch of NASA Ames Research Center. 

\end{acknowledgments}

\facilities{Spitzer (IRS), IRSA (Spitzer Heritage Archive, doi:~10.26131/IRSA433), NASA Planetary Data System} 

\software{CUBISM,  Astropy \citep{2018AJ....156..123A}, Numpy \citep{2020Natur.585..357H},
SciPy \citep{2020NatMe..17..261V}, scikit-learn \citep{2011JMLR...12.2825P} }

\clearpage

\appendix
\section{Akaike  Information Criterion (AICc)}
\label{sec:AIC-Appendix}

The Akaike Information Criterion (AIC) is well-used and can be derived from properties of the 
statistical distribution (cite derivation reference). The AIC is used to compare two sets of 
models fitted to data points and allows one to ask: Is the information content of one model 
fitted to one set of data points greater to or lesser than a second model fitted to a 
second (and possibly different) set of data points? What is the probability that $model_A$ 
for data set A is more robust than $model_B$ for data set A?

To define the AIC, a {\bf partition function} is required, which represents the volume of 
data space that has significant probability:

\begin{equation}
	Z_D(\alpha) =
	 \left( 2\pi \right)^{N/2} \prod_{i=1}^N \sigma_{i}(\alpha).
\end{equation}

\noindent To maximize the Likelihood $L(\alpha)$ for a model dependent upon a vector of 
parameters $\alpha$ or ${\rm v}_{param}$, we minimize  $-2\ln L(\alpha)$ 
\citep[see Eqn. 4.17,][]{2014sdmm.book.....I}. The Likelihood $L(\alpha)$ also 
can be defined as:

\begin{equation}
L\equiv \prod_{i=1}^{N_{data}} \{(2\pi)^{-0.5} \sigma_i^{-1} \exp{ ({-((x_i-model_i)/(\sqrt{2}\sigma_i))^2)}}. 
\end{equation}

\vspace{0.5cm}
\noindent Maximizing the Likelihood is equivalent to minimizing \chisq{} for data with Gaussian 
distributed uncertainties $\sigma$ \citep{2014sdmm.book.....I}. The Likelihood is related to AIC by 
$-2\ln L(\alpha) = \rm{AIC} + 2~k$. Expanding,

\begin{equation}
\begin{split}
AIC  = -2 \ln{L(\alpha)} + 2~k \\ 
=  \chi^2({\alpha}) + 2  \ln{Z_D} + 2~k \\
=  \sum_{i=1}^{N_{data}} \left( \frac{x_i - {model}_i(\alpha) }{\sigma_i(\alpha) } \right)^2 \\
+~2 \sum_{i=1}^{N_{data}} \ln{\sigma_i(\alpha)}
+ \frac{N_{data}}{2}\ln{2\pi} +2~k
\ .
\end{split}
\end{equation}

\noindent yeilding

\begin{equation}
\begin{split}
AIC= \sum_{i=1}^{N_{data}} \left( 
	\frac{x_i - {model}_i }{\sigma_i}
	\right)^2\\
+~2 \sum_{i=1}^{N_{data}-N_{outliers}} \ln{\sigma_i(\alpha)}\\
+~2 \sum_{i=1}^{N_{outliers}} \ln{\sigma_{i,outlier}(\alpha)}
+ \frac{N_{data}}{2}\ln{2\pi} +2~k
\ .
\end{split}
\end{equation}
\vspace{0.5cm}

\noindent If outliers are defined as having $\sigma_{i,outlier}=f_{mult}\times \sigma_{i}$, 
then the  $\sum_{i=1}^{N_{outliers}}\ln ( \sigma_{i,outlier} ) = N_{outliers} \ln (f_{mult} )+\sum_{i=1}^{N_{outliers}}\ln(\sigma_{i})$, 
produces an additive term in AIC such that

\begin{equation}
\begin{split}
AIC= \chi^2
+ 2 \sum_{i=1}^{N_{data}} \ln{\sigma_i(\alpha)}
+ 2~\ln(f_{mult})~{N_{outliers}}\\
+ \frac{N_{data}}{2}\ln{2\pi} +2~k
\ .
\end{split}
\end{equation}

\noindent and therefore

\begin{equation}
\begin{split}
    AIC_A - AIC_B= (\chi_A^2 ~-~ \chi_B^2) \\ 
+ (2~\ln(f_{mult})~
(
{N_{A,~outliers}} - 
{N_{B,~outliers}}
))\\
+ (2~k_A - 2~k_B)
\end{split}
\end{equation}

If both models have the same number of parameters and both sets of data points are equal, 
meaning no distinction in outliers, then the AIC reduces to the difference in their \chisq{} values. 
For small numbers of parameters, a correction term is included such that 
$AICc =AIC+2k(k+1)/ (Ndata-k+1)$, which in our case contributes a small correction 
and is included for completeness.

When comparing 2 models fitted to the SED data for one comet at one epoch, model$_A$ 
and model$_B$, with different number of outlier points ($N_{A, outliers},~N_{B, outliers}$) 
and potentially different numbers of parameters $k_A$ and $k_B$, we define the 
difference $\Delta AICc \equiv AICc_{A}~-~AICc_{B}$ because the relative likelihood 
of model$_B$ with respect to model$_A$ is exp$((\Delta AICc)/2)$.  Hence,

\begin{equation}
\begin{split}
    \Delta AICc 
    = (\chi_A^2 ~-~ \chi_B^2) \\ 
+ (2~\ln(f_{mult})~
(
{N_{B,~outliers}} - 
{N_{A,~outliers}}
))\\
+ (2~k_A - 2~k_B)\\
+ \frac{(2~k_A(k_A+1)}{(N_{A, data}-k_A+1)} - \frac{(2~k_B(k_B+1)}{(N_{B, data}-k_B+1)}
\end{split}
\end{equation}

\noindent For $model_B$, we have found that $f_{mult}\approx 20$ yields approximately 
the same $\chi^2_B$ as omitting the points. Since we are using AIC that contains the partition 
function, we need to penalize the outlier points by increasing their uncertainties (an 
factor $f_{mult}$) that lessons the influence of these data points on the models).

Since outliers are decided from the original  data set that has the same number of data points 
then the difference in the correction terms depends on the number of parameters in the two 
models and if equal then the correction terms cancel between the two models. If only $model_B$ 
has outliers and both models have the same number of parameters ($k_B=k_B$), then 

\begin{equation}
\Delta AICc=(\chi^2_A) - (\chi^2_B+2~ln(f_{mult})N_{B, outliers})
\end{equation}

Considering a case where $\chi^2_B$ is less than $\chi^2_A$ by exactly the penalty term 
for the outliers of $2~ln(f_{mult})N_{B, outliers})$,  then $\Delta AICc$ is zero and the models 
have equal relative probability. In general, this functional form of $AICc$ assumes model$_A$ 
is associated with $AICc_{min}$. However, if $\chi^2_B$ is yet lower than $(\chi^2_A) - 2~ln(f_{mult})N_{B, outliers}$ 
then the $\Delta AICc >0$ and relative likelihood is greater than unity and model$_B$ is more 
likely to hold more information, even though it holds fewer data points. 

Let us compare comet 8P with $N_{data}$=304,  $model_A$=SLLL with $N_{A, outliers}=0$ 
and $\chi^2_A=229.69$, and $model_B$=SLLL--SL that is fitted to 159 data points 
such that  $N_{B, outliers}=145$ and $\chi^2_B=118.39$.  
Their \chisq{} differ and their reduced--$\chi^{2}$ are about equal (0.7708, 0.7738).  
We find $-\Delta  AICc=(\chi^2_B+2~ln(f_{mult})N_{B, outliers}) - (\chi^2_A) = 
(118.39+2~ln(20)\times(145))-(229.69)=(118.39+2\times3.00\times145)-229.69=(118.39+870)-229.69=758.7$. 
For 8P, the probability of $model_B$ fitted to only the mid-infrared data SL compared to $model_A$ 
fitted to the full-wavelength data SLLL is exp(-$\Delta AICc/2$) = exp(-379), which is really small.
Omitting 145 data points can achieve a similar reduced--$\chi^{2}$ but the information 
contained in the model fitted to data with 145 outliers (SLLL--SL that is chosen so as to only 
include SL1 and SL2) has less information content in comparison to the model fitted 
to all the available data points.

The AIC analysis is applied to the selection of data sets to be fitted by the thermal models for analyses as discussed in 
Section~\ref{sec:waves_SLLL_vs_SLLLdashSL_vs_SL} and representative
outcomes highlight in Table~\ref{tab:A1}. \pagebreak

\input{table_Appendix_A1}
\clearpage

\input{table_Appendix_A2}
\clearpage

\input{table_Appendix_A3}


\bibliography{zzz2}{}
\bibliographystyle{aasjournal}

\end{document}

%% file: table_1.tex
\startlongtable
\begin{deluxetable*}{@{\extracolsep{0pt}}lcccccccccc}
    \tabletypesize{\footnotesize}
    \tablenum{1}
    \setlength{\tabcolsep}{2pt} 
    \tablewidth{0pc}
    \tablecaption{Spitzer IRS observational summary\label{tab:obssum}}
    \tablehead{\colhead{} \\
        \colhead{Comet} & \colhead{Date} & \colhead{Start time\tablenotemark{a}} &
        \colhead{End time\tablenotemark{b}} & \multicolumn{4}{c}{Module} &
        \colhead{$r_h$} & \colhead{$\Delta$} & \colhead{Phase}\\
        & \colhead{(UTC)} & \colhead{(hr:min:sec)} &
        \colhead{(hr:min:sec)} &&&&&
        \colhead{(au)} & \colhead{(au)} & \colhead{($^{\circ}$)}
    }
    \startdata
    6P/d'Arrest                     & 2008 Sep 12 & 07:05:47 & 07:41:54 & SL2       & SL1       & \nodata   & \nodata   & 1.392 & 0.596 & 41.5 \\
    8P/Tuttle                       & 2007 Nov 02 & 18:11:18 & 18:18:33 & SL2       & SL1       & LL2       & LL1       & 1.606 & 1.322 & 39.4 \\
    9P/Tempel 1                     & 2005 Jul 02 & 12:22:18 & 13:19:35 & SL2       & SL1       & LL2       & LL1       & 1.506 & 0.711 & 35.8 \\
    9P/Tempel 1                     & 2005 Jul 03 & 06:48:18 & 07:47:29 & SL2       & SL1       & LL2       & LL1       & 1.506 & 0.714 & 36.0 \\
    17P/Holmes                      & 2007 Nov 10 & 19:51:12 & 20:16:06 & SL2       & SL1       & LL2       & LL1       & 2.505 & 1.867 & 21.3 \\
    21P/Giacobini-Zinner            & 2005 Dec 18 & 20:18:04 & 21:12:09 & SL2       & SL1       & LL2       & LL1       & 2.292 & 1.907 & 25.9 \\
    29P/Schwassmann-Wachmann 1      & 2003 Nov 23 & 07:17:55 & 07:25:05 & \nodata   & SL1       & LL2       & LL1       & 5.734 & 5.540 & 10.1 \\
    37P/Forbes                      & 2005 Oct 14 & 15:11:40 & 16:09:38 & SL2       & SL1       & LL2       & LL1       & 1.735 & 1.185 & 35.0 \\
    41P/Tuttle-Giacobini-Kres\'ak   & 2006 Apr 18 & 02:13:37 & 02:24:02 & SL2       & SL1       & \nodata   & \nodata   & 1.275 & 0.866 & 51.8 \\
    46P/Wirtanen                    & 2007 Dec 08 & 11:44:10 & 11:48:25 & SL2       & SL1       & \nodata   & \nodata   & 1.290 & 0.585 & 50.0 \\
    46P/Wirtanen                    & 2008 Jan 17 & 20:41:48 & 20:45:11 & SL2       & SL1       & \nodata   & \nodata   & 1.078 & 0.446 & 69.1 \\
    46P/Wirtanen                    & 2008 Apr 24 & 15:27:14 & 15:41:21 & SL2       & SL1       & LL2       & LL1       & 1.494 & 0.807 & 39.0 \\
    46P/Wirtanen                    & 2008 May 24 & 01:26:01 & 03:05:22 & SL2       & SL1       & LL2       & LL1       & 1.742 & 1.227 & 34.5 \\
    46P/Wirtanen                    & 2008 Jul 02 & 13:08:13 & 14:36:56 & \nodata   & SL1       & LL2       & LL1       & 2.076 & 1.912 & 29.1 \\
    48P/Johnson                     & 2004 Oct 04 & 14:47:42 & 15:01:34 & \nodata   & SL1       & LL2       & LL1       & 2.311 & 1.740 & 24.4 \\
    62P/Tsuchinshan 1               & 2005 Feb 13 & 08:38:28 & 08:43:34 & SL2       & SL1       & \nodata   & \nodata   & 1.650 & 0.921 & 32.4 \\
    65P/Gunn                        & 2004 Aug 10 & 15:53:16 & 16:09:31 & \nodata   & SL1       & LL2       & LL1       & 3.478 & 3.003 & 16.1 \\
    67P/Churyumov-Gerasimenko       & 2008 Jun 29 & 19:41:18 & 20:05:08 & \nodata   & \nodata   & LL2       & LL1       & 2.784 & 2.492 & 21.3 \\
    67P/Churyumov-Gerasimenko       & 2008 Nov 28 & 09:03:56 & 09:57:57 & SL2       & SL1       & LL2       & LL1       & 1.652 & 1.005 & 35.7 \\
    67P/Churyumov-Gerasimenko       & 2008 Nov 29 & 02:36:16 & 03:30:17 & SL2       & SL1       & LL2       & LL1       & 1.647 & 1.006 & 35.9 \\
    67P/Churyumov-Gerasimenko       & 2008 Nov 29 & 16:49:13 & 17:43:13 & SL2       & SL1       & LL2       & LL1       & 1.643 & 1.007 & 36.1 \\
    67P/Churyumov-Gerasimenko       & 2008 Nov 30 & 11:41:08 & 12:35:09 & SL2       & SL1       & LL2       & LL1       & 1.637 & 1.008 & 36.4 \\
    71P/Clark                       & 2006 May 27 & 13:29:37 & 14:41:21 & SL2       & SL1       & LL2       & LL1       & 1.566 & 0.910 & 37.6 \\
    73P/Schwassmann-Wachmann 3 B    & 2006 Apr 17 & 14:07:47 & 14:32:37 & SL2       & SL1       & \nodata   & \nodata   & 1.192 & 0.441 & 54.8 \\
    73P/Schwassmann-Wachmann 3 B    & 2006 Aug 06 & 04:14:47 & 07:09:09 & SL2       & SL1       & LL2       & LL1       & 1.259 & 0.846 & 53.9 \\
    73P/Schwassmann-Wachmann 3 C    & 2006 Mar 17 & 00:58:00 & 01:22:50 & SL2       & SL1       & \nodata   & \nodata   & 1.466 & 0.784 & 40.0 \\
    73P/Schwassmann-Wachmann 3 C    & 2006 Aug 06 & 01:14:07 & 01:52:14 & SL2       & SL1       & \nodata   & \nodata   & 1.267 & 0.864 & 53.4 \\
    78P/Gehrels 2                   & 2004 Sep 01 & 14:22:55 & 14:36:45 & \nodata   & SL1       & LL2       & LL1       & 2.059 & 1.531 & 28.6 \\
    88P/Howell                      & 2004 Sep 01 & 14:47:34 & 15:01:26 & \nodata   & SL1       & LL2       & LL1       & 1.996 & 1.391 & 28.7 \\
    105P/Singer Brewster            & 2005 Jul 14 & 13:37:52 & 13:51:26 & \nodata   & SL1       & \nodata   & \nodata   & 2.091 & 1.918 & 29.1 \\
    121P/Shoemaker-Holt 2           & 2005 Apr 22 & 22:16:45 & 22:23:26 & \nodata   & SL1       & \nodata   & \nodata   & 2.989 & 2.357 & 17.0 \\
    123P/West-Hartley               & 2004 Feb 04 & 08:00:33 & 08:04:28 & \nodata   & \nodata   & LL2       & LL1       & 2.176 & 1.600 & 25.2 \\
    132P/Helin-Roman-Alu 2          & 2005 Nov 20 & 03:18:28 & 04:16:28 & SL2       & SL1       & LL2       & LL1       & 2.074 & 1.480 & 27.3 \\
    144P/Kushida                    & 2008 Dec 06 & 05:56:00 & 06:48:03 & SL2       & SL1       & \nodata   & \nodata   & 1.552 & 0.827 & 36.9 \\
    C/2001 Q4 (NEAT)                & 2004 Oct 23 & 09:21:43 & 09:26:54 & SL2       & SL1       & \nodata   & \nodata   & 2.622 & 2.528 & 22.7 \\
    C/2003 K4 (LINEAR)              & 2004 Jul 16 & 05:01:23 & 05:06:33 & SL2       & SL1       & \nodata   & \nodata   & 1.760 & 1.409 & 35.4 \\
    C/2003 K4 (LINEAR)              & 2005 Sep 13 & 06:01:33 & 06:59:49 & \nodata   & SL1       & LL2       & LL1       & 4.501 & 4.283 & 13.1 \\
    C/2003 T3 (Tabur)               & 2005 Jan 11 & 18:04:17 & 18:27:57 & \nodata   & SL1       & LL2       & LL1       & 3.560 & 2.961 & 14.3 \\
    C/2003 T4 (LINEAR)              & 2005 Nov 22 & 17:29:15 & 19:14:02 & \nodata   & SL1       & LL2       & LL1       & 3.514 & 3.265 & 16.7 \\
    C/2003 T4 (LINEAR)              & 2006 Jan 29 & 19:56:50 & 21:08:35 & \nodata   & SL1       & LL2       & LL1       & 4.257 & 3.774 & 12.6 \\
    C/2003 T4 (LINEAR)              & 2006 Mar 07 & 07:18:09 & 07:48:34 & \nodata   & \nodata   & LL2       & LL1       & 4.636 & 4.385 & 12.4 \\
    C/2004 B1 (LINEAR)              & 2005 Oct 15 & 05:35:43 & 06:27:38 & SL2       & SL1       & \nodata   & \nodata   & 2.210 & 2.031 & 27.5 \\
    C/2004 B1 (LINEAR)              & 2006 May 16 & 05:34:25 & 05:45:32 & SL2       & SL1       & \nodata   & \nodata   & 2.058 & 1.608 & 28.7 \\
    C/2004 B1 (LINEAR)              & 2006 Jul 26 & 03:51:11 & 05:07:13 & \nodata   & SL1       & LL2       & LL1       & 2.689 & 2.234 & 21.5 \\
    C/2004 B1 (LINEAR)              & 2007 Mar 08 & 08:19:30 & 09:24:57 & \nodata   & SL1       & LL2       & LL1       & 4.814 & 4.766 & 12.0 \\
    C/2004 B1 (LINEAR)              & 2007 Jun 09 & 20:14:04 & 21:44:38 & \nodata   & \nodata   & LL2       & LL1       & 5.643 & 5.395 & 10.2 \\
    C/2004 Q2 (Machholz)            & 2005 Jul 01 & 07:52:33 & 08:05:48 & SL2       & SL1       & \nodata   & \nodata   & 2.547 & 2.297 & 23.6 \\
    C/2006 P1 (McNaught)            & 2007 May 04 & 19:58:18 & 20:17:35 & SL2       & SL1       & LL2       & LL1       & 2.398 & 2.224 & 24.8 \\
    C/2006 P1 (McNaught)            & 2007 Aug 02 & 17:01:50 & 19:32:18 & SL2       & SL1       & LL2       & LL1       & 3.624 & 3.137 & 15.3 \\
    C/2006 P1 (McNaught)            & 2007 Sep 06 & 07:41:38 & 10:22:07 & \nodata   & SL1       & LL2       & LL1       & 4.044 & 3.849 & 14.7 \\
    C/2006 Q1 (McNaught)            & 2007 Mar 22 & 02:05:15 & 03:30:47 & \nodata   & \nodata   & LL2       & LL1       & 5.279 & 5.174 & 10.9 \\
    C/2006 Q1 (McNaught)            & 2008 Jan 17 & 14:30:12 & 15:10:15 & \nodata   & SL1       & LL2       & LL1       & 3.253 & 3.214 & 18.0 \\
    C/2006 Q1 (McNaught)            & 2008 Feb 25 & 07:56:36 & 08:01:45 & \nodata   & SL1       & \nodata   & \nodata   & 3.066 & 2.673 & 18.6 \\
    C/2006 Q1 (McNaught)            & 2008 Jul 02 & 02:41:55 & 02:54:30 & SL2       & SL1       & LL2       & LL1       & 2.764 & 2.375 & 21.1 \\
    C/2006 Q1 (McNaught)            & 2008 Jul 09 & 16:50:43 & 17:31:30 & SL2       & SL1       & \nodata   & \nodata   & 2.764 & 2.475 & 21.5 \\
    C/2007 N3 (Lulin)               & 2008 Oct 04 & 06:45:36 & 07:40:09 & SL2       & SL1       & \nodata   & \nodata   & 1.901 & 1.673 & 32.6 \\
    C/2008 T2 (Cardinal)            & 2009 Apr 04 & 08:25:04 & 09:25:08 & SL2       & SL1       & LL2       & LL1       & 1.607 & 1.088 & 37.9
    \enddata
    \tablenotetext{a}{ \ Start time of integration.}
    \tablenotetext{b}{ \ End time of integration.}
\end{deluxetable*}

%% file: table_2.tex

\startlongtable
\begin{deluxetable}{@{\extracolsep{0pt}}lcccc}

\setlength{\tabcolsep}{2pt} 

\tabletypesize{\footnotesize}
\tablenum{2}
\tablewidth{0pc}
\tablecaption{Color Temperature and Silicate Strength \label{tab:colex}}
\tablehead{
\colhead{} & \\
\colhead{Comet} &  \colhead{T$_{fit}$\tablenotemark{a}} & 
\colhead{T$_{BB}$\tablenotemark{b}} & & \\
\colhead{Name} & \colhead{(K)} & \colhead{(K)} & \colhead{T$_{fit}$/T$_{BB}$} & 
\colhead{F$_{10.5}$/F$_{c}$}
}
\startdata
6P & $261 \pm 1$ & $236$ & $1.106 \pm 0.003$ & $1.134 \pm 0.046$ \\
8P & $291 \pm 2$ & $219$ & $1.328 \pm 0.009$ & $1.061 \pm 0.065$ \\
9P-20050702 & $265 \pm 1$ & $227$ & $1.169 \pm 0.003$ & $1.194 \pm 0.044$ \\
9P-20050703 & $258 \pm 1$ & $227$ & $1.140 \pm 0.003$ & $1.189 \pm 0.045$ \\
17P & $196 \pm 1$ & $176$ & $1.113 \pm 0.006$ & $2.113 \pm 0.122$ \\
21P & $222 \pm 9$ & $184$ & $1.209 \pm 0.050$ & $1.213 \pm 0.411$ \\
29P & $214 \pm 15$ & $116$ & $1.841 \pm 0.126$ & $1.603 \pm 0.926$ \\
37P & $252 \pm 1$ & $211$ & $1.195 \pm 0.004$ & $1.242 \pm 0.043$ \\
41P & $276 \pm 3$ & $246$ & $1.120 \pm 0.011$ & $1.104 \pm 0.083$ \\
46P-20071208 & $267 \pm 0$ & $245$ & $1.093 \pm 0.002$ & $1.123 \pm 0.025$ \\
46P-20080117 & $285 \pm 0$ & $268$ & $1.063 \pm 0.001$ & $1.124 \pm 0.023$ \\
46P-20080424 & $252 \pm 0$ & $227$ & $1.109 \pm 0.002$ & $1.107 \pm 0.026$ \\
46P-20080524 & $232 \pm 0$ & $211$ & $1.100 \pm 0.001$ & $1.096 \pm 0.024$ \\
46P-20080702 & $218 \pm 2$ & $193$ & $1.128 \pm 0.010$ & $1.056 \pm 0.084$ \\
48P & $219 \pm 2$ & $183$ & $1.196 \pm 0.013$ & $1.310 \pm 0.128$ \\
62P & $247 \pm 3$ & $216$ & $1.140 \pm 0.013$ & $1.150 \pm 0.107$ \\
65P & $149 \pm 5$ & $149$ & $0.999 \pm 0.034$ & $1.272 \pm 0.525$ \\
67P-20080629.82 & $166 \pm 0$ & $167$ & $0.996 \pm 0.002$ & $\ldots$ \\
67P-20081128.38 & $229 \pm 1$ & $216$ & $1.057 \pm 0.004$ & $1.229 \pm 0.057$ \\
67P-20081129.11 & $230 \pm 1$ & $217$ & $1.063 \pm 0.003$ & $1.201 \pm 0.042$ \\
67P-20081129.70 & $228 \pm 1$ & $217$ & $1.050 \pm 0.004$ & $1.217 \pm 0.054$ \\
67P-20081130.48 & $233 \pm 1$ & $217$ & $1.074 \pm 0.004$ & $1.190 \pm 0.050$ \\
71P & $256 \pm 0$ & $222$ & $1.151 \pm 0.001$ & $1.107 \pm 0.021$ \\
73P(B)-20060417.59 & $286 \pm 0$ & $255$ & $1.122 \pm 0.001$ & $1.235 \pm 0.024$ \\
73P(B)-20060806.18 & $280 \pm 1$ & $248$ & $1.132 \pm 0.006$ & $1.178 \pm 0.051$ \\
73P(C)-20060317.04 & $263 \pm 0$ & $230$ & $1.144 \pm 0.001$ & $1.336 \pm 0.027$ \\
73P(C)-20060806.05 & $268 \pm 0$ & $247$ & $1.085 \pm 0.001$ & $1.239 \pm 0.025$ \\
78P & $209 \pm 1$ & $194$ & $1.081 \pm 0.003$ & $1.234 \pm 0.039$ \\
88P & $208 \pm 0$ & $197$ & $1.059 \pm 0.002$ & $1.104 \pm 0.032$ \\
105P & $237 \pm 2$ & $192$ & $1.231 \pm 0.008$ & $1.084 \pm 0.065$ \\
121P & $187 \pm 5$ & $161$ & $1.166 \pm 0.028$ & $1.222 \pm 0.297$ \\
123P & $205 \pm 1$ & $188$ & $1.086 \pm 0.007$ & $\ldots$ \\
132P & $227 \pm 4$ & $193$ & $1.176 \pm 0.019$ & $1.258 \pm 0.200$ \\
144P & $260 \pm 1$ & $223$ & $1.167 \pm 0.002$ & $1.052 \pm 0.026$ \\
C2001Q4 & $200 \pm 1$ & $172$ & $1.162 \pm 0.003$ & $1.250 \pm 0.046$ \\
C2003K4-20040716 & $234 \pm 0$ & $210$ & $1.116 \pm 0.002$ & $1.095 \pm 0.028$ \\
C2003K4-20050913 & $132 \pm 2$ & $131$ & $1.010 \pm 0.019$ & $1.160 \pm 0.299$ \\
C2003T3 & $167 \pm 2$ & $147$ & $1.136 \pm 0.013$ & $1.113 \pm 0.140$ \\
C2003T4-20051122 & $181 \pm 2$ & $148$ & $1.218 \pm 0.012$ & $1.081 \pm 0.109$ \\
C2003T4-20060129 & $153 \pm 4$ & $135$ & $1.136 \pm 0.030$ & $1.151 \pm 0.360$ \\
C2003T4-20060307 & $139 \pm 1$ & $129$ & $1.076 \pm 0.007$ & $\ldots$ \\
C2004B1-20051015 & $203 \pm 0$ & $187$ & $1.088 \pm 0.002$ & $1.097 \pm 0.034$ \\
C2004B1-20060516 & $210 \pm 0$ & $194$ & $1.081 \pm 0.002$ & $1.111 \pm 0.029$ \\
C2004B1-20060726 & $171 \pm 1$ & $170$ & $1.010 \pm 0.007$ & $1.201 \pm 0.089$ \\
C2004B1-20070308 & $148 \pm 5$ & $127$ & $1.168 \pm 0.036$ & $1.044 \pm 0.399$ \\
C2004B1-20070609 & $134 \pm 3$ & $117$ & $1.141 \pm 0.027$ & $\ldots$ \\
C2004Q2 & $188 \pm 0$ & $174$ & $1.080 \pm 0.003$ & $1.190 \pm 0.038$ \\
C2006P1-20070504 & $200 \pm 1$ & $180$ & $1.114 \pm 0.003$ & $1.127 \pm 0.042$ \\
C2006P1-20070802 & $157 \pm 1$ & $146$ & $1.077 \pm 0.008$ & $1.219 \pm 0.118$ \\
C2006P1-20070906 & $251 \pm 62$ & $138$ & $1.816 \pm 0.451$ & $0.682 \pm 1.218$ \\
C2006Q1-20070322 & $124 \pm 1$ & $121$ & $1.024 \pm 0.005$ & $\ldots$ \\
C2006Q1-20080117 & $170 \pm 1$ & $154$ & $1.103 \pm 0.007$ & $1.036 \pm 0.078$ \\
C2006Q1-20080225 & $164 \pm 2$ & $159$ & $1.034 \pm 0.015$ & $1.141 \pm 0.190$ \\
C2006Q1-20080702 & $178 \pm 1$ & $167$ & $1.064 \pm 0.005$ & $1.138 \pm 0.059$ \\
C2006Q1-20080709 & $173 \pm 0$ & $167$ & $1.032 \pm 0.001$ & $1.183 \pm 0.030$ \\
C2007N3 & $220 \pm 0$ & $202$ & $1.090 \pm 0.001$ & $1.084 \pm 0.023$ \\
C2008T2 & $246 \pm 0$ & $219$ & $1.123 \pm 0.002$ & $1.069 \pm 0.024$
\enddata
\tablenotetext{a}{ \ Temperature of BB fit to local continuum points defined as between \\ 7.8--8.0 \micron{} and 12.4--12.8 \micron.}
\tablenotetext{b}{ \ T$_{BB} = 278/\sqrt{r_h}$; where the heliocentric distance, $r_h$, is in AU.} 
\end{deluxetable}

%% file: table_3.tex
\begin{longrotatetable}
\begin{deluxetable*}{@{\extracolsep{0pt}}lcccccccccc}
\setlength{\tabcolsep}{2pt} 
\tablenum{3}
\tabletypesize{\footnotesize}
\tablewidth{0pc}
\tablecaption{All Comets Observed in Survey: Best-Fit Thermal Emission Model Parameters \label{tab:modparams}}
\tablehead{
 &  & & & & \multicolumn{5}{c}{\underbar{$N_p (\times 10^{16}$)}}  \\
\colhead{Comet} & & &
 &  & \colhead{Amorphous} &
\colhead{Amorphous} & \colhead{Amorphous} & \colhead{Crystalline} & 
\colhead{Crystalline} & \\
\colhead{Name} &  \colhead{N} & \colhead{M} & \colhead{$a_p$ \tablenotemark{a}} & \colhead{D} & \colhead{Pyroxene} & \colhead{Olivine} &
\colhead{Carbon} & \colhead{Olivine} & \colhead{Orthopyroxene} & \colhead{$\chi^{2}_{\nu}$}
}
\startdata
& \\[-0.9mm]
6P & 4.2 & 29.4 & 0.8 & 2.727 & ${\bf 6.508e+01^{+9.103e+00}_{-2.072e+01}}$ & $2.651e+00^{+1.119e+01}_{-2.651e+00}$ & ${\bf 2.104e+02^{+2.757e+00}_{-2.504e+00}}$ & $1.785e+01^{+2.098e+01}_{-1.785e+01}$ & $0.000e+00^{+3.748e+01}_{-0.000e+00}$ & 0.54 \\[0.9mm]
8P & 3.4 & 6.8 & 0.3 & 2.857 & $0.000e+00^{+0.000e+00}_{-0.000e+00}$ & ${\bf 1.730e+02^{+9.477e+01}_{-1.228e+02}}$ & ${\bf 6.842e+03^{+1.083e+02}_{-9.815e+01}}$ & $7.847e+01^{+1.263e+02}_{-7.847e+01}$ & $3.132e+01^{+2.081e+02}_{-3.132e+01}$ & 0.77 \\[0.9mm]
9P-20050702 & 3.6 & 14.4 & 0.5 & 2.857 & ${\bf 1.268e+03^{+1.598e+02}_{-2.005e+02}}$ & $9.497e+01^{+1.342e+02}_{-9.497e+01}$ & ${\bf 3.948e+03^{+3.893e+01}_{-3.912e+01}}$ & ${\bf 1.684e+02^{+1.120e+02}_{-1.134e+02}}$ & $0.000e+00^{+6.723e+01}_{-0.000e+00}$ & 1.78 \\[0.9mm]
9P-20050703 & 3.9 & 19.5 & 0.6 & 2.857 & ${\bf 1.197e+03^{+5.168e+01}_{-5.202e+01}}$ & $0.000e+00^{+0.000e+00}_{-0.000e+00}$ & ${\bf 3.427e+03^{+3.364e+01}_{-3.297e+01}}$ & ${\bf 1.224e+02^{+1.023e+02}_{-1.009e+02}}$ & $0.000e+00^{+0.000e+00}_{-0.000e+00}$ & 3.54 \\[0.9mm]
17P & 3.6 & 7.2 & 0.3 & 2.857 & ${\bf 2.841e+05^{+2.605e+04}_{-2.603e+04}}$ & ${\bf 3.954e+05^{+1.604e+04}_{-1.579e+04}}$ & ${\bf 2.526e+05^{+3.791e+03}_{-3.701e+03}}$ & ${\bf 1.180e+05^{+1.026e+04}_{-1.069e+04}}$ & $0.000e+00^{+3.425e+03}_{-0.000e+00}$ & 7.55 \\[0.9mm]
21P & 4.2 & 25.2 & 0.7 & 2.727 & ${\bf 1.059e+02^{+4.959e+01}_{-5.158e+01}}$ & ${\bf 1.312e+02^{+2.638e+01}_{-2.711e+01}}$ & ${\bf 7.992e+02^{+8.661e+00}_{-8.103e+00}}$ & $0.000e+00^{+2.012e+01}_{-0.000e+00}$ & $4.399e+01^{+5.952e+01}_{-4.399e+01}$ & 2.83 \\[0.9mm]
29P & 3.3 & 3.3 & 0.2 & 2.857 & ${\bf 7.407e+05^{+1.444e+05}_{-1.502e+05}}$ & ${\bf 6.095e+05^{+1.033e+05}_{-1.036e+05}}$ & ${\bf 1.173e+06^{+2.990e+04}_{-2.978e+04}}$ & ${\bf 4.293e+05^{+4.576e+04}_{-4.699e+04}}$ & $2.763e+04^{+7.237e+04}_{-2.763e+04}$ & 2.04 \\[0.9mm]
37P & 3.9 & 19.5 & 0.6 & 2.727 & ${\bf 6.007e+01^{+1.641e+01}_{-1.812e+01}}$ & ${\bf 4.846e+01^{+9.374e+00}_{-9.073e+00}}$ & ${\bf 2.473e+02^{+2.934e+00}_{-2.676e+00}}$ & ${\bf 1.628e+01^{+9.254e+00}_{-9.257e+00}}$ & $4.603e+00^{+1.731e+01}_{-4.603e+00}$ & 1.86 \\[0.9mm]
41P & 3.4 & 10.2 & 0.4 & 3.0 & $0.000e+00^{+2.360e+01}_{-0.000e+00}$ & ${\bf 3.011e+01^{+4.046e+00}_{-1.318e+01}}$ & ${\bf 2.408e+02^{+3.833e+00}_{-5.369e+00}}$ & ${\bf 1.238e+01^{+9.276e+00}_{-8.913e+00}}$ & $0.000e+00^{+0.000e+00}_{-0.000e+00}$ & 1.40 \\[0.9mm]
46P-20071208 & 4.1 & 28.7 & 0.8 & 2.727 & ${\bf 1.930e+01^{+7.201e+00}_{-7.268e+00}}$ & ${\bf 6.492e+00^{+4.185e+00}_{-4.481e+00}}$ & ${\bf 1.418e+02^{+9.967e-01}_{-9.827e-01}}$ & $5.801e+00^{+9.428e+00}_{-5.801e+00}$ & $1.025e+01^{+1.802e+01}_{-1.025e+01}$ & 0.47 \\[0.9mm]
46P-20080117 & 3.8 & 22.8 & 0.7 & 2.727 & ${\bf 2.300e+01^{+9.855e+00}_{-9.575e+00}}$ & ${\bf 9.890e+00^{+6.208e+00}_{-6.315e+00}}$ & ${\bf 2.370e+02^{+1.376e+00}_{-1.440e+00}}$ & ${\bf 1.389e+01^{+1.228e+01}_{-1.239e+01}}$ & ${\bf 3.851e+01^{+2.342e+01}_{-2.386e+01}}$ & 0.52 \\[0.9mm]
46P-20080424 & 3.5 & 17.5 & 0.6 & 2.727 & ${\bf 2.930e+01^{+7.532e+00}_{-7.567e+00}}$ & ${\bf 6.290e+00^{+4.493e+00}_{-4.627e+00}}$ & ${\bf 1.691e+02^{+1.179e+00}_{-1.160e+00}}$ & $0.000e+00^{+4.952e+00}_{-0.000e+00}$ & $0.000e+00^{+2.089e+00}_{-0.000e+00}$ & 1.48 \\[0.9mm]
46P-20080524 & 4.0 & 36.0 & 1.0 & 2.727 & ${\bf 1.542e+01^{+1.887e+00}_{-2.353e+00}}$ & $0.000e+00^{+1.006e+00}_{-0.000e+00}$ & ${\bf 1.369e+02^{+8.240e-01}_{-8.361e-01}}$ & $0.000e+00^{+5.258e+00}_{-0.000e+00}$ & ${\bf 2.687e+01^{+2.127e+01}_{-2.139e+01}}$ & 2.08 \\[0.9mm]
46P-20080702 & 3.4 & 17.0 & 0.6 & 3.0 & ${\bf 1.374e+01^{+6.005e+00}_{-5.970e+00}}$ & ${\bf 1.312e+01^{+5.174e+00}_{-5.583e+00}}$ & ${\bf 1.680e+02^{+2.807e+00}_{-2.618e+00}}$ & $0.000e+00^{+0.000e+00}_{-0.000e+00}$ & $0.000e+00^{+1.002e+01}_{-0.000e+00}$ & 2.02 \\[0.9mm]
48P & 3.3 & 9.9 & 0.4 & 3.0 & $0.000e+00^{+0.000e+00}_{-0.000e+00}$ & ${\bf 1.678e+03^{+1.108e+02}_{-1.069e+02}}$ & ${\bf 4.171e+03^{+1.103e+02}_{-1.122e+02}}$ & ${\bf 6.141e+02^{+1.583e+02}_{-1.525e+02}}$ & ${\bf 4.646e+02^{+2.163e+02}_{-2.157e+02}}$ & 1.22 \\[0.9mm]
62P & 3.3 & 9.9 & 0.4 & 3.0 & ${\bf 3.879e+01^{+7.478e+00}_{-2.145e+01}}$ & $0.000e+00^{+9.366e+00}_{-0.000e+00}$ & ${\bf 5.807e+01^{+6.528e+00}_{-4.562e+00}}$ & $1.750e+00^{+4.118e+00}_{-1.750e+00}$ & $7.777e+00^{+8.414e+00}_{-7.777e+00}$ & 0.61 \\[0.9mm]
65P & 3.7 & 37.0 & 1.1 & 3.0 & $0.000e+00^{+0.000e+00}_{-0.000e+00}$ & ${\bf 1.120e+02^{+1.074e+01}_{-1.089e+01}}$ & ${\bf 3.778e+02^{+1.113e+01}_{-1.085e+01}}$ & ${\bf 1.719e+02^{+1.023e+02}_{-1.003e+02}}$ & ${\bf 4.544e+02^{+1.503e+02}_{-1.566e+02}}$ & 2.95 \\[0.9mm]
67P-20080629.82 & 3.3 & 16.5 & 0.6 & 3.0 & ${\bf 1.622e+02^{+3.101e+01}_{-4.886e+01}}$ & $0.000e+00^{+5.428e+01}_{-0.000e+00}$ & ${\bf 9.351e+01^{+2.864e+01}_{-4.318e+01}}$ & $5.980e+01^{+5.811e+01}_{-5.980e+01}$ & $0.000e+00^{+0.000e+00}_{-0.000e+00}$ & 0.84 \\[0.9mm]
67P-20081128.38 & 4.2 & 37.8 & 1.0 & 2.727 & ${\bf 2.798e+01^{+1.447e+01}_{-1.550e+01}}$ & ${\bf 4.097e+01^{+8.921e+00}_{-8.667e+00}}$ & ${\bf 2.473e+02^{+2.926e+00}_{-2.732e+00}}$ & $0.000e+00^{+0.000e+00}_{-0.000e+00}$ & $1.598e+01^{+5.437e+01}_{-1.598e+01}$ & 3.09 \\[0.9mm]
67P-20081129.11 & 3.7 & 25.9 & 0.8 & 2.727 & $0.000e+00^{+2.481e+00}_{-0.000e+00}$ & ${\bf 5.960e+01^{+3.635e+00}_{-4.640e+00}}$ & ${\bf 2.569e+02^{+3.267e+00}_{-3.118e+00}}$ & $0.000e+00^{+1.519e+01}_{-0.000e+00}$ & $2.290e+01^{+3.764e+01}_{-2.290e+01}$ & 1.90 \\[0.9mm]
67P-20081129.70 & 4.2 & 42.0 & 1.1 & 2.727 & ${\bf 1.069e+01^{+1.008e+01}_{-1.031e+01}}$ & ${\bf 3.593e+01^{+6.001e+00}_{-6.293e+00}}$ & ${\bf 1.712e+02^{+1.879e+00}_{-1.824e+00}}$ & $0.000e+00^{+0.000e+00}_{-0.000e+00}$ & $0.000e+00^{+2.744e+01}_{-0.000e+00}$ & 2.78 \\[0.9mm]
67P-20081130.48 & 3.5 & 21.0 & 0.7 & 2.727 & ${\bf 4.739e+01^{+1.873e+01}_{-1.876e+01}}$ & ${\bf 4.762e+01^{+1.155e+01}_{-1.173e+01}}$ & ${\bf 3.042e+02^{+3.542e+00}_{-3.505e+00}}$ & $0.000e+00^{+6.446e+00}_{-0.000e+00}$ & $0.000e+00^{+0.000e+00}_{-0.000e+00}$ & 2.78 \\[0.9mm]
71P & 3.7 & 18.5 & 0.6 & 2.727 & ${\bf 4.766e+02^{+9.243e+01}_{-9.227e+01}}$ & ${\bf 6.911e+01^{+5.567e+01}_{-5.622e+01}}$ & ${\bf 2.491e+03^{+1.498e+01}_{-1.495e+01}}$ & ${\bf 8.001e+01^{+6.996e+01}_{-6.772e+01}}$ & $0.000e+00^{+0.000e+00}_{-0.000e+00}$ & 2.27 \\[0.9mm]
73P(B)-20060417.59 & 4.3 & 25.8 & 0.7 & 2.727 & ${\bf 5.319e+01^{+1.638e+00}_{-2.764e+00}}$ & $0.000e+00^{+0.000e+00}_{-0.000e+00}$ & ${\bf 1.258e+02^{+6.590e-01}_{-5.952e-01}}$ & ${\bf 9.576e+00^{+4.200e+00}_{-4.171e+00}}$ & $2.018e-01^{+8.586e+00}_{-2.018e-01}$ & 1.16 \\[0.9mm]
73P(B)-20060806.18 & 3.8 & 15.2 & 0.5 & 2.857 & ${\bf 1.038e+02^{+3.019e+00}_{-3.013e+00}}$ & $0.000e+00^{+0.000e+00}_{-0.000e+00}$ & ${\bf 2.816e+02^{+1.796e+00}_{-1.807e+00}}$ & ${\bf 5.965e+00^{+4.670e+00}_{-4.585e+00}}$ & $0.000e+00^{+0.000e+00}_{-0.000e+00}$ & 2.52 \\[0.9mm]
73P(C)-20060317.04 & 4.3 & 25.8 & 0.7 & 2.727 & ${\bf 3.134e+02^{+1.263e+01}_{-1.274e+01}}$ & $0.000e+00^{+0.000e+00}_{-0.000e+00}$ & ${\bf 5.054e+02^{+2.459e+00}_{-2.456e+00}}$ & ${\bf 5.649e+01^{+2.121e+01}_{-2.073e+01}}$ & ${\bf 6.845e+01^{+4.133e+01}_{-4.242e+01}}$ & 2.08 \\[0.9mm]
73P(C)-20060806.05 & 3.3 & 72.6 & 2.3 & 2.609 & ${\bf 1.181e+00^{+2.887e-02}_{-4.355e-02}}$ & $0.000e+00^{+0.000e+00}_{-0.000e+00}$ & ${\bf 1.973e+00^{+9.653e-03}_{-9.049e-03}}$ & $8.511e+01^{+1.234e+02}_{-8.511e+01}$ & $0.000e+00^{+2.079e+02}_{-0.000e+00}$ & 1.56 \\[0.9mm]
78P & 3.8 & 22.8 & 0.7 & 3.0 & ${\bf 1.275e+03^{+2.051e+02}_{-1.986e+02}}$ & ${\bf 1.576e+03^{+1.562e+02}_{-1.580e+02}}$ & ${\bf 5.128e+03^{+5.983e+01}_{-5.775e+01}}$ & ${\bf 1.442e+03^{+2.491e+02}_{-2.440e+02}}$ & $2.749e+02^{+3.728e+02}_{-2.749e+02}$ & 1.83 \\[0.9mm]
88P & 3.4 & 20.4 & 0.7 & 3.0 & ${\bf 1.380e+02^{+2.538e+01}_{-2.511e+01}}$ & ${\bf 1.412e+02^{+2.288e+01}_{-2.344e+01}}$ & ${\bf 6.439e+02^{+1.118e+01}_{-1.129e+01}}$ & ${\bf 6.478e+01^{+4.476e+01}_{-4.641e+01}}$ & $0.000e+00^{+0.000e+00}_{-0.000e+00}$ & 1.23 \\[0.9mm]
105P & 3.3 & 9.9 & 0.4 & 3.0 & ${\bf 3.560e+02^{+7.779e+01}_{-1.174e+02}}$ & $0.000e+00^{+4.371e+01}_{-0.000e+00}$ & ${\bf 9.173e+02^{+4.657e+01}_{-4.096e+01}}$ & $0.000e+00^{+1.857e+01}_{-0.000e+00}$ & ${\bf 1.034e+02^{+9.495e+01}_{-9.569e+01}}$ & 1.04 \\[0.9mm]
121P & 3.3 & 16.5 & 0.6 & 3.0 & $4.769e+02^{+1.350e+02}_{-4.769e+02}$ & $0.000e+00^{+1.968e+02}_{-0.000e+00}$ & ${\bf 2.846e+02^{+1.055e+02}_{-6.917e+01}}$ & ${\bf 1.748e+02^{+1.224e+02}_{-1.331e+02}}$ & $1.792e+02^{+2.310e+02}_{-1.792e+02}$ & 0.77 \\[0.9mm]
123P & 3.3 & 9.9 & 0.4 & 3.0 & ${\bf 1.572e+03^{+3.445e+02}_{-3.423e+02}}$ & ${\bf 9.833e+02^{+3.249e+02}_{-3.361e+02}}$ & ${\bf 6.517e+03^{+2.692e+02}_{-2.577e+02}}$ & ${\bf 9.647e+02^{+2.367e+02}_{-2.439e+02}}$ & $2.009e+02^{+3.712e+02}_{-2.009e+02}$ & 1.04 \\[0.9mm]
132P & 3.3 & 33.0 & 1.1 & 2.609 & $6.374e+00^{+3.423e+00}_{-6.374e+00}$ & $6.364e-01^{+2.627e+00}_{-6.364e-01}$ & ${\bf 9.691e+00^{+8.333e-01}_{-7.153e-01}}$ & $0.000e+00^{+3.036e+01}_{-0.000e+00}$ & $6.645e+01^{+9.299e+01}_{-6.645e+01}$ & 0.61 \\[0.9mm]
144P & 3.3 & 9.9 & 0.4 & 2.727 & ${\bf 1.481e+02^{+9.298e+00}_{-9.444e+00}}$ & $0.000e+00^{+0.000e+00}_{-0.000e+00}$ & ${\bf 5.666e+02^{+4.506e+00}_{-4.539e+00}}$ & $0.000e+00^{+0.000e+00}_{-0.000e+00}$ & $0.000e+00^{+0.000e+00}_{-0.000e+00}$ & 2.05 \\[0.9mm]
C2001Q4 & 3.3 & 13.2 & 0.5 & 3.0 & ${\bf 9.086e+03^{+1.135e+03}_{-1.150e+03}}$ & ${\bf 2.267e+03^{+7.569e+02}_{-7.557e+02}}$ & ${\bf 1.016e+04^{+1.138e+02}_{-1.139e+02}}$ & ${\bf 2.950e+03^{+8.416e+02}_{-8.461e+02}}$ & ${\bf 2.826e+03^{+1.416e+03}_{-1.437e+03}}$ & 1.99 \\[0.9mm]
C2003K4-20040716 & 3.6 & 21.6 & 0.7 & 2.727 & ${\bf 8.842e+02^{+6.539e+01}_{-7.802e+01}}$ & $0.000e+00^{+0.000e+00}_{-0.000e+00}$ & ${\bf 3.640e+03^{+2.259e+01}_{-2.210e+01}}$ & $0.000e+00^{+1.490e+02}_{-0.000e+00}$ & $3.256e+02^{+4.489e+02}_{-3.256e+02}$ & 2.07 \\[0.9mm]
C2003K4-20050913 & 3.3 & 39.6 & 1.3 & 3.0 & ${\bf 7.149e+02^{+1.910e+02}_{-1.843e+02}}$ & ${\bf 2.049e+03^{+2.229e+02}_{-2.319e+02}}$ & ${\bf 2.825e+03^{+1.193e+02}_{-1.159e+02}}$ & $0.000e+00^{+6.095e+02}_{-0.000e+00}$ & $0.000e+00^{+0.000e+00}_{-0.000e+00}$ & 2.70 \\[0.9mm]
C2003T3 & 3.5 & 31.5 & 1.0 & 2.857 & ${\bf 5.670e+02^{+3.161e+01}_{-3.694e+01}}$ & $0.000e+00^{+0.000e+00}_{-0.000e+00}$ & ${\bf 7.589e+02^{+2.457e+01}_{-2.345e+01}}$ & ${\bf 4.062e+02^{+2.504e+02}_{-2.677e+02}}$ & $0.000e+00^{+3.703e+02}_{-0.000e+00}$ & 1.41 \\[0.9mm]
C2003T4-20051122 & 3.6 & 28.8 & 0.9 & 3.0 & ${\bf 1.337e+03^{+9.425e+01}_{-9.456e+01}}$ & ${\bf 2.054e+02^{+9.407e+01}_{-9.392e+01}}$ & ${\bf 1.444e+03^{+3.780e+01}_{-3.921e+01}}$ & ${\bf 1.337e+03^{+2.570e+02}_{-2.561e+02}}$ & ${\bf 1.036e+03^{+3.918e+02}_{-3.705e+02}}$ & 5.45 \\[0.9mm]
C2003T4-20060129 & 3.7 & 40.7 & 1.2 & 2.727 & $0.000e+00^{+0.000e+00}_{-0.000e+00}$ & ${\bf 1.295e+02^{+1.786e+01}_{-2.153e+01}}$ & ${\bf 1.118e+03^{+1.895e+01}_{-1.727e+01}}$ & ${\bf 1.319e+03^{+4.511e+02}_{-4.531e+02}}$ & $0.000e+00^{+4.933e+02}_{-0.000e+00}$ & 3.59 \\[0.9mm]
C2003T4-20060307 & 3.3 & 16.5 & 0.6 & 2.857 & ${\bf 2.196e+03^{+1.261e+02}_{-1.253e+02}}$ & $0.000e+00^{+0.000e+00}_{-0.000e+00}$ & ${\bf 3.002e+03^{+8.862e+01}_{-8.929e+01}}$ & ${\bf 8.969e+02^{+2.926e+02}_{-3.003e+02}}$ & ${\bf 6.357e+02^{+4.511e+02}_{-4.566e+02}}$ & 4.72 \\[0.9mm]
C2004B1-20051015 & 4.2 & 46.2 & 1.2 & 2.727 & ${\bf 2.222e+02^{+1.689e+01}_{-2.362e+01}}$ & $0.000e+00^{+0.000e+00}_{-0.000e+00}$ & ${\bf 9.621e+02^{+6.291e+00}_{-6.104e+00}}$ & $0.000e+00^{+2.111e+02}_{-0.000e+00}$ & $1.984e+02^{+6.119e+02}_{-1.984e+02}$ & 3.42 \\[0.9mm]
C2004B1-20060516 & 4.2 & 42.0 & 1.1 & 2.727 & ${\bf 1.780e+02^{+1.528e+01}_{-2.453e+01}}$ & $0.000e+00^{+7.585e+00}_{-0.000e+00}$ & ${\bf 6.609e+02^{+5.986e+00}_{-5.528e+00}}$ & $0.000e+00^{+1.546e+02}_{-0.000e+00}$ & $1.630e+02^{+3.030e+02}_{-1.630e+02}$ & 1.08 \\[0.9mm]
C2004B1-20060726 & 3.5 & 31.5 & 1.0 & 2.727 & $0.000e+00^{+0.000e+00}_{-0.000e+00}$ & ${\bf 5.426e+01^{+5.314e+00}_{-6.661e+00}}$ & ${\bf 3.363e+02^{+5.320e+00}_{-4.718e+00}}$ & $4.315e+00^{+7.078e+01}_{-4.315e+00}$ & $8.920e+01^{+1.206e+02}_{-8.920e+01}$ & 1.67 \\[0.9mm]
C2004B1-20070308 & 3.6 & 32.4 & 1.0 & 2.857 & $0.000e+00^{+0.000e+00}_{-0.000e+00}$ & $0.000e+00^{+0.000e+00}_{-0.000e+00}$ & ${\bf 2.310e+03^{+7.643e+00}_{-7.767e+00}}$ & $0.000e+00^{+8.210e+01}_{-0.000e+00}$ & $0.000e+00^{+0.000e+00}_{-0.000e+00}$ & 5.46 \\[0.9mm]
C2004B1-20070609 & 3.3 & 19.8 & 0.7 & 3.0 & ${\bf 1.912e+03^{+1.318e+02}_{-1.276e+02}}$ & $0.000e+00^{+0.000e+00}_{-0.000e+00}$ & ${\bf 1.980e+03^{+1.076e+02}_{-1.149e+02}}$ & $0.000e+00^{+0.000e+00}_{-0.000e+00}$ & ${\bf 5.937e+02^{+5.146e+02}_{-5.193e+02}}$ & 27.07 \\[0.9mm]
C2004Q2 & 4.3 & 60.2 & 1.5 & 2.727 & ${\bf 8.320e+02^{+2.727e+01}_{-2.795e+01}}$ & $0.000e+00^{+0.000e+00}_{-0.000e+00}$ & ${\bf 1.199e+03^{+1.438e+01}_{-1.443e+01}}$ & $0.000e+00^{+0.000e+00}_{-0.000e+00}$ & $0.000e+00^{+5.452e+02}_{-0.000e+00}$ & 4.96 \\[0.9mm]
C2006P1-20070504 & 3.6 & 28.8 & 0.9 & 2.727 & ${\bf 2.159e+03^{+9.194e+01}_{-1.195e+02}}$ & $0.000e+00^{+2.079e+01}_{-0.000e+00}$ & ${\bf 8.451e+03^{+5.688e+01}_{-5.334e+01}}$ & $1.981e+02^{+9.666e+02}_{-1.981e+02}$ & $0.000e+00^{+0.000e+00}_{-0.000e+00}$ & 2.47 \\[0.9mm]
C2006P1-20070802 & 3.6 & 32.4 & 1.0 & 3.0 & $0.000e+00^{+0.000e+00}_{-0.000e+00}$ & ${\bf 1.923e+03^{+6.671e+01}_{-6.883e+01}}$ & ${\bf 2.819e+03^{+7.241e+01}_{-7.225e+01}}$ & $1.826e+02^{+6.355e+02}_{-1.826e+02}$ & $0.000e+00^{+0.000e+00}_{-0.000e+00}$ & 4.01 \\[0.9mm]
C2006P1-20070906 & 4.0 & 48.0 & 1.3 & 3.0 & $8.127e+00^{+1.137e+02}_{-8.127e+00}$ & ${\bf 4.505e+02^{+5.352e+01}_{-1.280e+02}}$ & ${\bf 1.999e+03^{+6.401e+01}_{-5.219e+01}}$ & $6.600e+02^{+9.028e+02}_{-6.600e+02}$ & ${\bf 3.097e+03^{+1.265e+03}_{-1.239e+03}}$ & 7.32 \\[0.9mm]
C2006Q1-20070322 & 3.7 & 48.1 & 1.4 & 2.857 & ${\bf 5.170e+02^{+7.135e+01}_{-1.453e+02}}$ & $0.000e+00^{+1.233e+02}_{-0.000e+00}$ & ${\bf 2.751e+03^{+5.992e+01}_{-6.826e+01}}$ & $1.735e+03^{+2.131e+03}_{-1.735e+03}$ & $0.000e+00^{+1.100e+03}_{-0.000e+00}$ & 2.93 \\[0.9mm]
C2006Q1-20080117 & 4.0 & 44.0 & 1.2 & 2.857 & ${\bf 3.305e+02^{+8.840e+01}_{-1.536e+02}}$ & $4.224e+01^{+1.149e+02}_{-4.224e+01}$ & ${\bf 3.030e+03^{+4.637e+01}_{-4.444e+01}}$ & $1.402e+02^{+6.824e+02}_{-1.402e+02}$ & $0.000e+00^{+7.945e+02}_{-0.000e+00}$ & 0.90 \\[0.9mm]
C2006Q1-20080225 & 3.4 & 27.2 & 0.9 & 3.0 & ${\bf 1.915e+03^{+3.372e+02}_{-5.831e+02}}$ & $0.000e+00^{+2.369e+02}_{-0.000e+00}$ & ${\bf 1.917e+03^{+2.365e+02}_{-2.000e+02}}$ & $2.227e+02^{+7.057e+02}_{-2.227e+02}$ & $0.000e+00^{+1.367e+02}_{-0.000e+00}$ & 1.18 \\[0.9mm]
C2006Q1-20080702 & 3.5 & 28.0 & 0.9 & 3.0 & $0.000e+00^{+1.658e+01}_{-0.000e+00}$ & ${\bf 1.001e+03^{+4.165e+01}_{-4.650e+01}}$ & ${\bf 2.738e+03^{+4.120e+01}_{-4.060e+01}}$ & $2.536e+02^{+3.453e+02}_{-2.536e+02}$ & $4.930e+02^{+5.056e+02}_{-4.930e+02}$ & 1.39 \\[0.9mm]
C2006Q1-20080709 & 3.3 & 23.1 & 0.8 & 3.0 & ${\bf 2.439e+03^{+4.492e+02}_{-4.467e+02}}$ & ${\bf 4.807e+02^{+2.945e+02}_{-3.072e+02}}$ & ${\bf 2.113e+03^{+4.296e+01}_{-4.152e+01}}$ & $3.281e+02^{+5.855e+02}_{-3.281e+02}$ & $7.939e+02^{+9.225e+02}_{-7.939e+02}$ & 1.24 \\[0.9mm]
C2007N3 & 4.0 & 36.0 & 1.0 & 2.727 & ${\bf 8.039e+02^{+6.671e+01}_{-9.628e+01}}$ & $0.000e+00^{+1.859e+01}_{-0.000e+00}$ & ${\bf 4.229e+03^{+2.448e+01}_{-2.249e+01}}$ & $6.668e+01^{+6.136e+02}_{-6.668e+01}$ & $6.034e+02^{+1.094e+03}_{-6.034e+02}$ & 1.26 \\[0.9mm]
C2008T2 & 3.6 & 18.0 & 0.6 & 2.857 & ${\bf 1.620e+02^{+4.142e+01}_{-4.281e+01}}$ & ${\bf 4.025e+01^{+2.714e+01}_{-2.662e+01}}$ & ${\bf 1.138e+03^{+9.314e+00}_{-9.569e+00}}$ & $0.000e+00^{+0.000e+00}_{-0.000e+00}$ & $0.000e+00^{+0.000e+00}_{-0.000e+00}$ & 1.95 \\[0.9mm]
\enddata
\tablecomments{\, Bold values indicate values constrained to a confidence level of 95\%.}
\tablenotetext{a}{ \ Derived parameter.}
\end{deluxetable*}
\end{longrotatetable}

%% file: table_4.tex
\startlongtable
\begin{deluxetable*}{@{\extracolsep{0pt}}lcccccccc}
\setlength{\tabcolsep}{2pt} 
\tablenum{4}
\tabletypesize{\footnotesize}
\tablewidth{0pc}
\tablecaption{All Comets Observed in Survey: Mass of Submicron Grains \label{tab:minmass}}
\tablehead{
 & & \multicolumn{5}{c}{\underbar{Mass Relative to Total Mass}} & & \\
\colhead{Comet} & \colhead{Total Mass} & \colhead{Amorphous} & \colhead{Amorphous} & \colhead{Amorphous} & \colhead{Crystalline} & \colhead{Crystalline} & \colhead{Silicate/Carbon} & \\
\colhead{Name} & \colhead{($\times 10^{7}$ kg)} & \colhead{Pyroxene} & \colhead{Olivine} & \colhead{Carbon} & \colhead{Olivine} & 
\colhead{Orthopyroxene} & \colhead{Ratio} &  \colhead{$f_{cryst}$\tablenotemark{a}}
}
\startdata
6P & $3.393e+03^{+1.267e+03}_{-3.393e+03}$ & $0.334^{+0.078}_{-0.126}$ & $0.014^{+0.064}_{-0.014}$ & $0.491^{+0.098}_{-0.133}$ & $0.162^{+0.128}_{-0.162}$ & $0.000^{+0.257}_{-0.000}$ & $1.039^{+0.761}_{-0.341}$ & $0.318^{+0.325}_{-0.318}$ \\[0.9mm]
8P & $2.398e+04^{+1.683e+03}_{-2.398e+04}$ & $0.000^{+0.000}_{-0.000}$ & $0.050^{+0.028}_{-0.036}$ & $0.908^{+0.039}_{-0.058}$ & $0.030^{+0.045}_{-0.030}$ & $0.012^{+0.075}_{-0.012}$ & $0.102^{+0.076}_{-0.045}$ & $0.454^{+0.421}_{-0.454}$ \\[0.9mm]
9P-20050702 & $5.740e+04^{+2.683e+03}_{-5.740e+04}$ & $0.375^{+0.046}_{-0.055}$ & $0.028^{+0.042}_{-0.028}$ & $0.531^{+0.025}_{-0.025}$ & $0.066^{+0.040}_{-0.044}$ & $0.000^{+0.026}_{-0.000}$ & $0.885^{+0.091}_{-0.086}$ & $0.141^{+0.084}_{-0.089}$ \\[0.9mm]
9P-20050703 & $5.907e+04^{+2.375e+03}_{-5.907e+04}$ & $0.410^{+0.027}_{-0.026}$ & $0.000^{+0.000}_{-0.000}$ & $0.534^{+0.023}_{-0.022}$ & $0.056^{+0.043}_{-0.046}$ & $0.000^{+0.000}_{-0.000}$ & $0.873^{+0.079}_{-0.077}$ & $0.120^{+0.084}_{-0.097}$ \\[0.9mm]
17P & $6.212e+06^{+9.948e+04}_{-6.212e+06}$ & $0.299^{+0.024}_{-0.025}$ & $0.417^{+0.022}_{-0.022}$ & $0.121^{+0.003}_{-0.003}$ & $0.163^{+0.012}_{-0.013}$ & $0.000^{+0.005}_{-0.000}$ & $7.263^{+0.204}_{-0.195}$ & $0.185^{+0.015}_{-0.015}$ \\[0.9mm]
21P & $1.150e+04^{+1.630e+03}_{-1.150e+04}$ & $0.156^{+0.078}_{-0.079}$ & $0.194^{+0.052}_{-0.049}$ & $0.536^{+0.060}_{-0.065}$ & $0.000^{+0.050}_{-0.000}$ & $0.114^{+0.123}_{-0.114}$ & $0.866^{+0.259}_{-0.189}$ & $0.246^{+0.221}_{-0.246}$ \\[0.9mm]
29P & $6.037e+06^{+1.993e+05}_{-6.037e+06}$ & $0.300^{+0.054}_{-0.058}$ & $0.246^{+0.047}_{-0.046}$ & $0.216^{+0.008}_{-0.008}$ & $0.224^{+0.021}_{-0.023}$ & $0.014^{+0.037}_{-0.014}$ & $3.638^{+0.175}_{-0.158}$ & $0.304^{+0.044}_{-0.034}$ \\[0.9mm]
37P & $3.998e+03^{+4.529e+02}_{-3.998e+03}$ & $0.233^{+0.063}_{-0.073}$ & $0.188^{+0.047}_{-0.042}$ & $0.437^{+0.036}_{-0.043}$ & $0.110^{+0.053}_{-0.060}$ & $0.031^{+0.105}_{-0.031}$ & $1.290^{+0.251}_{-0.173}$ & $0.251^{+0.158}_{-0.126}$ \\[0.9mm]
41P & $2.511e+03^{+2.181e+02}_{-2.511e+03}$ & $0.000^{+0.146}_{-0.000}$ & $0.198^{+0.030}_{-0.095}$ & $0.720^{+0.037}_{-0.066}$ & $0.081^{+0.053}_{-0.058}$ & $0.000^{+0.000}_{-0.000}$ & $0.388^{+0.141}_{-0.068}$ & $0.291^{+0.141}_{-0.200}$ \\[0.9mm]
46P-20071208 & $2.070e+03^{+6.781e+02}_{-2.070e+03}$ & $0.163^{+0.072}_{-0.062}$ & $0.055^{+0.053}_{-0.041}$ & $0.543^{+0.156}_{-0.134}$ & $0.087^{+0.109}_{-0.087}$ & $0.153^{+0.177}_{-0.153}$ & $0.841^{+0.602}_{-0.412}$ & $0.524^{+0.213}_{-0.490}$ \\[0.9mm]
46P-20080117 & $3.998e+03^{+8.501e+02}_{-3.998e+03}$ & $0.099^{+0.045}_{-0.040}$ & $0.042^{+0.039}_{-0.029}$ & $0.463^{+0.125}_{-0.081}$ & $0.105^{+0.075}_{-0.092}$ & $0.291^{+0.106}_{-0.153}$ & $1.161^{+0.461}_{-0.459}$ & $0.737^{+0.089}_{-0.201}$ \\[0.9mm]
46P-20080424 & $1.787e+03^{+1.531e+02}_{-1.787e+03}$ & $0.261^{+0.056}_{-0.065}$ & $0.056^{+0.042}_{-0.042}$ & $0.683^{+0.020}_{-0.054}$ & $0.000^{+0.071}_{-0.000}$ & $0.000^{+0.031}_{-0.000}$ & $0.463^{+0.126}_{-0.042}$ & $0.000^{+0.216}_{-0.000}$ \\[0.9mm]
46P-20080524 & $1.966e+03^{+5.928e+02}_{-1.966e+03}$ & $0.123^{+0.066}_{-0.039}$ & $0.000^{+0.008}_{-0.000}$ & $0.495^{+0.202}_{-0.114}$ & $0.000^{+0.070}_{-0.000}$ & $0.382^{+0.145}_{-0.272}$ & $1.019^{+0.605}_{-0.586}$ & $0.757^{+0.105}_{-0.378}$ \\[0.9mm]
46P-20080702 & $2.880e+03^{+2.602e+02}_{-2.880e+03}$ & $0.133^{+0.054}_{-0.058}$ & $0.127^{+0.050}_{-0.057}$ & $0.740^{+0.020}_{-0.060}$ & $0.000^{+0.000}_{-0.000}$ & $0.000^{+0.089}_{-0.000}$ & $0.352^{+0.120}_{-0.035}$ & $0.000^{+0.283}_{-0.000}$ \\[0.9mm]
48P & $7.841e+04^{+3.219e+03}_{-7.841e+04}$ & $0.000^{+0.000}_{-0.000}$ & $0.361^{+0.033}_{-0.031}$ & $0.408^{+0.019}_{-0.018}$ & $0.132^{+0.032}_{-0.031}$ & $0.100^{+0.042}_{-0.045}$ & $1.454^{+0.115}_{-0.112}$ & $0.391^{+0.057}_{-0.064}$ \\[0.9mm]
62P & $1.259e+03^{+9.606e+01}_{-1.259e+03}$ & $0.519^{+0.110}_{-0.253}$ & $0.000^{+0.147}_{-0.000}$ & $0.353^{+0.099}_{-0.037}$ & $0.023^{+0.054}_{-0.023}$ & $0.104^{+0.106}_{-0.104}$ & $1.830^{+0.333}_{-0.620}$ & $0.197^{+0.196}_{-0.180}$ \\[0.9mm]
65P & $2.321e+04^{+4.028e+03}_{-2.321e+04}$ & $0.000^{+0.000}_{-0.000}$ & $0.123^{+0.034}_{-0.024}$ & $0.189^{+0.039}_{-0.027}$ & $0.189^{+0.104}_{-0.106}$ & $0.499^{+0.104}_{-0.124}$ & $4.299^{+0.903}_{-0.913}$ & $0.848^{+0.033}_{-0.051}$ \\[0.9mm]
67P-20080629.82 & $7.420e+03^{+1.561e+03}_{-7.420e+03}$ & $0.613^{+0.183}_{-0.189}$ & $0.000^{+0.216}_{-0.000}$ & $0.161^{+0.077}_{-0.078}$ & $0.226^{+0.148}_{-0.226}$ & $0.000^{+0.000}_{-0.000}$ & $5.222^{+5.864}_{-2.012}$ & $0.269^{+0.173}_{-0.269}$ \\[0.9mm]
67P-20081128.38 & $3.260e+03^{+1.442e+03}_{-3.260e+03}$ & $0.133^{+0.084}_{-0.082}$ & $0.195^{+0.070}_{-0.068}$ & $0.536^{+0.097}_{-0.162}$ & $0.000^{+0.000}_{-0.000}$ & $0.136^{+0.280}_{-0.136}$ & $0.867^{+0.812}_{-0.287}$ & $0.293^{+0.372}_{-0.293}$ \\[0.9mm]
67P-20081129.11 & $3.815e+03^{+1.224e+03}_{-3.815e+03}$ & $0.000^{+0.011}_{-0.000}$ & $0.275^{+0.075}_{-0.077}$ & $0.538^{+0.122}_{-0.129}$ & $0.000^{+0.115}_{-0.000}$ & $0.187^{+0.191}_{-0.187}$ & $0.857^{+0.585}_{-0.343}$ & $0.405^{+0.259}_{-0.405}$ \\[0.9mm]
67P-20081129.70 & $1.719e+03^{+6.593e+02}_{-1.719e+03}$ & $0.086^{+0.074}_{-0.083}$ & $0.289^{+0.057}_{-0.089}$ & $0.625^{+0.024}_{-0.173}$ & $0.000^{+0.000}_{-0.000}$ & $0.000^{+0.285}_{-0.000}$ & $0.599^{+0.610}_{-0.058}$ & $0.000^{+0.523}_{-0.000}$ \\[0.9mm]
67P-20081130.48 & $4.044e+03^{+2.146e+02}_{-4.044e+03}$ & $0.203^{+0.072}_{-0.077}$ & $0.204^{+0.056}_{-0.056}$ & $0.593^{+0.024}_{-0.031}$ & $0.000^{+0.046}_{-0.000}$ & $0.000^{+0.000}_{-0.000}$ & $0.687^{+0.094}_{-0.064}$ & $0.000^{+0.108}_{-0.000}$ \\[0.9mm]
71P & $2.856e+04^{+1.900e+03}_{-2.856e+04}$ & $0.262^{+0.048}_{-0.047}$ & $0.038^{+0.033}_{-0.031}$ & $0.623^{+0.043}_{-0.039}$ & $0.077^{+0.059}_{-0.064}$ & $0.000^{+0.000}_{-0.000}$ & $0.605^{+0.107}_{-0.104}$ & $0.204^{+0.126}_{-0.167}$ \\[0.9mm]
73P(B)-20060417.59 & $2.158e+03^{+2.505e+02}_{-2.158e+03}$ & $0.417^{+0.028}_{-0.060}$ & $0.000^{+0.000}_{-0.000}$ & $0.448^{+0.023}_{-0.047}$ & $0.132^{+0.046}_{-0.055}$ & $0.003^{+0.107}_{-0.003}$ & $1.231^{+0.259}_{-0.107}$ & $0.244^{+0.157}_{-0.083}$ \\[0.9mm]
73P(B)-20060806.18 & $3.984e+03^{+8.683e+01}_{-3.984e+03}$ & $0.433^{+0.017}_{-0.016}$ & $0.000^{+0.000}_{-0.000}$ & $0.534^{+0.013}_{-0.012}$ & $0.033^{+0.025}_{-0.025}$ & $0.000^{+0.000}_{-0.000}$ & $0.873^{+0.044}_{-0.044}$ & $0.071^{+0.051}_{-0.054}$ \\[0.9mm]
73P(C)-20060317.04 & $1.290e+04^{+1.228e+03}_{-1.290e+04}$ & $0.411^{+0.059}_{-0.048}$ & $0.000^{+0.000}_{-0.000}$ & $0.301^{+0.032}_{-0.026}$ & $0.130^{+0.044}_{-0.044}$ & $0.158^{+0.076}_{-0.092}$ & $2.321^{+0.314}_{-0.320}$ & $0.412^{+0.087}_{-0.116}$ \\[0.9mm]
73P(C)-20060806.05 & $2.295e+02^{+6.318e+02}_{-2.295e+02}$ & $0.006^{+0.565}_{-0.004}$ & $0.000^{+0.000}_{-0.000}$ & $0.004^{+0.425}_{-0.003}$ & $0.990^{+0.006}_{-0.990}$ & $0.000^{+0.976}_{-0.000}$ & $2.263e+02^{+6.248e+02}_{-2.249e+02}$ & $0.994^{+0.004}_{-0.994}$ \\[0.9mm]
78P & $2.084e+05^{+1.197e+04}_{-2.084e+05}$ & $0.185^{+0.029}_{-0.029}$ & $0.228^{+0.031}_{-0.030}$ & $0.338^{+0.019}_{-0.018}$ & $0.209^{+0.030}_{-0.032}$ & $0.040^{+0.050}_{-0.040}$ & $1.960^{+0.171}_{-0.156}$ & $0.376^{+0.056}_{-0.060}$ \\[0.9mm]
88P & $1.946e+04^{+1.333e+03}_{-1.946e+04}$ & $0.217^{+0.038}_{-0.038}$ & $0.222^{+0.046}_{-0.042}$ & $0.460^{+0.036}_{-0.030}$ & $0.102^{+0.059}_{-0.071}$ & $0.000^{+0.000}_{-0.000}$ & $1.175^{+0.151}_{-0.158}$ & $0.188^{+0.096}_{-0.127}$ \\[0.9mm]
105P & $1.477e+04^{+1.057e+03}_{-1.477e+04}$ & $0.406^{+0.098}_{-0.128}$ & $0.000^{+0.052}_{-0.000}$ & $0.476^{+0.057}_{-0.040}$ & $0.000^{+0.021}_{-0.000}$ & $0.118^{+0.098}_{-0.109}$ & $1.102^{+0.191}_{-0.226}$ & $0.225^{+0.191}_{-0.202}$ \\[0.9mm]
121P & $2.694e+04^{+5.065e+03}_{-2.694e+04}$ & $0.497^{+0.213}_{-0.497}$ & $0.000^{+0.290}_{-0.000}$ & $0.135^{+0.118}_{-0.034}$ & $0.182^{+0.133}_{-0.131}$ & $0.187^{+0.200}_{-0.187}$ & $6.423^{+2.495}_{-3.465}$ & $0.426^{+0.278}_{-0.270}$ \\[0.9mm]
123P & $1.126e+05^{+5.295e+03}_{-1.126e+05}$ & $0.235^{+0.051}_{-0.052}$ & $0.147^{+0.051}_{-0.051}$ & $0.443^{+0.026}_{-0.025}$ & $0.144^{+0.032}_{-0.035}$ & $0.030^{+0.053}_{-0.030}$ & $1.256^{+0.134}_{-0.123}$ & $0.313^{+0.085}_{-0.076}$ \\[0.9mm]
132P & $1.865e+03^{+2.634e+03}_{-1.865e+03}$ & $0.039^{+0.589}_{-0.039}$ & $0.004^{+0.099}_{-0.004}$ & $0.027^{+0.313}_{-0.015}$ & $0.000^{+0.476}_{-0.000}$ & $0.931^{+0.045}_{-0.931}$ & $36.293^{+49.018}_{-34.352}$ & $0.956^{+0.035}_{-0.956}$ \\[0.9mm]
144P & $4.003e+03^{+8.483e+01}_{-4.003e+03}$ & $0.365^{+0.016}_{-0.018}$ & $0.000^{+0.000}_{-0.000}$ & $0.635^{+0.017}_{-0.017}$ & $0.000^{+0.000}_{-0.000}$ & $0.000^{+0.000}_{-0.000}$ & $0.575^{+0.044}_{-0.040}$ & $0.000^{+0.006}_{-0.000}$ \\[0.9mm]
C2001Q4 & $5.081e+05^{+4.183e+04}_{-5.081e+05}$ & $0.418^{+0.043}_{-0.042}$ & $0.104^{+0.044}_{-0.039}$ & $0.212^{+0.020}_{-0.017}$ & $0.136^{+0.033}_{-0.034}$ & $0.130^{+0.053}_{-0.062}$ & $3.708^{+0.399}_{-0.402}$ & $0.337^{+0.063}_{-0.077}$ \\[0.9mm]
C2003K4-20040716 & $5.377e+04^{+1.349e+04}_{-5.377e+04}$ & $0.284^{+0.077}_{-0.073}$ & $0.000^{+0.000}_{-0.000}$ & $0.532^{+0.112}_{-0.106}$ & $0.000^{+0.078}_{-0.000}$ & $0.184^{+0.168}_{-0.184}$ & $0.880^{+0.470}_{-0.327}$ & $0.393^{+0.238}_{-0.393}$ \\[0.9mm]
C2003K4-20050913 & $8.181e+04^{+1.226e+04}_{-8.181e+04}$ & $0.177^{+0.046}_{-0.048}$ & $0.506^{+0.054}_{-0.085}$ & $0.317^{+0.016}_{-0.042}$ & $0.000^{+0.131}_{-0.000}$ & $0.000^{+0.000}_{-0.000}$ & $2.152^{+0.485}_{-0.149}$ & $0.000^{+0.183}_{-0.000}$ \\[0.9mm]
C2003T3 & $3.072e+04^{+1.129e+04}_{-3.072e+04}$ & $0.388^{+0.114}_{-0.113}$ & $0.000^{+0.000}_{-0.000}$ & $0.236^{+0.065}_{-0.063}$ & $0.376^{+0.114}_{-0.227}$ & $0.000^{+0.268}_{-0.000}$ & $3.239^{+1.557}_{-0.916}$ & $0.493^{+0.174}_{-0.207}$ \\[0.9mm]
C2003T4-20051122 & $1.388e+05^{+1.336e+04}_{-1.388e+05}$ & $0.292^{+0.031}_{-0.026}$ & $0.045^{+0.024}_{-0.022}$ & $0.144^{+0.014}_{-0.012}$ & $0.293^{+0.047}_{-0.048}$ & $0.227^{+0.063}_{-0.069}$ & $5.967^{+0.648}_{-0.634}$ & $0.606^{+0.043}_{-0.053}$ \\[0.9mm]
C2003T4-20060129 & $3.749e+04^{+1.372e+04}_{-3.749e+04}$ & $0.000^{+0.000}_{-0.000}$ & $0.043^{+0.019}_{-0.015}$ & $0.169^{+0.057}_{-0.045}$ & $0.788^{+0.045}_{-0.197}$ & $0.000^{+0.235}_{-0.000}$ & $4.921^{+2.125}_{-1.489}$ & $0.948^{+0.020}_{-0.028}$ \\[0.9mm]
C2003T4-20060307 & $1.176e+05^{+1.213e+04}_{-1.176e+05}$ & $0.391^{+0.062}_{-0.050}$ & $0.000^{+0.000}_{-0.000}$ & $0.243^{+0.028}_{-0.023}$ & $0.214^{+0.064}_{-0.066}$ & $0.152^{+0.089}_{-0.105}$ & $3.114^{+0.421}_{-0.423}$ & $0.483^{+0.077}_{-0.103}$ \\[0.9mm]
C2004B1-20051015 & $1.213e+04^{+1.375e+04}_{-1.213e+04}$ & $0.219^{+0.133}_{-0.125}$ & $0.000^{+0.000}_{-0.000}$ & $0.430^{+0.236}_{-0.228}$ & $0.000^{+0.291}_{-0.000}$ & $0.351^{+0.336}_{-0.351}$ & $1.325^{+2.624}_{-0.824}$ & $0.617^{+0.267}_{-0.617}$ \\[0.9mm]
C2004B1-20060516 & $1.065e+04^{+8.753e+03}_{-1.065e+04}$ & $0.231^{+0.155}_{-0.117}$ & $0.000^{+0.010}_{-0.000}$ & $0.390^{+0.232}_{-0.175}$ & $0.000^{+0.283}_{-0.000}$ & $0.379^{+0.246}_{-0.379}$ & $1.566^{+2.098}_{-0.959}$ & $0.622^{+0.233}_{-0.622}$ \\[0.9mm]
C2004B1-20060726 & $5.965e+03^{+4.012e+03}_{-5.965e+03}$ & $0.000^{+0.000}_{-0.000}$ & $0.145^{+0.126}_{-0.067}$ & $0.409^{+0.313}_{-0.163}$ & $0.021^{+0.274}_{-0.021}$ & $0.426^{+0.207}_{-0.426}$ & $1.447^{+1.623}_{-1.062}$ & $0.755^{+0.141}_{-0.755}$ \\[0.9mm]
C2004B1-20070308 & $2.198e+04^{+2.318e+03}_{-2.198e+04}$ & $0.000^{+0.000}_{-0.000}$ & $0.000^{+0.000}_{-0.000}$ & $1.000^{+0.000}_{-0.096}$ & $0.000^{+0.096}_{-0.000}$ & $0.000^{+0.000}_{-0.000}$ & $0.000^{+0.106}_{-0.000}$ & $ nan^{+ nan}_{ nan}$ \\[0.9mm]
C2004B1-20070609 & $1.044e+05^{+1.480e+04}_{-1.044e+05}$ & $0.561^{+0.110}_{-0.085}$ & $0.000^{+0.000}_{-0.000}$ & $0.264^{+0.044}_{-0.036}$ & $0.000^{+0.000}_{-0.000}$ & $0.174^{+0.111}_{-0.149}$ & $2.783^{+0.587}_{-0.545}$ & $0.237^{+0.137}_{-0.200}$ \\[0.9mm]
C2004Q2 & $9.059e+03^{+6.473e+03}_{-9.059e+03}$ & $0.604^{+0.010}_{-0.253}$ & $0.000^{+0.000}_{-0.000}$ & $0.396^{+0.010}_{-0.165}$ & $0.000^{+0.000}_{-0.000}$ & $0.000^{+0.418}_{-0.000}$ & $1.527^{+1.806}_{-0.065}$ & $0.000^{+0.542}_{-0.000}$ \\[0.9mm]
C2006P1-20070504 & $1.084e+05^{+2.880e+04}_{-1.084e+05}$ & $0.340^{+0.029}_{-0.079}$ & $0.000^{+0.003}_{-0.000}$ & $0.605^{+0.043}_{-0.126}$ & $0.055^{+0.204}_{-0.055}$ & $0.000^{+0.000}_{-0.000}$ & $0.654^{+0.437}_{-0.110}$ & $0.140^{+0.359}_{-0.140}$ \\[0.9mm]
C2006P1-20070802 & $9.605e+04^{+1.755e+04}_{-9.605e+04}$ & $0.000^{+0.000}_{-0.000}$ & $0.568^{+0.043}_{-0.093}$ & $0.378^{+0.031}_{-0.058}$ & $0.054^{+0.150}_{-0.054}$ & $0.000^{+0.000}_{-0.000}$ & $1.643^{+0.483}_{-0.199}$ & $0.087^{+0.213}_{-0.087}$ \\[0.9mm]
C2006P1-20070906 & $9.516e+04^{+2.593e+04}_{-9.516e+04}$ & $0.002^{+0.022}_{-0.002}$ & $0.088^{+0.035}_{-0.033}$ & $0.177^{+0.056}_{-0.036}$ & $0.129^{+0.160}_{-0.129}$ & $0.604^{+0.133}_{-0.173}$ & $4.639^{+1.435}_{-1.346}$ & $0.891^{+0.033}_{-0.055}$ \\[0.9mm]
C2006Q1-20070322 & $4.917e+04^{+3.700e+04}_{-4.917e+04}$ & $0.125^{+0.181}_{-0.064}$ & $0.000^{+0.036}_{-0.000}$ & $0.303^{+0.397}_{-0.129}$ & $0.572^{+0.179}_{-0.572}$ & $0.000^{+0.283}_{-0.000}$ & $2.305^{+2.452}_{-1.876}$ & $0.821^{+0.101}_{-0.821}$ \\[0.9mm]
C2006Q1-20080117 & $3.115e+04^{+2.121e+04}_{-3.115e+04}$ & $0.170^{+0.058}_{-0.094}$ & $0.022^{+0.061}_{-0.022}$ & $0.710^{+0.095}_{-0.286}$ & $0.098^{+0.288}_{-0.098}$ & $0.000^{+0.358}_{-0.000}$ & $0.409^{+0.952}_{-0.167}$ & $0.339^{+0.478}_{-0.339}$ \\[0.9mm]
C2006Q1-20080225 & $9.176e+04^{+1.927e+04}_{-9.176e+04}$ & $0.636^{+0.097}_{-0.200}$ & $0.000^{+0.084}_{-0.000}$ & $0.290^{+0.075}_{-0.058}$ & $0.074^{+0.187}_{-0.074}$ & $0.000^{+0.042}_{-0.000}$ & $2.454^{+0.875}_{-0.712}$ & $0.104^{+0.261}_{-0.104}$ \\[0.9mm]
C2006Q1-20080702 & $9.104e+04^{+1.605e+04}_{-9.104e+04}$ & $0.000^{+0.005}_{-0.000}$ & $0.335^{+0.073}_{-0.057}$ & $0.416^{+0.081}_{-0.062}$ & $0.085^{+0.105}_{-0.085}$ & $0.165^{+0.131}_{-0.165}$ & $1.405^{+0.421}_{-0.390}$ & $0.427^{+0.141}_{-0.233}$ \\[0.9mm]
C2006Q1-20080709 & $1.570e+05^{+3.622e+04}_{-1.570e+05}$ & $0.488^{+0.103}_{-0.088}$ & $0.096^{+0.085}_{-0.066}$ & $0.192^{+0.048}_{-0.036}$ & $0.066^{+0.100}_{-0.066}$ & $0.159^{+0.130}_{-0.159}$ & $4.208^{+1.200}_{-1.039}$ & $0.278^{+0.155}_{-0.224}$ \\[0.9mm]
C2007N3 & $6.137e+04^{+3.569e+04}_{-6.137e+04}$ & $0.205^{+0.101}_{-0.087}$ & $0.000^{+0.005}_{-0.000}$ & $0.490^{+0.207}_{-0.180}$ & $0.030^{+0.223}_{-0.030}$ & $0.275^{+0.240}_{-0.275}$ & $1.041^{+1.183}_{-0.607}$ & $0.598^{+0.231}_{-0.598}$ \\[0.9mm]
C2008T2 & $1.481e+04^{+3.171e+02}_{-1.481e+04}$ & $0.225^{+0.052}_{-0.056}$ & $0.056^{+0.039}_{-0.037}$ & $0.719^{+0.020}_{-0.019}$ & $0.000^{+0.000}_{-0.000}$ & $0.000^{+0.000}_{-0.000}$ & $0.391^{+0.038}_{-0.037}$ & $0.000^{+0.000}_{-0.000}$ \\[0.9mm]
\enddata
\tablenotetext{a}{Defined as $f_{cryst}^{silicates} \equiv$ (crystalline)/(crystalline + amorphous).}
\end{deluxetable*}

%% file: table_5.tex
\begin{deluxetable*}{@{\extracolsep{0pt}}lccccccccccc}
\tablenum{5}
\tabletypesize{\small}
\tablewidth{0pc}
\tablecaption{Statistical summary of model parameters for different spectral data sets.\label{tab:KStest_all_models}}
\tablehead{
\colhead{Statistic} & \colhead{Set 0} & \colhead{Set 1} & \colhead{Set 2} & \colhead{f(ap50)} & \colhead{f(ao50)} 
& \colhead{f(co)} & \colhead{f(cp)} & \colhead{f(ac)} & \colhead{AS} & \colhead{CS} & \colhead{$f_{cryst}$}
}
\startdata
Median  & SL    &&& 0.409 & 0.000 & 0.070 & 0.124 & 0.439 & 0.409 & 0.173 & 0.305 \\
Mean    & SL    &&& 0.367 & 0.025 & 0.067 & 0.125 & 0.416 & 0.393 & 0.191 & 0.326 \\
Std.    & SL    &&& 0.173 & 0.052 & 0.063 & 0.126 & 0.145 & 0.156 & 0.130 & 0.217 \\
\hline
Median  && SLLL && 0.194 & 0.136 & 0.044 & 0.000 & 0.535 & 0.337 & 0.088 & 0.196 \\
Mean    && SLLL && 0.186 & 0.147 & 0.056 & 0.067 & 0.544 & 0.333 & 0.123 & 0.250 \\
Std     && SLLL && 0.137 & 0.132 & 0.060 & 0.124 & 0.164 & 0.137 & 0.125 & 0.224 \\
\hline
Median  &&& SLLL-SL & 0.328 & 0.000 & 0.079 & 0.188 & 0.345 & 0.352 & 0.321 & 0.496 \\
Mean    &&& SLLL-SL & 0.332 & 0.017 & 0.083 & 0.210 & 0.358 & 0.349 & 0.293 & 0.465 \\
Std     &&& SLLL-SL & 0.180 & 0.037 & 0.061 & 0.154 & 0.142 & 0.191 & 0.171 & 0.260 \\
\hline
Median  &&& SLLL-SL \& SL & 0.358 & 0.000 & 0.076 & 0.158 & 0.393 & 0.374 & 0.239 & 0.387 \\
Mean    &&& SLLL-SL \& SL & 0.349 & 0.021 & 0.075 & 0.169 & 0.386 & 0.370 & 0.245 & 0.399 \\
Std     &&& SLLL-SL \& SL & 0.175 & 0.045 & 0.062 & 0.146 & 0.145 & 0.174 & 0.159 & 0.248 \\
\hline
KS\tablenotemark{a} pair &  & SLLL & SLLL-SL & 0.020 & 0.001 & 0.394 & 0.001 & 0.007 & 0.872 & 0.007 & 0.007 \\
KS pair &  & SLLL & SL           & 0.012 & 0.001 & 0.869 & 0.080 & 0.009 & 0.109 & 0.133 & 0.246 \\
KS pair &  & SLLL-SL & SL     & 0.518 & 0.995 & 0.820 & 0.216 & 0.312 & 0.518 & 0.125 & 0.133 \\
KS pair &  & SLLL & SLLL-SL \& SL & 0.003 & 0.000 & 0.472 & 0.002 & 0.001 & 0.221 & 0.013 & 0.056 \\
\hline
KS pair &  & SLLL (JFC) & SLLL (OCC) & 0.961 & 0.705 & 0.803 & 0.841 & 0.754 & 0.094 & 0.656 & 0.754 \\
KS pair &  & SLLL-SL (JFC) & SLLL-SL (OCC) & 0.325 & 1.000 & 0.371 & 0.878 & 0.239 & 0.754 & 0.878 & 0.705 \\
KS pair &  & SL (JFC) & SL (OCC) & 0.897 & 0.998 & 0.091 & 0.214 & 0.324 & 0.324 & 0.324 & 0.743 \\
KS pair &  & SLLL-SL \& SL (JFC) & SLLL-SL \& SL (OCC) & 0.300 & 0.935 & 0.273 & 0.273 & 0.386 & 0.386 & 0.700 & 0.508 \\
\hline
KS pair &  & SLLL (JFC) & SLLL-SL \& SL (JFC) & 0.002 & 0.000 & 0.258 & 0.009 & 0.006 & 0.602 & 0.034 & 0.097 \\
KS pair &  & SLLL (OCC) & SLLL-SL \& SL (OCC) & 0.211 & 0.136 & 0.800 & 0.373 & 0.087 & 0.319 & 0.438 & 0.681 \\
\hline
KS pair &  & SLLL (JFC) & SLLL-SL (JFC) & 0.016 & 0.001 & 0.751 & 0.005 & 0.016 & 0.963 & 0.016 & 0.045 \\
KS pair &  & SLLL (OCC) & SLLL-SL (OCC) & 0.873 & 0.357 & 0.357 & 0.357 & 0.357 & 0.873 & 0.357 & 0.357 \\
KS pair &  & SLLL (JFC) & SL (JFC) & 0.011 & 0.011 & 0.216 & 0.372 & 0.076 & 0.556 & 0.237 & 0.237 \\
KS pair &  & SLLL (OCC) & SL (OCC) & 0.165 & 0.235 & 0.851 & 0.769 & 0.126 & 0.316 & 0.769 & 0.997 \\
KS pair &  & SLLL-SL (JFC) & SL (JFC) & 0.766 & 0.996 & 0.641 & 0.086 & 0.216 & 0.766 & 0.147 & 0.237 \\
KS pair &  & SLLL-SL (OCC) & SL (OCC) & 0.586 & 0.963 & 0.235 & 0.851 & 0.769 & 0.586 & 0.586 & 0.500 \\
KS pair &  & SLLL (JFC) & SLLL-SL \& SL (JFC) & 0.002 & 0.000 & 0.258 & 0.009 & 0.006 & 0.602 & 0.034 & 0.097 \\
KS pair &  & SLLL (OCC) & SLLL-SL \& SL (OCC) & 0.211 & 0.136 & 0.800 & 0.373 & 0.087 & 0.319 & 0.438 & 0.681 \\
\enddata
\tablenotetext{a}{Kolmogorov-Smirnov (KS) test pair probability or pvalue (see text Section 4.6.1).}
\end{deluxetable*}

%% file: table_6.tex
\begin{deluxetable}{@{\extracolsep{15pt}}lcccc}
\tablenum{6}
\tablewidth{0pc}
\tablecaption{Spearman Rank Correlation Values\tablenotemark{a} \label{tab:SLLL_spearmanr}}
\centering
\tablehead{
\colhead{} && \colhead{f(ac)} &&\colhead{f(cp)}
}
\startdata
f(co)  && -0.62 && $\ldots$       \\ 
CS     && -0.77 && 0.79  \\ 
f(ap50) && $\ldots$     && -0.70 \\ 
\enddata
\tablenotetext{a}{Spearman Rank correlation values considered not 
attributable to random chance at 3$\sigma$ for parameters in the set of SLLL models.}
\end{deluxetable}

%% file: table_7.tex

\startlongtable
\begin{deluxetable}{@{\extracolsep{0pt}}lcc}
\tablenum{7}
\tablewidth{0pc}
\tablecaption{C/Si Ratio \label{tab:c_si_tab}}
\tablehead{
\colhead{Comet} &  \multicolumn{2}{c}{C/Si} 
}
\startdata
6P/d'Arrest & $4.255 \pm 1.384$ \\
8P/Tuttle & $86.146 \pm 13.179$ \\
9P/Tempel 1 & $4.716 \pm 1.715$ \\
17P/Holmes & $0.672 \pm 0.133$ \\
21P/Giacobini-Zinner & $6.353 \pm 1.904$ \\
29P/Schwassmann-Wachmann 1 & $1.791 \pm 0.464$ \\
37P/Forbes & $3.835 \pm 1.017$ \\
41P/Tuttle-Giacobini-Kresak & $20.276 \pm 3.961$ \\
46P/Wirtanen & $10.075 \pm 7.152$ \\
48P/Johnson & $2.558 \pm 0.606$ \\
62P/Tsuchinshan 1 & $3.515 \pm 1.599$ \\
65P/Gunn & $3.214 \pm 1.002$ \\
67P/Churyumov-Gerasimenko & $9.087 \pm 5.426$ \\
71P/Clark & $6.757 \pm 1.953$ \\
73P/Schwassmann-Wachmann 3 [B] & $4.552 \pm 1.716$ \\
73P/Schwassmann-Wachmann 3 [C] & $2.318 \pm 0.509$ \\
78P/Gehrels 2 & $2.703 \pm 0.642$ \\
88P/Howell & $4.222 \pm 1.120$ \\
105P/Singer Brewster & $5.059 \pm 1.850$ \\
121P/Shoemaker-Holt 2 & $1.025 \pm 0.473$ \\
144P/Kushida & $7.194 \pm 1.999$ \\
C/2001 Q4 (NEAT) & $1.475 \pm 0.347$ \\
C/2003 K4 (LINEAR) & $6.264 \pm 2.230$ \\
C/2003 T3 (Tabur) & $1.250 \pm 0.360$ \\
C/2003 T4 (LINEAR) & $1.086 \pm 0.428$ \\
C/2004 B1 (LINEAR) & $6.391 \pm 4.178$ \\
C/2004 Q2 Machholz & $2.709 \pm 0.593$ \\
C/2006 P1 (McNaught) & $3.972 \pm 1.955$ \\
C/2006 Q1 (McNaught) & $4.471 \pm 2.260$ \\
C/2007 N3 (Lulin) & $8.826 \pm 2.896$ \\
C/2008 T2 (Cardinal) & $10.936 \pm 3.470$ \\
\enddata
\end{deluxetable}

%% file: table_Appendix_A1.tex
\begin{deluxetable*}{@{\extracolsep{0pt}}llr}
\tablenum{A1}
\tablewidth{0pc}
\tablecaption{$\Delta AICc = AICc$ (SLLL) - $AICc$ (SLLL-SL)  \label{tab:A1}}
\tablehead{\colhead{Comet Name} & \colhead{Date Obs (UT)} & \colhead{$\Delta AICc$} 
}
\startdata
8P/Tuttle & 2007-11-02 & -757.5 \\
9P/Tempel 1 & 2005-07-02 & -486.5 \\
9P/Tempel 1 & 2005-07-03 & 4.5 \\
17P/Holmes & 2007-11-10 & 468.2 \\
37P/Forbes & 2005-10-14 & -566.5 \\
46P/Wirtanen & 2008-04-24 & -578.9 \\
46P/Wirtanen & 2008-05-24 & -594.7 \\
46P/Wirtanen & 2008-07-02 & -597.6 \\
48P/Johnson & 2004-10-04 & -775.5 \\
67P/Churyumov-Gerasimenko & 2008-11-30 & -188.6 \\
67P/Churyumov-Gerasimenko & 2008-11-29 & -213.9 \\
67P/Churyumov-Gerasimenko & 2008-11-29 & -471.1 \\
67P/Churyumov-Gerasimenko & 2008-11-28 & -198.8 \\
71P/Clark & 2006-05-27 & -436.3 \\
73P/Schwassmann-Wachmann 3 [B] & 2006-08-06 & -431.6 \\
78P/Gehrels 2 & 2004-09-01 & -494.0 \\
88P/Howell & 2004-09-01 & -607.7 \\
C/2004 B1 (LINEAR) & 2006-07-26 & -559.9 \\
C/2006 P1 (McNaught) & 2007-05-04 & -301.8 \\
C/2006 Q1 (McNaught) & 2008-07-02 & -725.7 \\
C/2006 Q1 (McNaught) & 2008-01-17 & -824.6 \\
C/2008 T2 (Cardinal) & 2009-04-04 & -470.7 \\
\enddata
\end{deluxetable*}

%% file: table_Appendix_A2.tex
\begin{longrotatetable}
\begin{deluxetable*}{@{\extracolsep{0pt}}lcccccccccc}
\tablenum{A2}

\setlength{\tabcolsep}{2pt} 
\tablewidth{0pc}
\tabletypesize{\footnotesize}
\tablecaption{Comets Observed with SLLL and Only Modeled Over (SLLL-SL): Best-Fit Thermal Emission Model Parameters \label{tab:tab2_appendix_modparams}}
\tablehead{
 &  & & & & \multicolumn{5}{c}{\underbar{$N_p (\times 10^{16}$)}}  \\
\colhead{Comet} & & &
 &  & \colhead{Amorphous} &
\colhead{Amorphous} & \colhead{Amorphous} & \colhead{Crystalline} & 
\colhead{Crystalline} & \\
\colhead{Name} &  \colhead{N} & \colhead{M} & \colhead{$a_p$ \tablenotemark{a}} & \colhead{D} & \colhead{Pyroxene} & \colhead{Olivine} &
\colhead{Carbon} & \colhead{Olivine} & \colhead{Orthopyroxene} & \colhead{$\chi^{2}_{\nu}$}
}
\startdata
8P & 3.4 & 6.8 & 0.3 & 2.857 & $6.659e+01^{+3.461e+02}_{-6.659e+01}$ & $0.000e+00^{+2.023e+02}_{-0.000e+00}$ & ${\bf 6.845e+03^{+8.637e+01}_{-1.432e+02}}$ & ${\bf 2.880e+02^{+2.199e+02}_{-2.533e+02}}$ & ${\bf 1.010e+03^{+3.514e+02}_{-5.529e+02}}$ & 0.77 \\
9P-20050702 & 3.4 & 10.2 & 0.4 & 2.727 & ${\bf 2.525e+03^{+2.306e+02}_{-4.887e+02}}$ & $4.135e+01^{+3.077e+02}_{-4.135e+01}$ & ${\bf 5.054e+03^{+5.989e+01}_{-5.520e+01}}$ & ${\bf 3.283e+02^{+2.572e+02}_{-2.989e+02}}$ & $4.364e+02^{+5.046e+02}_{-4.364e+02}$ & 0.73 \\
9P-20050703 & 3.5 & 14.0 & 0.5 & 2.727 & ${\bf 1.604e+03^{+1.273e+02}_{-1.836e+02}}$ & $0.000e+00^{+9.972e+01}_{-0.000e+00}$ & ${\bf 3.393e+03^{+3.886e+01}_{-3.905e+01}}$ & $1.941e+02^{+2.215e+02}_{-1.941e+02}$ & $3.783e+02^{+4.296e+02}_{-3.783e+02}$ & 0.59 \\
17P & 3.5 & 7.0 & 0.3 & 2.857 & ${\bf 6.970e+05^{+4.803e+04}_{-4.895e+04}}$ & ${\bf 1.798e+05^{+2.620e+04}_{-2.581e+04}}$ & ${\bf 1.783e+05^{+5.546e+03}_{-5.569e+03}}$ & ${\bf 1.080e+05^{+1.653e+04}_{-1.695e+04}}$ & ${\bf 3.154e+04^{+3.116e+04}_{-3.127e+04}}$ & 5.36 \\
21P & 3.6 & 14.4 & 0.5 & 2.727 & ${\bf 7.449e+02^{+3.591e+01}_{-1.044e+02}}$ & $0.000e+00^{+4.500e+01}_{-0.000e+00}$ & ${\bf 1.138e+03^{+2.004e+01}_{-1.482e+01}}$ & ${\bf 1.078e+02^{+5.713e+01}_{-5.985e+01}}$ & $0.000e+00^{+1.160e+02}_{-0.000e+00}$ & 0.50 \\
29P & 4.0 & 8.0 & 0.3 & 2.727 & ${\bf 2.093e+06^{+3.186e+05}_{-6.946e+05}}$ & $0.000e+00^{+2.742e+05}_{-0.000e+00}$ & ${\bf 4.355e+05^{+3.982e+04}_{-3.554e+04}}$ & ${\bf 2.695e+05^{+1.385e+05}_{-1.581e+05}}$ & ${\bf 5.067e+05^{+3.815e+05}_{-4.167e+05}}$ & 0.60 \\
37P & 3.4 & 10.2 & 0.4 & 2.727 & ${\bf 3.279e+02^{+3.169e+01}_{-5.311e+01}}$ & $9.612e+00^{+2.814e+01}_{-9.612e+00}$ & ${\bf 4.322e+02^{+7.490e+00}_{-6.710e+00}}$ & ${\bf 4.130e+01^{+2.447e+01}_{-2.621e+01}}$ & $3.140e+01^{+4.523e+01}_{-3.140e+01}$ & 1.79 \\
46P-20080424 & 4.0 & 28.0 & 0.8 & 2.727 & ${\bf 1.563e+01^{+7.437e+00}_{-7.286e+00}}$ & ${\bf 5.736e+00^{+4.397e+00}_{-4.677e+00}}$ & ${\bf 1.263e+02^{+9.310e-01}_{-9.317e-01}}$ & $7.830e+00^{+9.879e+00}_{-7.830e+00}$ & $1.410e+01^{+1.870e+01}_{-1.410e+01}$ & 0.63 \\
46P-20080524 & 4.3 & 43.0 & 1.1 & 2.727 & ${\bf 1.923e+01^{+2.699e+00}_{-2.919e+00}}$ & $0.000e+00^{+0.000e+00}_{-0.000e+00}$ & ${\bf 1.310e+02^{+8.747e-01}_{-8.525e-01}}$ & $0.000e+00^{+1.526e+01}_{-0.000e+00}$ & ${\bf 6.564e+01^{+4.201e+01}_{-4.154e+01}}$ & 1.55 \\
46P-20080702 & 3.4 & 17.0 & 0.6 & 3.0 & ${\bf 4.189e+01^{+8.498e+00}_{-2.836e+01}}$ & $0.000e+00^{+1.430e+01}_{-0.000e+00}$ & ${\bf 1.613e+02^{+6.176e+00}_{-4.859e+00}}$ & $0.000e+00^{+2.626e+00}_{-0.000e+00}$ & $1.542e+01^{+1.940e+01}_{-1.542e+01}$ & 2.35 \\
48P & 3.8 & 19.0 & 0.6 & 2.727 & ${\bf 1.277e+03^{+3.784e+02}_{-5.467e+02}}$ & $1.036e+02^{+2.389e+02}_{-1.036e+02}$ & ${\bf 1.992e+03^{+9.253e+01}_{-8.327e+01}}$ & ${\bf 3.746e+02^{+2.357e+02}_{-2.518e+02}}$ & $4.036e+02^{+5.502e+02}_{-4.036e+02}$ & 0.66 \\
65P & 3.3 & 26.4 & 0.9 & 3.0 & ${\bf 6.573e+02^{+5.040e+01}_{-6.402e+01}}$ & $0.000e+00^{+0.000e+00}_{-0.000e+00}$ & ${\bf 1.316e+02^{+3.471e+01}_{-2.903e+01}}$ & $0.000e+00^{+0.000e+00}_{-0.000e+00}$ & $0.000e+00^{+1.714e+02}_{-0.000e+00}$ & 2.48 \\
67P-20081128.38 & 4.3 & 43.0 & 1.1 & 2.727 & ${\bf 1.184e+02^{+5.697e+00}_{-7.838e+00}}$ & $0.000e+00^{+0.000e+00}_{-0.000e+00}$ & ${\bf 1.942e+02^{+3.105e+00}_{-2.806e+00}}$ & $0.000e+00^{+3.306e+01}_{-0.000e+00}$ & $0.000e+00^{+9.067e+01}_{-0.000e+00}$ & 1.38 \\
67P-20081129.11 & 4.2 & 42.0 & 1.1 & 2.727 & ${\bf 7.895e+01^{+5.708e+00}_{-1.290e+01}}$ & $0.000e+00^{+6.333e+00}_{-0.000e+00}$ & ${\bf 1.617e+02^{+2.886e+00}_{-2.549e+00}}$ & $2.307e+01^{+5.032e+01}_{-2.307e+01}$ & $3.606e+01^{+1.063e+02}_{-3.606e+01}$ & 0.49 \\
67P-20081129.70 & 4.3 & 47.3 & 1.2 & 2.727 & ${\bf 7.663e+01^{+4.425e+00}_{-6.112e+00}}$ & $0.000e+00^{+0.000e+00}_{-0.000e+00}$ & ${\bf 1.402e+02^{+2.008e+00}_{-1.925e+00}}$ & $0.000e+00^{+5.189e+01}_{-0.000e+00}$ & $4.907e+01^{+1.222e+02}_{-4.907e+01}$ & 0.88 \\
67P-20081130.48 & 4.2 & 42.0 & 1.1 & 2.727 & ${\bf 8.305e+01^{+4.480e+00}_{-8.009e+00}}$ & $0.000e+00^{+0.000e+00}_{-0.000e+00}$ & ${\bf 1.718e+02^{+2.653e+00}_{-2.253e+00}}$ & $5.067e+00^{+5.545e+01}_{-5.067e+00}$ & $1.129e+01^{+1.052e+02}_{-1.129e+01}$ & 0.78 \\
71P & 4.3 & 30.1 & 0.8 & 2.727 & ${\bf 2.225e+02^{+9.586e+01}_{-9.636e+01}}$ & $5.810e+01^{+5.672e+01}_{-5.810e+01}$ & ${\bf 1.833e+03^{+1.250e+01}_{-1.251e+01}}$ & $7.017e+01^{+1.088e+02}_{-7.017e+01}$ & ${\bf 2.381e+02^{+2.038e+02}_{-2.043e+02}}$ & 0.89 \\
73P(B)-20060806.18 & 3.5 & 10.5 & 0.4 & 2.727 & ${\bf 1.569e+02^{+6.229e+00}_{-6.780e+00}}$ & $0.000e+00^{+0.000e+00}_{-0.000e+00}$ & ${\bf 3.498e+02^{+2.443e+00}_{-2.387e+00}}$ & ${\bf 1.150e+01^{+9.362e+00}_{-9.496e+00}}$ & $0.000e+00^{+8.785e+00}_{-0.000e+00}$ & 1.75 \\
78P & 4.3 & 43.0 & 1.1 & 2.727 & ${\bf 1.169e+03^{+8.371e+01}_{-9.710e+01}}$ & $0.000e+00^{+3.097e+01}_{-0.000e+00}$ & ${\bf 2.099e+03^{+2.858e+01}_{-2.802e+01}}$ & ${\bf 5.010e+02^{+4.749e+02}_{-4.953e+02}}$ & ${\bf 1.199e+03^{+9.488e+02}_{-9.682e+02}}$ & 1.00 \\
88P & 3.3 & 19.8 & 0.7 & 3.0 & ${\bf 2.924e+02^{+2.888e+01}_{-8.383e+01}}$ & $0.000e+00^{+4.504e+01}_{-0.000e+00}$ & ${\bf 5.520e+02^{+2.146e+01}_{-1.744e+01}}$ & $2.157e+01^{+6.102e+01}_{-2.157e+01}$ & ${\bf 2.265e+02^{+9.263e+01}_{-1.017e+02}}$ & 0.79 \\
132P & 3.3 & 33.0 & 1.1 & 2.609 & $6.374e+00^{+3.495e+00}_{-6.374e+00}$ & $6.364e-01^{+2.661e+00}_{-6.364e-01}$ & ${\bf 9.691e+00^{+8.236e-01}_{-7.315e-01}}$ & $0.000e+00^{+3.044e+01}_{-0.000e+00}$ & $6.645e+01^{+9.345e+01}_{-6.645e+01}$ & 0.61 \\
C2003K4-20050913 & 3.4 & 40.8 & 1.3 & 3.0 & ${\bf 4.892e+03^{+3.149e+02}_{-3.542e+02}}$ & $0.000e+00^{+0.000e+00}_{-0.000e+00}$ & ${\bf 2.858e+03^{+1.992e+02}_{-1.828e+02}}$ & $0.000e+00^{+1.755e+03}_{-0.000e+00}$ & $0.000e+00^{+2.627e+02}_{-0.000e+00}$ & 3.45 \\
C2003T3 & 3.6 & 36.0 & 1.1 & 2.727 & ${\bf 2.164e+02^{+6.347e+01}_{-1.068e+02}}$ & $0.000e+00^{+3.920e+01}_{-0.000e+00}$ & ${\bf 6.243e+02^{+3.110e+01}_{-2.640e+01}}$ & $1.629e+02^{+4.262e+02}_{-1.629e+02}$ & $8.271e+02^{+9.748e+02}_{-8.271e+02}$ & 0.81 \\
C2003T4-20051122 & 3.4 & 23.8 & 0.8 & 2.857 & $0.000e+00^{+0.000e+00}_{-0.000e+00}$ & ${\bf 4.913e+02^{+7.849e+01}_{-7.805e+01}}$ & ${\bf 2.057e+03^{+5.366e+01}_{-5.325e+01}}$ & ${\bf 1.450e+03^{+3.605e+02}_{-3.559e+02}}$ & ${\bf 1.423e+03^{+5.538e+02}_{-5.735e+02}}$ & 5.04 \\
C2003T4-20060129 & 3.3 & 29.7 & 1.0 & 2.857 & ${\bf 6.020e+02^{+2.255e+02}_{-2.493e+02}}$ & $0.000e+00^{+0.000e+00}_{-0.000e+00}$ & ${\bf 1.303e+03^{+1.346e+02}_{-1.225e+02}}$ & $3.139e+02^{+6.151e+02}_{-3.139e+02}$ & ${\bf 4.487e+03^{+1.215e+03}_{-1.175e+03}}$ & 3.54 \\
C2004B1-20060726 & 3.3 & 23.1 & 0.8 & 3.0 & ${\bf 5.372e+02^{+3.374e+01}_{-4.588e+01}}$ & $0.000e+00^{+0.000e+00}_{-0.000e+00}$ & ${\bf 3.306e+02^{+2.541e+01}_{-2.027e+01}}$ & $2.460e+00^{+7.997e+01}_{-2.460e+00}$ & $3.747e+01^{+1.226e+02}_{-3.747e+01}$ & 1.39 \\
C2004B1-20070308 & 3.3 & 33.0 & 1.1 & 2.857 & $0.000e+00^{+2.127e+02}_{-0.000e+00}$ & $6.743e+00^{+1.498e+02}_{-6.743e+00}$ & ${\bf 1.533e+03^{+2.528e+01}_{-1.530e+02}}$ & $0.000e+00^{+0.000e+00}_{-0.000e+00}$ & $1.919e+03^{+1.739e+03}_{-1.919e+03}$ & 1.26 \\
C2006P1-20070504 & 3.9 & 35.1 & 1.0 & 2.727 & ${\bf 2.313e+03^{+2.109e+02}_{-2.826e+02}}$ & $0.000e+00^{+1.375e+02}_{-0.000e+00}$ & ${\bf 8.186e+03^{+6.031e+01}_{-5.916e+01}}$ & $1.173e+03^{+1.539e+03}_{-1.173e+03}$ & ${\bf 3.946e+03^{+2.773e+03}_{-2.811e+03}}$ & 0.97 \\
C2006P1-20070802 & 3.3 & 26.4 & 0.9 & 3.0 & ${\bf 2.762e+03^{+1.179e+03}_{-1.203e+03}}$ & $6.453e+02^{+6.356e+02}_{-6.453e+02}$ & ${\bf 1.954e+03^{+1.971e+02}_{-1.952e+02}}$ & ${\bf 1.911e+03^{+9.285e+02}_{-9.537e+02}}$ & ${\bf 1.524e+03^{+1.436e+03}_{-1.509e+03}}$ & 2.08 \\
C2006P1-20070906 & 3.3 & 36.3 & 1.2 & 3.0 & $0.000e+00^{+5.046e+02}_{-0.000e+00}$ & ${\bf 1.657e+03^{+1.549e+01}_{-3.393e+02}}$ & $0.000e+00^{+0.000e+00}_{-0.000e+00}$ & $0.000e+00^{+8.857e+02}_{-0.000e+00}$ & $1.466e+03^{+1.972e+03}_{-1.466e+03}$ & 8.44 \\
C2006Q1-20080117 & 3.3 & 26.4 & 0.9 & 2.857 & $7.906e+02^{+1.990e+02}_{-7.906e+02}$ & $0.000e+00^{+4.059e+02}_{-0.000e+00}$ & ${\bf 2.943e+03^{+2.117e+02}_{-1.175e+02}}$ & $3.815e+02^{+7.693e+02}_{-3.815e+02}$ & $4.240e+02^{+1.229e+03}_{-4.240e+02}$ & 1.08 \\
C2006Q1-20080702 & 3.3 & 23.1 & 0.8 & 3.0 & ${\bf 1.628e+03^{+5.376e+02}_{-5.509e+02}}$ & ${\bf 3.930e+02^{+3.314e+02}_{-3.305e+02}}$ & ${\bf 2.471e+03^{+7.833e+01}_{-7.808e+01}}$ & ${\bf 6.524e+02^{+4.946e+02}_{-4.928e+02}}$ & ${\bf 1.331e+03^{+7.770e+02}_{-7.930e+02}}$ & 1.16 \\
C2008T2 & 3.9 & 27.3 & 0.8 & 2.727 & ${\bf 8.959e+01^{+1.729e+01}_{-3.597e+01}}$ & $2.387e+00^{+1.866e+01}_{-2.387e+00}$ & ${\bf 6.202e+02^{+6.374e+00}_{-5.568e+00}}$ & $1.885e+01^{+4.254e+01}_{-1.885e+01}$ & $5.143e+01^{+8.455e+01}_{-5.143e+01}$ & 0.79 \\
\enddata
\tablecomments{\, Bold values indicate values constrained to a confidence level of 95\%.}
\tablenotetext{a}{ \ Derived parameter.}
\end{deluxetable*}
\end{longrotatetable}

%% file: table_Appendix_A3.tex
%
\startlongtable
\begin{deluxetable*}{@{\extracolsep{0pt}}lcccccccc}
\tablenum{A3}
\setlength{\tabcolsep}{2pt} 
\tabletypesize{\footnotesize}
\tablewidth{0pc}
\tablecaption{Comets Observed with SLLL and Only Modeled Over (SLLL-SL): Mass of Submicron Grains \label{tab:tab3_appendix_minmass}}
\tablehead{
 & & \multicolumn{5}{c}{\underbar{Mass Relative to Total Mass}} & & \\
\colhead{Comet} & \colhead{Total Mass} & \colhead{Amorphous} & \colhead{Amorphous} & \colhead{Amorphous} & \colhead{Crystalline} & \colhead{Crystalline} & \colhead{Silicate/Carbon} & \\
\colhead{Name} & \colhead{($\times 10^{7}$ kg)} & \colhead{Pyroxene} & \colhead{Olivine} & \colhead{Carbon} & \colhead{Olivine} & 
\colhead{Orthopyroxene} & \colhead{Ratio} &  \colhead{$f_{cryst}$\tablenotemark{a}}
}
\startdata
8P & $3.414e+04^{+3.094e+03}_{-3.414e+04}$ & $0.014^{+0.076}_{-0.014}$ & $0.000^{+0.046}_{-0.000}$ & $0.638^{+0.093}_{-0.057}$ & $0.077^{+0.057}_{-0.067}$ & $0.271^{+0.072}_{-0.134}$ & $0.568^{+0.154}_{-0.200}$ & $0.962^{+0.038}_{-0.255}$ \\
9P-20050702 & $5.974e+04^{+8.005e+03}_{-5.974e+04}$ & $0.409^{+0.081}_{-0.081}$ & $0.007^{+0.057}_{-0.007}$ & $0.372^{+0.071}_{-0.044}$ & $0.091^{+0.063}_{-0.082}$ & $0.121^{+0.111}_{-0.121}$ & $1.685^{+0.358}_{-0.432}$ & $0.337^{+0.149}_{-0.240}$ \\
9P-20050703 & $5.470e+04^{+9.963e+03}_{-5.470e+04}$ & $0.388^{+0.102}_{-0.086}$ & $0.000^{+0.026}_{-0.000}$ & $0.373^{+0.075}_{-0.057}$ & $0.081^{+0.082}_{-0.081}$ & $0.158^{+0.134}_{-0.158}$ & $1.682^{+0.487}_{-0.450}$ & $0.381^{+0.171}_{-0.265}$ \\
17P & $7.719e+06^{+3.457e+05}_{-7.719e+06}$ & $0.611^{+0.033}_{-0.032}$ & $0.158^{+0.030}_{-0.028}$ & $0.071^{+0.005}_{-0.004}$ & $0.124^{+0.016}_{-0.017}$ & $0.036^{+0.033}_{-0.036}$ & $13.072^{+0.900}_{-0.883}$ & $0.173^{+0.038}_{-0.040}$ \\
21P & $1.896e+04^{+2.460e+03}_{-1.896e+04}$ & $0.514^{+0.039}_{-0.099}$ & $0.000^{+0.032}_{-0.000}$ & $0.357^{+0.029}_{-0.040}$ & $0.129^{+0.055}_{-0.068}$ & $0.000^{+0.124}_{-0.000}$ & $1.800^{+0.349}_{-0.210}$ & $0.200^{+0.179}_{-0.091}$ \\
29P & $1.621e+07^{+1.971e+06}_{-1.621e+07}$ & $0.585^{+0.157}_{-0.165}$ & $0.000^{+0.093}_{-0.000}$ & $0.055^{+0.017}_{-0.007}$ & $0.125^{+0.063}_{-0.068}$ & $0.235^{+0.147}_{-0.186}$ & $17.071^{+2.720}_{-4.238}$ & $0.381^{+0.147}_{-0.191}$ \\
37P & $6.372e+03^{+7.870e+02}_{-6.372e+03}$ & $0.498^{+0.078}_{-0.082}$ & $0.015^{+0.048}_{-0.015}$ & $0.299^{+0.044}_{-0.032}$ & $0.107^{+0.054}_{-0.064}$ & $0.081^{+0.099}_{-0.081}$ & $2.350^{+0.408}_{-0.430}$ & $0.269^{+0.133}_{-0.161}$ \\
46P-20080424 & $2.055e+03^{+7.159e+02}_{-2.055e+03}$ & $0.133^{+0.075}_{-0.062}$ & $0.049^{+0.061}_{-0.041}$ & $0.488^{+0.192}_{-0.126}$ & $0.118^{+0.112}_{-0.118}$ & $0.212^{+0.173}_{-0.212}$ & $1.049^{+0.710}_{-0.578}$ & $0.645^{+0.161}_{-0.460}$ \\
46P-20080524 & $2.701e+03^{+1.079e+03}_{-2.701e+03}$ & $0.098^{+0.072}_{-0.037}$ & $0.000^{+0.000}_{-0.000}$ & $0.303^{+0.171}_{-0.086}$ & $0.000^{+0.127}_{-0.000}$ & $0.599^{+0.112}_{-0.261}$ & $2.302^{+1.311}_{-1.190}$ & $0.860^{+0.062}_{-0.182}$ \\
46P-20080702 & $3.644e+03^{+4.239e+02}_{-3.644e+03}$ & $0.321^{+0.081}_{-0.204}$ & $0.000^{+0.125}_{-0.000}$ & $0.561^{+0.117}_{-0.058}$ & $0.000^{+0.020}_{-0.000}$ & $0.118^{+0.122}_{-0.118}$ & $0.782^{+0.207}_{-0.308}$ & $0.269^{+0.246}_{-0.269}$ \\
48P & $5.693e+04^{+1.488e+04}_{-5.693e+04}$ & $0.350^{+0.164}_{-0.155}$ & $0.028^{+0.085}_{-0.028}$ & $0.249^{+0.083}_{-0.049}$ & $0.179^{+0.088}_{-0.108}$ & $0.193^{+0.182}_{-0.193}$ & $3.024^{+0.996}_{-1.008}$ & $0.496^{+0.190}_{-0.282}$ \\
65P & $2.192e+04^{+4.661e+03}_{-2.192e+04}$ & $0.917^{+0.021}_{-0.196}$ & $0.000^{+0.000}_{-0.000}$ & $0.083^{+0.025}_{-0.023}$ & $0.000^{+0.000}_{-0.000}$ & $0.000^{+0.198}_{-0.000}$ & $10.986^{+4.598}_{-2.754}$ & $0.000^{+0.215}_{-0.000}$ \\
67P-20081128.38 & $2.838e+03^{+2.275e+03}_{-2.838e+03}$ & $0.573^{+0.014}_{-0.270}$ & $0.000^{+0.000}_{-0.000}$ & $0.427^{+0.014}_{-0.189}$ & $0.000^{+0.210}_{-0.000}$ & $0.000^{+0.441}_{-0.000}$ & $1.341^{+1.860}_{-0.076}$ & $0.000^{+0.602}_{-0.000}$ \\
67P-20081129.11 & $3.572e+03^{+2.926e+03}_{-3.572e+03}$ & $0.305^{+0.223}_{-0.155}$ & $0.000^{+0.028}_{-0.000}$ & $0.284^{+0.193}_{-0.127}$ & $0.160^{+0.251}_{-0.160}$ & $0.250^{+0.335}_{-0.250}$ & $2.518^{+2.841}_{-1.421}$ & $0.573^{+0.246}_{-0.573}$ \\
67P-20081129.70 & $2.705e+03^{+2.864e+03}_{-2.705e+03}$ & $0.335^{+0.224}_{-0.183}$ & $0.000^{+0.000}_{-0.000}$ & $0.279^{+0.177}_{-0.142}$ & $0.000^{+0.315}_{-0.000}$ & $0.386^{+0.296}_{-0.386}$ & $2.590^{+3.757}_{-1.393}$ & $0.536^{+0.288}_{-0.536}$ \\
67P-20081130.48 & $2.632e+03^{+2.981e+03}_{-2.632e+03}$ & $0.436^{+0.092}_{-0.247}$ & $0.000^{+0.000}_{-0.000}$ & $0.410^{+0.082}_{-0.217}$ & $0.048^{+0.319}_{-0.048}$ & $0.106^{+0.440}_{-0.106}$ & $1.440^{+2.746}_{-0.405}$ & $0.261^{+0.504}_{-0.261}$ \\
71P & $2.882e+04^{+7.508e+03}_{-2.882e+04}$ & $0.134^{+0.068}_{-0.059}$ & $0.035^{+0.048}_{-0.035}$ & $0.502^{+0.162}_{-0.104}$ & $0.075^{+0.093}_{-0.075}$ & $0.254^{+0.134}_{-0.208}$ & $0.992^{+0.520}_{-0.485}$ & $0.661^{+0.137}_{-0.360}$ \\
73P(B)-20060806.18 & $3.189e+03^{+1.691e+02}_{-3.189e+03}$ & $0.468^{+0.030}_{-0.035}$ & $0.000^{+0.000}_{-0.000}$ & $0.474^{+0.021}_{-0.024}$ & $0.058^{+0.043}_{-0.048}$ & $0.000^{+0.043}_{-0.000}$ & $1.110^{+0.112}_{-0.088}$ & $0.111^{+0.096}_{-0.088}$ \\
78P & $7.107e+04^{+2.704e+04}_{-7.107e+04}$ & $0.226^{+0.157}_{-0.072}$ & $0.000^{+0.006}_{-0.000}$ & $0.184^{+0.112}_{-0.050}$ & $0.174^{+0.140}_{-0.171}$ & $0.416^{+0.163}_{-0.295}$ & $4.424^{+2.019}_{-2.049}$ & $0.723^{+0.099}_{-0.270}$ \\
88P & $2.427e+04^{+2.946e+03}_{-2.427e+04}$ & $0.369^{+0.071}_{-0.095}$ & $0.000^{+0.063}_{-0.000}$ & $0.317^{+0.052}_{-0.033}$ & $0.027^{+0.072}_{-0.027}$ & $0.286^{+0.081}_{-0.108}$ & $2.154^{+0.365}_{-0.446}$ & $0.459^{+0.113}_{-0.144}$ \\
132P & $1.865e+03^{+2.639e+03}_{-1.865e+03}$ & $0.039^{+0.598}_{-0.039}$ & $0.004^{+0.101}_{-0.004}$ & $0.027^{+0.313}_{-0.015}$ & $0.000^{+0.540}_{-0.000}$ & $0.931^{+0.044}_{-0.931}$ & $36.293^{+48.777}_{-34.353}$ & $0.956^{+0.035}_{-0.956}$ \\
C2003K4-20050913 & $1.236e+05^{+3.623e+04}_{-1.236e+05}$ & $0.790^{+0.020}_{-0.190}$ & $0.000^{+0.000}_{-0.000}$ & $0.210^{+0.023}_{-0.047}$ & $0.000^{+0.223}_{-0.000}$ & $0.000^{+0.040}_{-0.000}$ & $3.767^{+1.379}_{-0.464}$ & $0.000^{+0.280}_{-0.000}$ \\
C2003T3 & $3.256e+04^{+2.996e+04}_{-3.256e+04}$ & $0.095^{+0.288}_{-0.067}$ & $0.000^{+0.019}_{-0.000}$ & $0.125^{+0.277}_{-0.057}$ & $0.128^{+0.277}_{-0.128}$ & $0.652^{+0.184}_{-0.652}$ & $7.012^{+6.837}_{-5.523}$ & $0.891^{+0.075}_{-0.540}$ \\
C2003T4-20051122 & $1.231e+05^{+2.142e+04}_{-1.231e+05}$ & $0.000^{+0.000}_{-0.000}$ & $0.093^{+0.036}_{-0.025}$ & $0.176^{+0.036}_{-0.025}$ & $0.369^{+0.072}_{-0.067}$ & $0.362^{+0.082}_{-0.108}$ & $4.670^{+0.921}_{-0.968}$ & $0.887^{+0.032}_{-0.050}$ \\
C2003T4-20060129 & $1.630e+05^{+4.014e+04}_{-1.630e+05}$ & $0.078^{+0.054}_{-0.040}$ & $0.000^{+0.000}_{-0.000}$ & $0.077^{+0.017}_{-0.012}$ & $0.055^{+0.092}_{-0.055}$ & $0.790^{+0.066}_{-0.088}$ & $11.991^{+2.384}_{-2.312}$ & $0.915^{+0.044}_{-0.060}$ \\
C2004B1-20060726 & $2.282e+04^{+4.205e+03}_{-2.282e+04}$ & $0.739^{+0.057}_{-0.147}$ & $0.000^{+0.000}_{-0.000}$ & $0.207^{+0.023}_{-0.031}$ & $0.003^{+0.100}_{-0.003}$ & $0.052^{+0.137}_{-0.052}$ & $3.840^{+0.867}_{-0.483}$ & $0.069^{+0.205}_{-0.069}$ \\
C2004B1-20070308 & $6.364e+04^{+4.541e+04}_{-6.364e+04}$ & $0.000^{+0.136}_{-0.000}$ & $0.002^{+0.158}_{-0.002}$ & $0.211^{+0.724}_{-0.089}$ & $0.000^{+0.000}_{-0.000}$ & $0.787^{+0.090}_{-0.787}$ & $3.746^{+3.490}_{-3.677}$ & $0.997^{+ nan}_{ nan}$ \\
C2006P1-20070504 & $2.383e+05^{+8.932e+04}_{-2.383e+05}$ & $0.152^{+0.103}_{-0.051}$ & $0.000^{+0.010}_{-0.000}$ & $0.245^{+0.143}_{-0.066}$ & $0.138^{+0.150}_{-0.138}$ & $0.464^{+0.155}_{-0.269}$ & $3.079^{+1.517}_{-1.502}$ & $0.798^{+0.077}_{-0.219}$ \\
C2006P1-20070802 & $2.364e+05^{+6.120e+04}_{-2.364e+05}$ & $0.357^{+0.130}_{-0.130}$ & $0.083^{+0.123}_{-0.083}$ & $0.115^{+0.046}_{-0.028}$ & $0.247^{+0.100}_{-0.106}$ & $0.197^{+0.126}_{-0.195}$ & $7.703^{+2.826}_{-2.495}$ & $0.502^{+0.106}_{-0.171}$ \\
C2006P1-20070906 & $7.249e+04^{+5.180e+04}_{-7.249e+04}$ & $0.000^{+0.119}_{-0.000}$ & $0.531^{+0.427}_{-0.267}$ & $0.000^{+0.000}_{-0.000}$ & $0.000^{+0.241}_{-0.000}$ & $0.469^{+0.194}_{-0.469}$ & $ inf^{+ nan}_{ nan}$ & $0.469^{+0.210}_{-0.454}$ \\
C2006Q1-20080117 & $7.276e+04^{+4.321e+04}_{-7.276e+04}$ & $0.246^{+0.158}_{-0.246}$ & $0.000^{+0.173}_{-0.000}$ & $0.416^{+0.249}_{-0.145}$ & $0.160^{+0.223}_{-0.160}$ & $0.178^{+0.300}_{-0.178}$ & $1.405^{+1.283}_{-0.901}$ & $0.579^{+0.315}_{-0.579}$ \\
C2006Q1-20080702 & $1.609e+05^{+3.166e+04}_{-1.609e+05}$ & $0.317^{+0.090}_{-0.090}$ & $0.077^{+0.091}_{-0.066}$ & $0.219^{+0.058}_{-0.037}$ & $0.127^{+0.080}_{-0.092}$ & $0.260^{+0.098}_{-0.132}$ & $3.565^{+0.942}_{-0.961}$ & $0.495^{+0.098}_{-0.160}$ \\
C2008T2 & $8.726e+03^{+2.809e+03}_{-8.726e+03}$ & $0.180^{+0.082}_{-0.081}$ & $0.005^{+0.045}_{-0.005}$ & $0.566^{+0.179}_{-0.136}$ & $0.067^{+0.125}_{-0.067}$ & $0.183^{+0.197}_{-0.183}$ & $0.768^{+0.558}_{-0.425}$ & $0.575^{+0.220}_{-0.575}$ \\
\enddata
\tablenotetext{a}{Defined as $f_{cryst}^{silicates} \equiv$ (crystalline)/(crystalline + amorphous).}
\end{deluxetable*}